\documentclass[aps,prd,nofootinbib,onecolumn,superscriptaddress,preprintnumbers,letterpaper,longbibliography]{revtex4-1}
\usepackage{amssymb}
\usepackage{amsmath}
\usepackage{epsfig}
\usepackage{hyperref}
\usepackage{comment}
\usepackage{color}
\usepackage{cleveref}
\usepackage{multirow}
\usepackage{capt-of}
\usepackage{graphicx}
\usepackage{gensymb}
\usepackage[caption=false]{subfig}
\usepackage{slashed}
\usepackage{tikz-feynman}
\usepackage{tabularx}
\usepackage{aas_macros}

\makeatletter
\def\simgt{\mathrel{\lower2.5pt\vbox{\lineskip=0pt\baselineskip=0pt
           \hbox{$>$}\hbox{$\sim$}}}}
\def\simlt{\mathrel{\lower2.5pt\vbox{\lineskip=0pt\baselineskip=0pt
           \hbox{$<$}\hbox{$\sim$}}}}
\makeatother

\newcommand{\be}{\begin{equation}}
\newcommand{\ee}{\end{equation}}
\newcommand{\bea}{\begin{eqnarray}}
\newcommand{\nn}{\nonumber}
\newcommand{\eea}{\end{eqnarray}}
\def\sfrac#1#2{{\textstyle\frac{#1}{#2}}}
\def\R{{\sss R}}
\def\L{{\sss L}}
\def\sss{\scriptscriptstyle}

\begin{document}
\title{TASI Lectures on Early Universe Cosmology:\\
Inflation, Baryogenesis and Dark Matter}
\author{James M.\ Cline}
\affiliation{Dept.\ of Physics, McGill University, Montr\'eal,
Qu\'ebec, Canada}
\begin{abstract}
These lectures, presented at TASI 2018, provide a concise introduction
to inflation, baryogenesis, and aspects of dark matter not covered by
the other lectures.  The emphasis for inflation is an intuitive
understanding and techniques for constraining inflationary models.
For baryogenesis we focus on two examples, leptogenesis and electroweak
baryogenesis, with attention to singlet-assisted two-step phase 
transitions.   Concerning dark matter, we review different 
classes of models distinguished by their mechanisms for
obtaining the observed relic density, including thermal freeze-out,
asymmetric dark matter, freeze-in, SIMP dark matter, 
the misalignment mechanism for ultralight scalars and axions, and
production of primordial black holes during inflation.  Problem sets
are provided.

\end{abstract}

\maketitle

\tableofcontents

\newpage

\section{Introduction}
\begin{figure}
\centerline{\includegraphics[width=10cm]{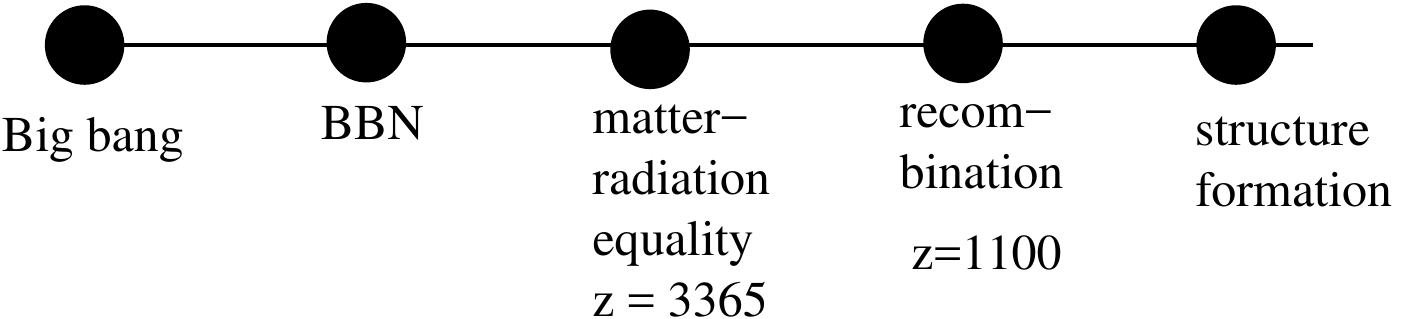}}
\caption{The cosmological timeline, {\it ca.} 1970.}
\label{fig:timeline1}
\end{figure}

\begin{figure}
\centerline{\includegraphics[width=15cm]{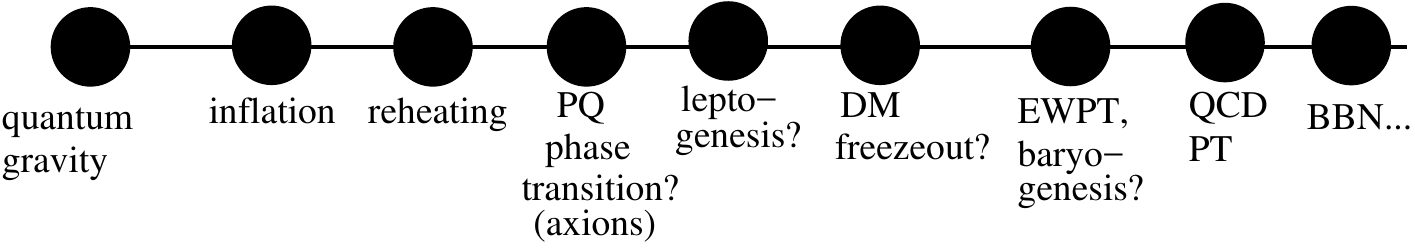}}
\caption{The cosmological timeline, {\it ca.} 2018.}
\label{fig:timeline2}
\end{figure}

For these lectures I was assigned the topic of ``Early Universe
Cosmology.''  If we could go back in time fifty years, this would 
seem like a more straightforward task, since the cosmological timeline
was relatively uncrowded by notable events; see 
fig.\ \ref{fig:timeline1}.
Following the big bang, there was nucleosynthesis (BBN),
matter-radiation equality, recombination, and formation of galactic
structure.  But our current understanding intersperses many more
likely or at least possible events of significance (fig.\
\ref{fig:timeline2}), replacing the big bang by inflation, introducing
leptogenesis or baryogenesis, several cosmological phase transitions, and some
kind of origin for the dark matter (DM) of the universe.  For these
lectures I have therefore chosen to discuss inflation, baryogenesis,
and aspects of dark matter not covered by other lecturers.  P.\ Fox
has included a nice introduction to BBN as well as thermal freeze-out
of dark matter in his lectures on SUSY WIMPs, while direct and
indirect detection of DM are treated respectively by T.\ Lin and
D.\ Hooper, and axions by A.\ Hook.  Structure formation will be 
introduced by M.\ Vogelsberger.  Hence I hope that the more general
picture of early universe cosmology will get a comprehensive treatment through
our combined lectures.


\subsection{Conventions and basics of big bang cosmology}

I will be using natural units,
\be
\hbar = c = k_B = 1,\qquad G = {1\over M_p^2}, \qquad 8\pi G = {1\over
m_p^2}
\ee
The reader should be warned that my choice of upper and lower case for
the unreduced Planck mass $M_p = 1.22\times
10^{19}\,$GeV and the reduced one, $m_p = 2.43\times 10^{18}\,$GeV, is not a standard
convention, even though it seems logical.  The Einstein equations then
read
\be
	G_{\mu\nu} = {1\over m_p^2} T_{\mu\nu}
	\quad \longleftrightarrow \quad 
	R_{\mu\nu} = {1\over m_p^2}
	\left(T_{\mu\nu} -\sfrac12g_{\mu\nu}T\right)
\ee
with
\be
	g_{\mu\nu} = \left(\begin{array}{cccc}1 & & &\\
	                                     & -a^2 & & \\
					     & & -a^2 & \\
					     & & & -a^2
	\end{array}\right),\qquad
	T_{\mu\nu} = \left(\begin{array}{cccc}\rho & & &\\
	                                     & p & & \\
					     & & p & \\
					     & & & p
	\end{array}\right)
\ee

Basic elements of cosmology are summarized in the PDG reviews
\cite{Patrignani:2016xqp}
or the textbook of Kolb and Turner \cite{Kolb:1990vq}; I recapitulate them here for convenience.  
The Friedmann-Robertson-Walker (FRW) line element is
\be
	ds^2 = dt^2 - a^2(t)\, dx^2
\ee
where $dx^2$ represents a unit 3D metric that can have curvature
$K = 0,\pm 1$.  Redshift is defined by $1+z = a_0/a$, where $a_0$ is
the present value of the scale factor, while the Hubble parameter is
\be
	H = {\dot a\over a}\qquad \left(\dot H = {\ddot a\over
a}-H^2\right)
\ee
Then the (00) and $(ij)$ components of the Einstein equations can
be written respectively as
\bea
	H^2 &=& {\rho\over 3 m_p^2} - {k\over a^2}
	2{\ddot a\over a}  + H^2 = -{p\over m_p^2}- {k\over a^2},\\
\eea
where $k$ has units of 1/(distance)$^2$ if $a$ is taken to be
dimensionless.  The first of these is the usual
Friedmann equation that together with the equation of state
fixing $\rho$ as a function of $a$ 
determines the evolution of a homogeneous universe.  Although the
present universe does not look very homogeneous at first glance,
the approximation starts to be valid when averaging over scales
$\gtrsim 70$\,Mpc \cite{Davis:1996sa}.  More importantly for these
lectures, the cosmic microwave background (CMB) shows that the
universe was very homogeneous, to a part in 20,000, at $z=1100$.

To complete the Friedmann equation we need the time- or $a$-dependence
of the energy density $\rho$, which is determined by the equation of 
state (EOS),
\be
	p = w\rho,\qquad w = \left\{\begin{array}{rl}\sfrac13,&
	\hbox{radiation}\\
	0,& \hbox{matter}\\
	-1,& \hbox{vacuum energy}
	\end{array}\right.
\label{eos}
\ee
The conservation of stress-energy, $\partial_\mu
T^{\mu\nu} = 0$, implies that $\dot\rho = -3H(\rho+p) = -3H\rho(1+w)$,
and combining this with (\ref{eos}) gives
\be
	\rho \sim \left\{\begin{array}{rl}{1/
a^4},&\hbox{radiation}\\
	{1/a^3},&\hbox{matter}\\
	\hbox{const.},& \hbox{vacuum energy}\end{array}\right.
\ee
Taking $a_0 = 1$, we can integrate the Friedmann equation,
\be
	\int dt = \pm\sqrt{3}\,m_p \int da\,\left({\rho_{r,0}\over a^2}
	+ {\rho_{m,0}\over a} + \rho_\Lambda a^2 -
\rho_k\right)^{-1/2}
\label{feq}
\ee
where $\rho_k$ is a fictitious energy density going as $K m_p^2
/R_0^2$, with $K=0,\pm 1$ and $R_0$ being a physical length scale for
the 3D curvature.  $\rho_{x,0}$ denotes the present density of
matter ($x=m$) or radiation ($x=r$).  Although eq.\ (\ref{feq})
cannot be usefully integrated in closed form in the general case, 
it is easy to do so when any of the individual terms dominate,
in particular
\be
	a\sim \left\{\begin{array}{rl} t^{1/2},& \hbox{radiation}\\
	t^{2/3},& \hbox{matter}\\
	\exp(t\sqrt{\rho_\Lambda/3}/m_p),& \hbox{vacuum energy}
	\end{array}\right.
\ee

Introducing the critical density $\rho_c = 3 m_p^2 H_0^2 \cong
(2.47\,{\rm meV})^4$, we can define the fractional contributions of
the various components to the total energy density of the universe,
\be
	\Omega_i = {\rho_i\over \rho_c} \cong\left\{
	\begin{array}{rl}5\times 10^{-5},& \gamma\\
	0.05,& \hbox{baryons}\\
	0.26,& \hbox{CDM}\\
	0.69,& \Lambda\\
	< 0.015,& k \hbox{\,(curvature)}\end{array}\right.
\ee
which identically satisfy $\sum_i\Omega_i = 1$ when including $\Omega_k$ in the
sum.  CDM stands for cold dark matter, and the upper limit on $\Omega_k$
applies (roughly) to its absolute value.

\section{Inflation}
Although the cosmological timeline looked simple in 1970, by later 
in the decade an awareness was building that all was not well with
the big bang picture; see for example the essay by Dicke and Peebles
in ref.\ \cite{Hawking:2010mca}.  This was due to the now famous 
horizon and flatness problems.

\begin{figure}
\centerline{\includegraphics[width=0.2\textwidth]{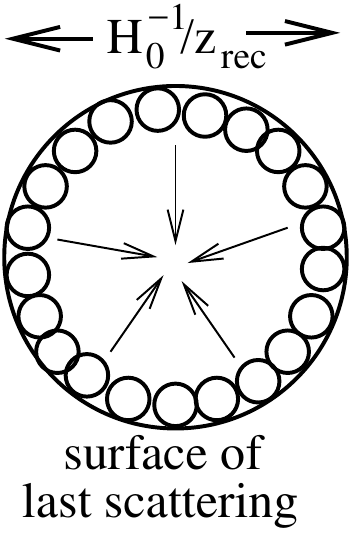}}
\caption{The horizon problem; we are at the center.}
\label{fig:horizon}
\end{figure}

\subsection{Horizon problem}

CMB photons have been free-streaming from the surface of last
scattering, representing the epoch when the universe became
transparent, around the time of electron-proton recombination,
fig.\ \ref{fig:horizon}.  To understand this picture we need the
idea of the {\it particle horizon} $d_H(t)$, which is the distance that
a photon could have traveled by a given time $t$.  Since photons
follow null worldlines,  $dt = a\, dx$, we have
\be
	d_H = a x = a\int {dt\over a} = \left\{\begin{array}{ll}
	2t\sim 2a^2,& \hbox{radiation domination}\\
	3t\sim 3a^{3/2},&\hbox{matter domination}\end{array}\right.
\ee 
$d_H(t)$ is therefore the maximum size of a causally connected region
at time $t$, within which one could expect any degree of uniformity 
due to thermal equilibration.  Our current horizon is $d_H(t_0)\sim
H_0^{-1}\sim 10^{26}$\,m, and the CMB radiation contained within it is
extremely uniform, to better than $10^{-4}$, suggesting that the
region containing the whole presently observable universe was already 
in causal contact at the time of recombination $t_{\rm rec}$.  
Yet if we consider how large this region was at that time, 
$\sim d_H(t_0)/z_{\rm rec}$, by rescaling it according to the Hubble
expansion, it is much larger than $d_H(t_{\rm rec})$.  This is
the conundrum illustrated in fig.\ \ref{fig:horizon}.  The small 
circles represent the largest regions that should have uniform
temperature.  We can estimate the number of them around the big
circle as
\be
	{2\pi d_H(t_0)/z_{\rm rec}\over 2 d_H(t_{\rm rec}}
	= \pi\sqrt{z_{\rm rec}}\cong 100
\ee
so that each one subtends an angle of $\Delta\theta\cong 3.5^\circ$.
How did the temperature come to be so uniform across all of these
regions?

\subsection{Flatness problem}  The inverse curvature radius $R^{-1}$
must be
tuned to an extremely small value in order for the $\rho_k$ term in
(\ref{feq}) to avoid dominating the current expansion of the universe. 
Recalling the relation $|\rho_k| = 3 m_p^2/R_0^2$, we see that
\be
	|\Omega_k| = \left|\rho_k\over\rho_c\right| = {1\over H_0^2
R_0^2} \qquad \Rightarrow\qquad R_0\gtrsim {10\over H_0} \sim
10^{27}\,{\rm m} \sim 30,000\, {\rm Mpc}
\ee
The curvature radius scales simply with the Hubble expansion.  Scaling
back to the Planck time $t_p\sim 1/m_p$, we get
\be
	R_p = R_0 a_p \sim R_0\,{T_{\gamma,0}\over m_p}
\sim {10\over H_0}\,{T_{\gamma}\over m_p}\sim {10 T_\gamma\over
\rho_c^{1/2}} = {2.4\times 10^{-3}\,{\rm eV}\over (2.5\times 10^{-3}\,
{\rm eV})^2 }\sim {1\over 3\times 10^{-3}\,{\rm eV}}
\ee 
This is to be compared to the natural value $R_p\sim 1/m_p$, since
$m_p$ is the only relevant dimensionful parameter.  Thus we see that a
tuning of one part in $10^{31}$ is required for the initial curvature
radius, if we are allowed to extrapolate back as far as $m_p$.  Limiting the earliest time to larger values only gives a modest
improvement unless we push that time all the way into the present.

\subsection{A little history}
In 1979, the hot topic was grand unified theories (GUTs). In the previous
year, Zeldovich and Khlopov  \cite{Zeldovich:1978wj} had estimated
that if the universe ever went through the GUT symmetry breaking
transition, then pointlike topological defects, magnetic mononpoles,
with masses of $M_{\rm GUT}\sim 10^{16}\,$GeV would have been so
copiously produced that the universe should have recollapsed shortly
thereafter.  J.\ Preskill, at that time a graduate student at Harvard,
corrected an important overestimate in their calculation but
nevertheless confirmed their conclusion 
\cite{Preskill:1979zi}.\footnote{It required some conviction
on his part, since his supervisor
S.\ Weinberg reportedly told him he was ``crazy'' to work on that problem.}
At that time A.\ Guth was a postdoc at Cornell, and with H.\ Tye
proposed some ideas for suppressing the density of monopoles 
\cite{Guth:1979bh}.  He then moved to SLAC, taking up his fourth
postdoctoral position, still thinking about the monopole problem.
It led him to propose inflation \cite{Guth:1980zm}, which he
recognized did much more than solve the monopole problem; it also
solved the horizon and flatness problems.  Soon thereafter, people
realized that it additionally explained the origin of density
perturbations leading to large scale structure.

\begin{figure}
\centerline{\includegraphics[width=0.3\textwidth]{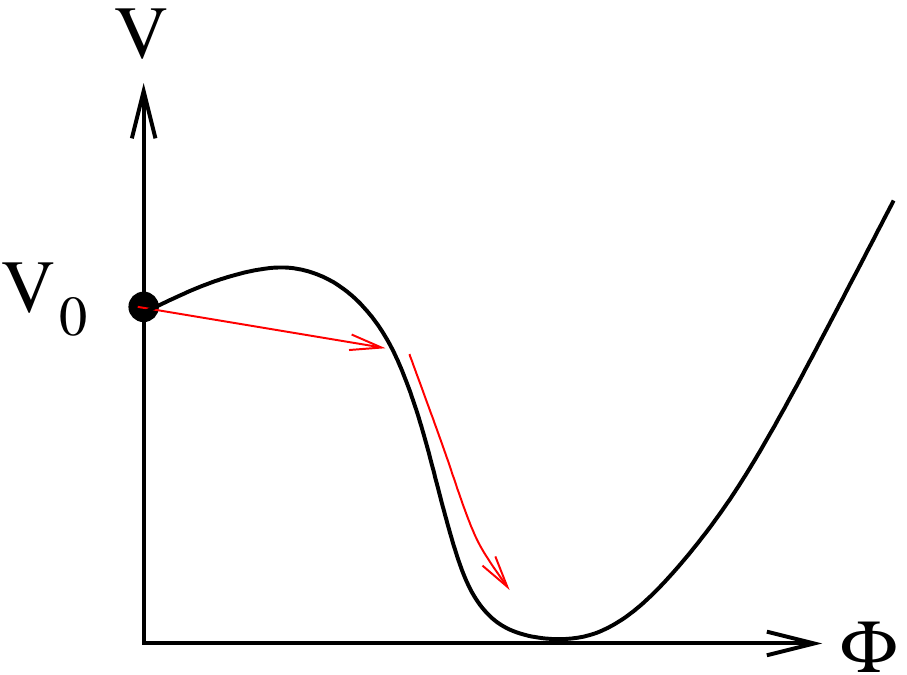}}
\caption{The SU(5) GUT potential near the critical temperature,
leading to a first order phase transition.}
\label{fig:su5}
\end{figure}

Guth's original idea, now called ``old inflation,'' was inspired by
the SU(5) GUT, which requires a Higgs field $\varphi$ to break SU(5) down
to the standard model gauge group.  At high temperatures, thermal
effects make $m_\varphi^2>0$ at the origin $\varphi=0$, but as $T$
decreases, a local minimum develops at $\varphi\neq 0$ as shown in fig.\
\ref{fig:su5}.  Initially $\varphi$ is trapped in the false minimum,
whose vacuum energy causes the universe to expand as
\be
	a \sim \exp(t\sqrt{V_0/3}/m_p)
\ee
Inflation ends when $\varphi$ tunnels through the barrier and rolls
down to the true minimum.  But as Guth realized even in the seminal
paper, this picture is flawed because the tunneling leads to
nucleation of bubbles of true vacuum that are cold and empty.
These universes could be heated up by the energy released by
collisions of walls from neighboring bubbles, but first of all these
collisions are exceedingly rare---the phase transition never
completes---and second, the resulting radiation would be quite
inhomogeneous, undoing all the smoothing of initial inhomogeneities
that occurred during inflation.  A further problem, seemingly not
noticed at the time, is that bubbles nucleated in this way have very 
large negative curvature \cite{Brown:1987dd}, undoing the solution of the flatness
problem afforded by the initial inflationary expansion.  However it
was soon realized that the potential in the post-nucleation phase could be
made sufficiently flat so that inflation would still take place, with
no need for the prior false-vacuum phase 
\cite{Linde:1981mu,Albrecht:1982wi}, except perhaps to justify the
initial condition by tunneling.  This was dubbed ``new
inflation.''

It is sometimes noted that the first model of inflation was published
by Starobinsky \cite{Starobinsky:1980te} before all of these 
developments, based upon an $R^2$ addition to the Einstein-Hilbert
gravitational action.  This is particularly interesting now because of
the preference given to this model for fitting current CMB data as observed by
Planck \cite{Ade:2015lrj}.  However Starobinsky was not aware at that
time that inflation was a general mechanism that could solve the
problems of big bang cosmology  as emphasized by Guth.

\subsection{Inflation in brief}

We now recapitulate the main features of inflation, settling for a
heuristic rather than a rigorous approach.  I will fill in some
details in the following subsection.  Inflation can be driven by any
scalar field $\varphi$ whose potential is sufficiently flat, as measured by the
potential slow roll parameters
\be
	\epsilon = {m_p^2\over 2}\left(V'\over V\right)^2,\quad
	\eta = m_p^2\, {V''\over V}
\ee
This parametrization is perhaps getting outdated since many
practitioners prefer the Hubble flow functions
\bea
	\epsilon_1 &=& -{\dot H\over H^2} \cong \epsilon \nonumber\\
	\epsilon_2 &=& {\dot\epsilon_1\over H\epsilon_1} \cong
	4\epsilon - 2\eta\nn\\
	\epsilon_3 &=& {\dot\epsilon_2\over H\epsilon_2} \dots
\eea
but for leading-order calculations (involving only $\epsilon$ and
$\eta$) either is sufficient.  When $\epsilon,\eta\ll 1$, we can
ignore the $\ddot\varphi$ term in the equation of motion (EOM) for the
homogeneous mode of $\varphi$, approximating it by the slow-roll EOM,
\be
	3H\dot\varphi \cong -V'(\varphi)
\label{sreq}
\ee
The Hubble damping term comes from varying the scalar field action in 
the presence of the background metric, 
$S = \int d^{\,4}x\, a^3( a^{-2}\dot\varphi^2/2 - V)$.  The kinetic
energy of $\varphi$ is then much smaller than $V$, and the Friedmann
equation can be integrated to find
\be
	a \sim \exp\left(\int H dt\right) \equiv e^{N}
\ee
with $H = \sqrt{V/3m_p^2}$ and $N$ being the number of e-foldings.  Eventually, as $\varphi$ reaches the minimum
of its potential, either $\epsilon$ or $\eta$ will exceed unity, and
$\ddot\varphi$ can no longer be ignored; instead $H\dot\varphi$ becomes
negligible,
\be
		\ddot\varphi \cong -V'(\varphi)
\ee
and $\varphi$ oscillates around the minimum of $V$.  These oscillations
lead to particle production and reheating of the universe.

A naive estimate of the reheating temperature is
\be
	T_{\rm rh}\sim \sqrt{m_p\Gamma_\varphi}
\label{Trh}
\ee
where $\Gamma_{\varphi}$ is the decay rate of the inflaton.  One can
easily derive (\ref{Trh}) from the usual relation between time and
temperature,
\be
	t \sim {1\over \Gamma_\varphi} \sim {1\over H} \sim {m_p\over
\sqrt{\rho}} \sim {m_p\over T_{\rm rh}^2}
\ee
which is valid as long as the computed $T_{\rm rh}$ does not exceed
the available energy scale from inflation, $V_i^{1/4}$ (where $V_i$
is the magnitude of $V$ during inflation).  It may seem
counterintuitive that $T_{\rm rh}$ would be independent of $V_i$, but
this occurs because the particles produced by the early decays are 
diluted by continuing quasi-exponential expansion of the universe;
only those produced near the end of the reheating phase dominate 
the final density.  See problem 2 of the inflation exercises.

Let's review how inflation solves the problems of big bang cosmology.
The flatness problem is solved by the stretching of space by the 
expansion, which increases an initial curvature radius $R_i$ as
\be
	{1\over R_i^2}\to {e^{-2N}\over R_i^2}
\ee
To reduce $1/R$ by a factor of $10^{31}$, as we motivated in our
example, would require $N=70$ e-foldings of inflation.  This is
actually an overestimate since $1/R_i$ must be somewhat smaller than
$m_p$ in order for inflation to get started (at least if the curvature
is positive): $H^2 = V(\varphi)/3 m_p^2 - 1/(a^2 R_i^2)$ must be
positive.  Then if $V_i \equiv \Lambda_i^4$, we require $1/R_i \lesssim
\Lambda_i^2/m_p$.  Its value today would be
\be
	{1\over R_0}\sim {e^{-N}\over R_i}\left(T_0\over T_{\rm
rh}\right)
\ee
which requires
\be
	N > \ln {R_0\, T_0\over R_i\, T_{\rm rh}} >
	\ln {10\, H_0^{-1} T_0 \Lambda_i^2\over m_p\, T_{\rm rh}}
	\sim \ln{10\, T_0 \Lambda_i^2\over\rho_c^{1/2}\, T_{\rm rh}}
\label{Nflat}
\ee
Taking for example $\Lambda_i = 10^{-3}\,m_p$ (we will see that CMB
data provide this as an upper limit) and $T_{\rm rh}\sim \Lambda_i$,
we find
\be
	N \gtrsim\ln {10^{-3}{\,\rm eV} \times 10^{22} {\,\rm eV}\over
	10^{-6} {\,\rm eV}^2} \cong 58
\ee

\begin{figure}
\centerline{\includegraphics[width=\textwidth]{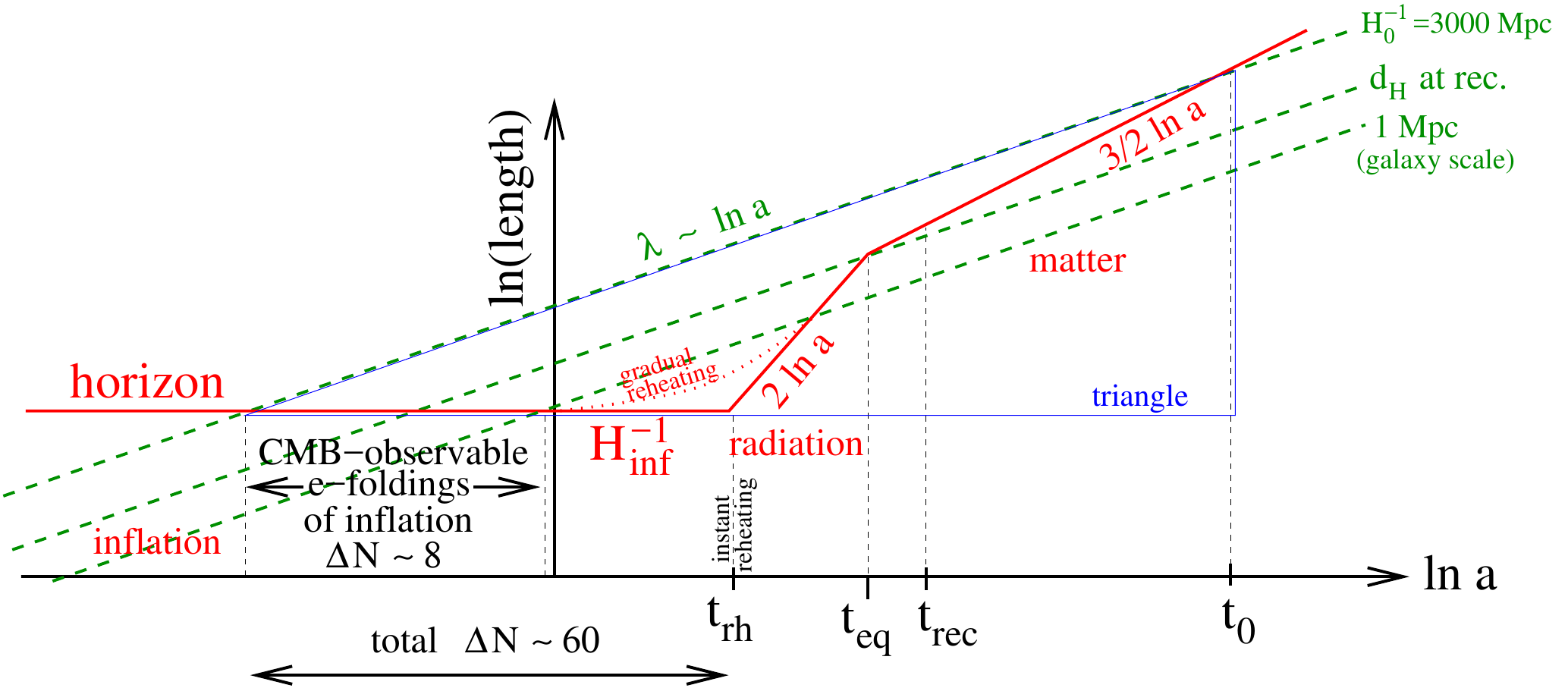}}
\caption{Evolution of comoving scales (green, dashed) and the 
particle horizon (red, solid) with scale factor as a proxy for
time.}
\label{fig:scales}
\end{figure}

For the horizon problem, fig.\ \ref{fig:scales} (inspired by fig.\ 8.4
of ref.\ \cite{Kolb:1990vq})  is helpful.  The
dashed line labeled ``$H_0^{-1}=3000{\rm Mpc}$'' is relevant
for the horizon problem: it shows how the current horizon scale
shrinks with the universe projected back in time, being much
larger than the maximum causally connected scale $d_H(t_{\rm rec})$ at the time of recombination. In
standard big bang cosmology, the $H_0^{-1}$ scale was always outside the
horizon at earlier times, signifying that its enclosed particles
could not have achieved significant causal equilibrium.  But with
inflation, provided it lasted long enough, there is an intersection 
such that at early times it was inside the horizon and causal
processes would have been able to make it homogeneous.

We can estimate the minimum number of $e$-foldings needed to solve the
horizon problem using this figure and a bit of trigonometry.  Consider
the right triangle (blue, labeled ``triangle'') whose base extends along the inflationary horizon, 
from where the $H_0^{-1}$ scale acting as hypotenuse
intersects it, until the present time.  The vertical leg of the
triangle has length $\ln(H_{\rm inf}/H_0)$, while the horizontal
one has height $\Delta N + \ln(T_{\rm rh}/T_0)$.  The slope is
unity since scales grow linearly with $a$. This gives the minimum 
number of $e$-foldings of inflation as
\be
	\Delta N = \ln {H_{\rm inf} T_0\over H_0 T_{\rm rh}}
\ee
This is a rather crude approximation, since we have assumed in the
picture that reheating happens instantaneously, which would imply a 
very efficient mechanism of reheating such that $T_{\rm rh}\sim
\Lambda_{\rm inf}$ (the energy scale of inflation).  Nevertheless,
we estimate $\Delta N \sim \ln(\Lambda_i T_0/\rho_c^{1/2})$,
parametrically the same as for the flatness problem, eq.\ (\ref{Nflat}).
In fig.\ \ref{fig:scales} we show by the dotted curve the more
realistic evolution of the horizon when reheating is more gradual.
It is clear that $\Delta N$ is reduced in this case since inflation
ends earlier.  A careful derivation \cite{Liddle:2003as} shows that the number of
$e$-foldings until the end of inflation, at the time when a scale of
comoving wave number $k_*$ crosses outside the horizon, is given by
\be
	N_* \cong 67 - \ln{k_*\over a_0 H_0} + 
	\sfrac14\ln {V_*^2\over m_p^4 \rho_{\rm end}}
	+\sfrac{1}{12}\ln {T_{\rm rh}^4\over \rho_{\rm end}}
\label{Nstar}
\ee
where $V_*$ is the value of $V(\varphi)$ at the time of horizon crossing
(see eq.\ (\ref{hceq}) below for the definition), and $\rho_{\rm end}$ is the value of $V$ when inflation ends.
Even this formula is simplified to the case where the equation of
state during reheating is that of radiation, $w=1/3$.  The more
general result can be found in \cite{Liddle:2003as,Ade:2015lrj}.

Soon after Guth's introduction of the inflationary universe, it was
realized that the quantum fluctuations of $\varphi$ during inflation  can
account for the density perturbations necessary for growth of
structure in the universe
\cite{Hawking:1982cz,Starobinsky:1982ee,Guth:1982ec,Bardeen:1983qw}.\footnote{
Of these references, I find the last one to be the most understandable
and still worth reading; it forms the basis for the presentation given
in the textbook \cite{Kolb:1990vq}.  There is incidentally an interesting story
about it; it originally appeared as a preprint by Steinhardt and
Turner, who were finding that inflation could not produce sufficiently
large fluctuations to explain the observed structure.  The corrected
version came out with J.M.\ Bardeen, an expert on cosmological
perturbations, as a coauthor.  He recognized that the curvature
invariant that is conserved while perturbations are outside the
horizon is given by $\zeta \sim {\delta\rho/(\rho+p)}$ during
inflation, rather than $\delta\rho/\rho$ as had been assumed in the
preprint version.  Since $\rho + p$ is very close to zero during
slow-roll inflation, this provides a huge enhancement, needed to get
the observed level of density perturbations.  Some copies of the
original preprint still exist.}  We will later show that the quantum
fluctuation of the Fourier mode of the inflaton during inflation is
\be
	\delta\varphi_k = \int d^{\,3}x\,e^{i k\cdot x} \delta\varphi(x)
	\sim {H\over 2\pi}
\ee
for any $k$, implying that the fluctuations are nearly {\it scale
invariant,} which is phenomenologically important for avoiding too
large fluctuations on small scales (that would produce too many black
holes) or on long ones (leading to inhomogeneity at large scales).

\begin{figure}
\centerline{\includegraphics[width=0.3\textwidth]{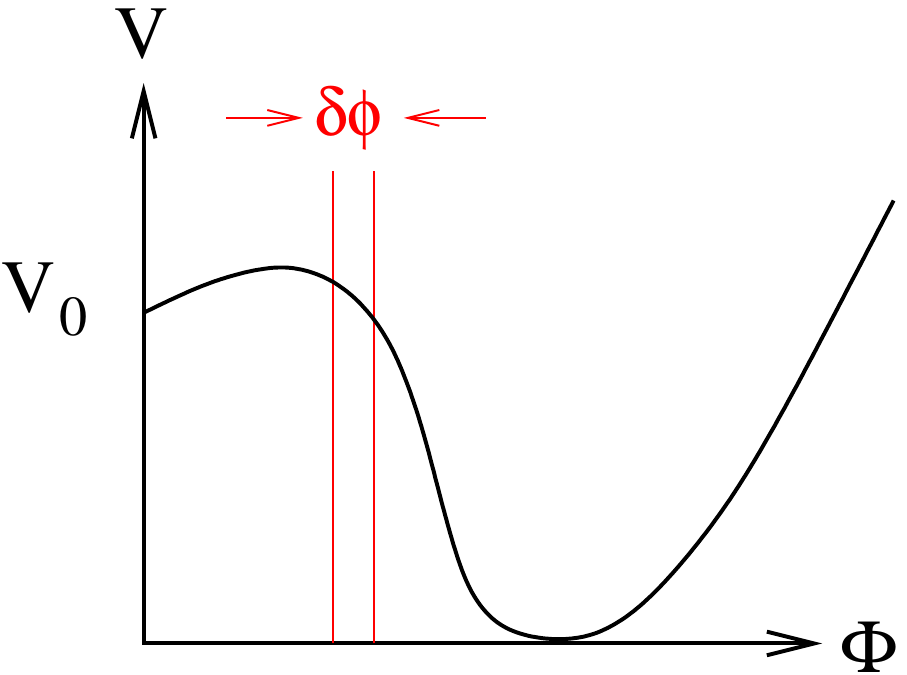}}
\caption{Quantum fluctuations of the inflaton that cause reheating to
occur at different times for different places in the universe.}
\label{fig:reheat}
\end{figure}

The inflaton fluctuations lead to density fluctuations, that can be
quickly understood at a heuristic level, referring to fig.\ 
\ref{fig:reheat}.  Consider two regions of the universe whose value of
$\varphi$ differs by $\delta\varphi$.  Inflation will end slightly later in
one region, with the time difference
\be
	\delta t \sim {\delta\varphi\over \dot\varphi}
\ee
The differing amounts of inflation cause local perturbations in the 
3D curvature of a surface at fixed time, that we can estimate as
\be
	{\cal R}_k\sim \delta\left(1\over a^2\right)\sim
	{\delta a\over a}\sim H\delta t\sim {H^2\over \dot\varphi}
\ee
again with an approximately scale-invariant spectrum since both $H$ 
and $\dot\varphi$ change very slowly during inflation.  We can relate
this quantity to the potential $V$ using the slow-roll equation of 
motion,
\be
	{H^2\over\dot\varphi} = {H^3\over H\dot\varphi} \sim 
	{V^{3/2}(\varphi)\over V'(\varphi)}
\ee
evaluated at the moment when the scale $k$ exits the horizon,
\be
	{k\over a} \cong  k e^{-Ht} = H = ke^{-N}
\label{hceq}
\ee
This is the important {\it horizon crossing} condition, showing 
the origin of the relation $N\sim \ln k/H$ that we observe
in eq.\ (\ref{Nstar}).  As always, $k$ is the comoving wave number
that does not change with the expansion, and whose value refers to the
present time, while $k/a$ is the physical, time-dependent wave number.

An important observable quantity is the correlation function of the 
3D curvature, giving rise to the scalar power spectrum $P_s$,
\be
	P_s = \int d^{\,3}x\,e^{ik\cdot x}\left\langle {\cal R}(0)
	{\cal R}(x)\right\rangle = \left|{\cal R}_k\right|^2 \sim
	{H^4\over \dot\varphi^2} \equiv A_s \left(k\over
k_*\right)^{n_s-1}
\label{Pseq}
\ee
For historical reasons, $n_s=1$ is the definition of scale-invariance,
known as the Harrison-Zeldovich spectrum, and from (\ref{Pseq})
we see that
\be
	n_s -1 = {d\ln P_s\over d\ln k}\cong {d\ln P_s\over dN}
\label{nsdef}
\ee
where we used $N = \ln k/H$ from the horizon-crossing condition
(\ref{hceq}) and approximated $H$ as being constant during inflation.
The deviation of $n_s$ from 1 is important since it impacts the 
correlations of CMB temperature fluctuations
\be
	\left\langle {\delta T\over T}(0)\, {\delta T\over T}(x)\right\rangle
\ee
which leads to constraints on models of inflation.

\begin{figure}
\centerline{\includegraphics[width=0.5\textwidth]{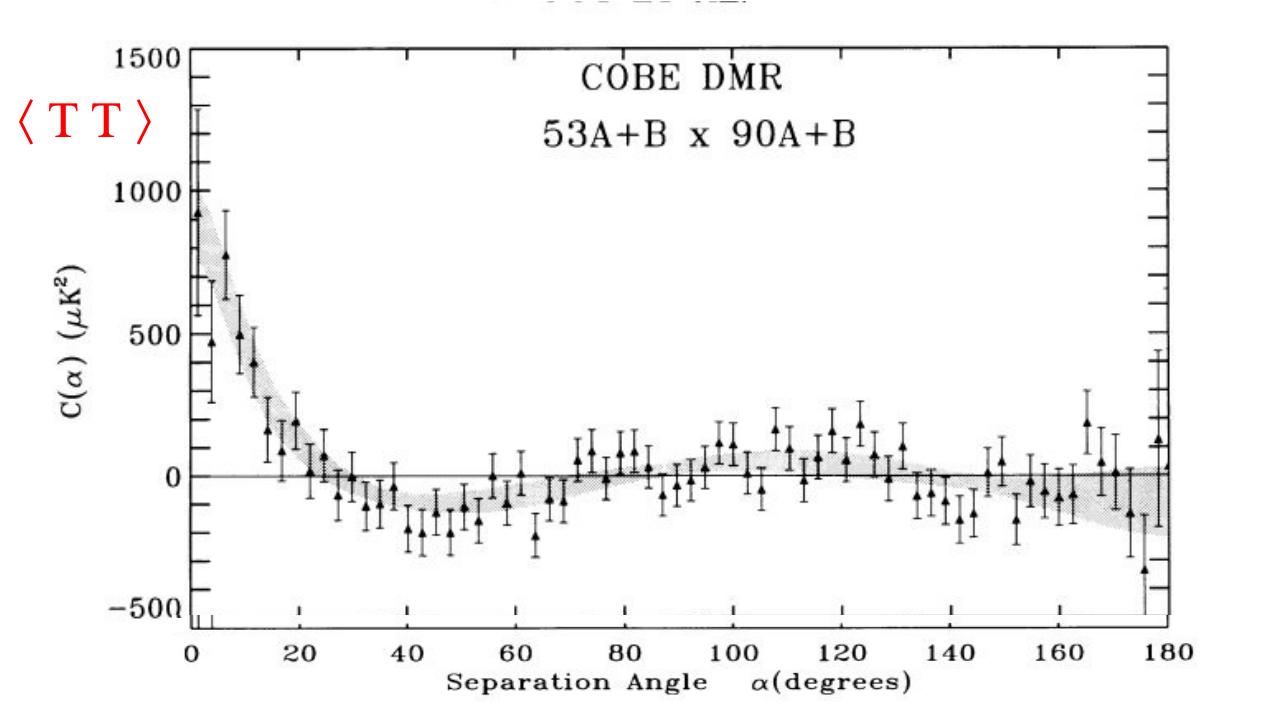}\hfil
\includegraphics[width=0.5\textwidth]{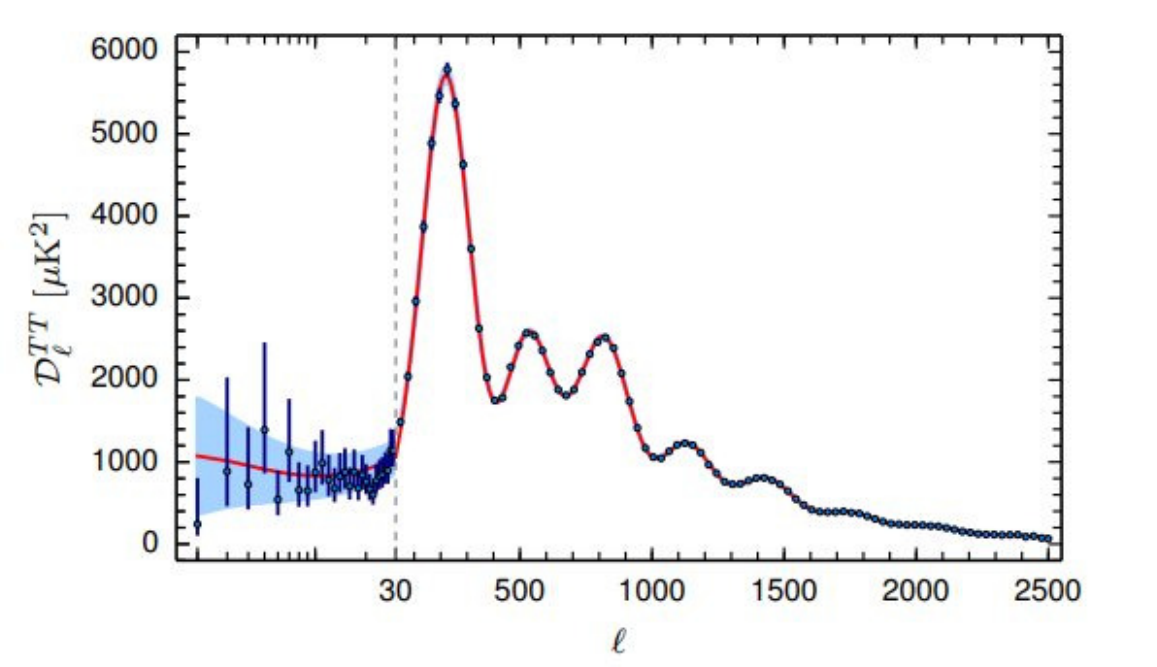}}
\caption{Left: angular correlations of the CMB temperature fluctuations, as
measured by COBE \cite{1992ApJ...396L...1S}.  Right: ${\cal D}_{\ell}
= \ell(\ell+1)C_\ell/2\pi$ versus $\ell$ as measured by Planck
\cite{Ade:2015lrj}.}
\label{fig:cobe}
\end{figure}

\begin{figure}
\centerline{\includegraphics[width=0.75\textwidth]{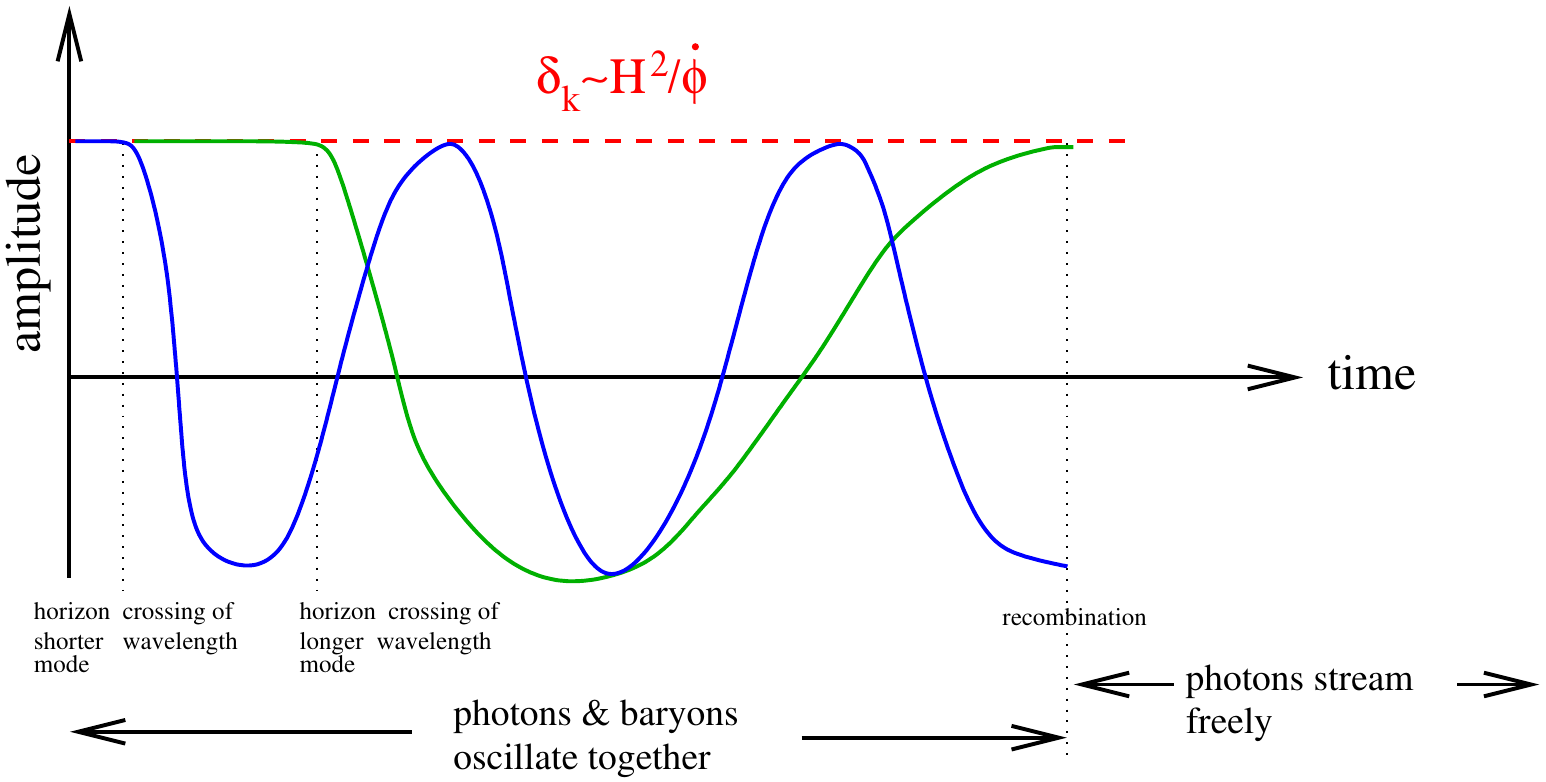}}
\caption{Illustration of the origin of CMB Doppler peaks.  Sound waves
of different wavelengths start oscillating 
with nearly the same amplitude, but at different times, when they
cross back inside the horizon, $k/a = H$.  Thus they are out of
phase with each other at $t_{\rm rec}$, when the universe becomes
transparent and the photons start freely streaming.}
\label{fig:sound}
\end{figure}

In 1992 the NASA experiment COBE first observed the CMB temperature
fluctuations at the level of $\delta T/T= 5\times 10^{-5}$
\cite{1992ApJ...396L...1S}, close to the value that was already
understood to be needed for consistency with structure formation.
COBE measured angular correlations obtaining an oscillatory pattern
as reproduced in fig.\ \ref{fig:cobe}(left).  These oscillations are better
visualized in $\ell$-space by expanding in spherical harmonics,
\be
	\delta T = \sum_{\ell,m} a_{\ell m} Y_{\ell m}(\theta,\varphi)
\ee
and plotting
\be
	C_\ell  = {1\over 2\ell + 1} \sum_m |a_{\ell m}|^2
\ee
versus $\ell$.  This reveals the famous acoustic peaks, shown in
fig.\ \ref{fig:cobe}(right).\footnote{One might wonder why the COBE
correlation rises at small angles while that of Planck becomes small
at large $\ell$. The angular resolution of COBE was much lower
than that of Planck, 
probing only $\ell \lesssim 25$ \cite{HKS}.}  They represent sound waves in the 
coupled photon-baryon plasma, at the time of recombination.  During
the tightly-coupled epoch, density perturbations
$(\delta\rho/\rho)_k$ at a scale $k$ undergo acoustic oscillations because of
the plasma pressure.  But these oscillations do not begin until
that scale has crossed back inside the particle horizon, which happens
at different times for different scales, leading to the sound waves at 
different scales being out of phase with each other at the ``moment''
of recombination, when they start to become visible in the CMB.  This
process is illustrated in fig.\ \ref{fig:sound}.

The temperature fluctuations arise from the CMB photons climbing out
of gravitational potential wells created by the density perturbations.
If the perturbations were static, the net gravitational redshift would
vanish, but they are evolving while the photons traverse them, leading
to a net change, known as the Sachs-Wolfe effect.  The detailed shape
of the peaks is hard to approximate analytically, and requires solving Boltzmann
equations that take into account the evolution of the density
perturbations.  This is all done in publicly available codes such as
 CAMB, part of the CosmoMC package \cite{Lewis:2002ah}.  Clearly,
the shape will be affected by the spectrum of the scalar power since
it determines the slope of the line bounding the oscillations,
shown as horizontal in fig.\ \ref{fig:sound}.  Comparison with Planck
data determines the spectral index as \cite{Ade:2015lrj}
\be
	n_s = 0.968 \pm 0.006\, .
\ee

Let's now consider the prediction for $n_s$ from slow roll inflation. 
From eq.\ (\ref{nsdef}) we find 
\be
	n_s-1 = {d\over dN}\left(3\ln V - 2\ln V'\right)
	 = \left(3{V'\over V} - 2{V''\over V}\right) {d\varphi\over dN}\, .
\ee
Since $dN = H dt$, it follows that $d\varphi/dN = \dot\varphi/H$.
Then using the slow roll equation (\ref{sreq}) we get
\be
	{d\varphi\over dN} = -{V'\over 3H^2} = - m_p^2 {V'\over V}
\ee
and 
\be
	n_s -1 = -6\epsilon + 2\eta + O(\epsilon^2, \eta^2,
\epsilon\eta,\dots)
\ee
with higher-order slow roll parameters included in the dots. 

The
amplitude of the scalar power is constrained by the magnitude of the
$C_\ell$'s:
\be
	A_s = {H^4\over (2\pi\dot\varphi)^2} = {V\over 24\pi^2
	m_p^4\epsilon} = e^{3.1}\times 10^{-10}
\label{Aseq}
\ee
where the manner of expressing the experimental value is Planck's
convention \cite{Ade:2015lrj}.  Implicit in this equation is the
choice of a reference scale
\be
	k_* = {1\over 20\,{\rm Mpc}}
\ee
at which all quantities are evaluated (using the horizon crossing
condition to associate $k_*$ with a field value $\varphi_*$).  One can
infer from (\ref{Aseq}) the constraint
\be
	\left|{V^{3/2}\over m_p^3\, V'}\right| = 5.1\times 10^{-4}
\label{cobenorm}
\ee
(again at $k_*$), which fixes the overall magnitude of $V$ in a given
model of inflation.

\begin{figure}
\centerline{\includegraphics[width=0.5\textwidth]{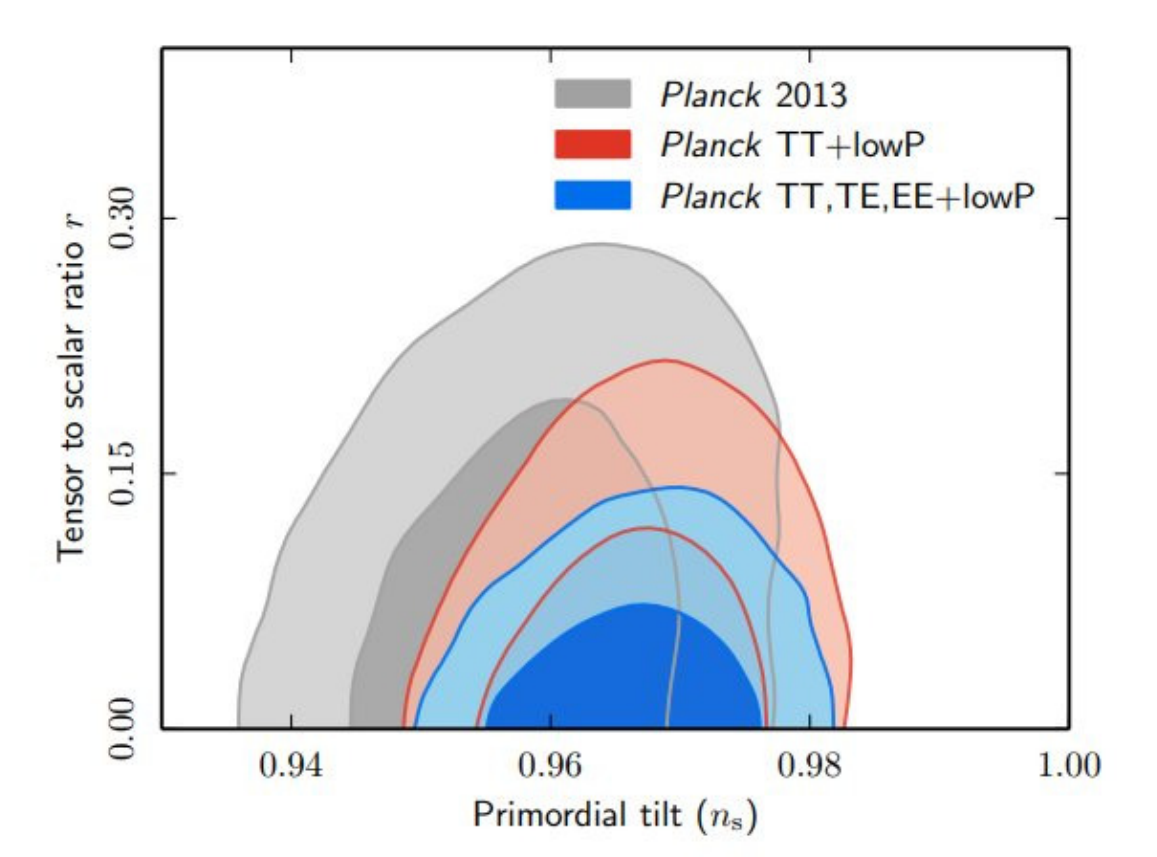}
\raisebox{10mm}{\includegraphics[width=0.5\textwidth]{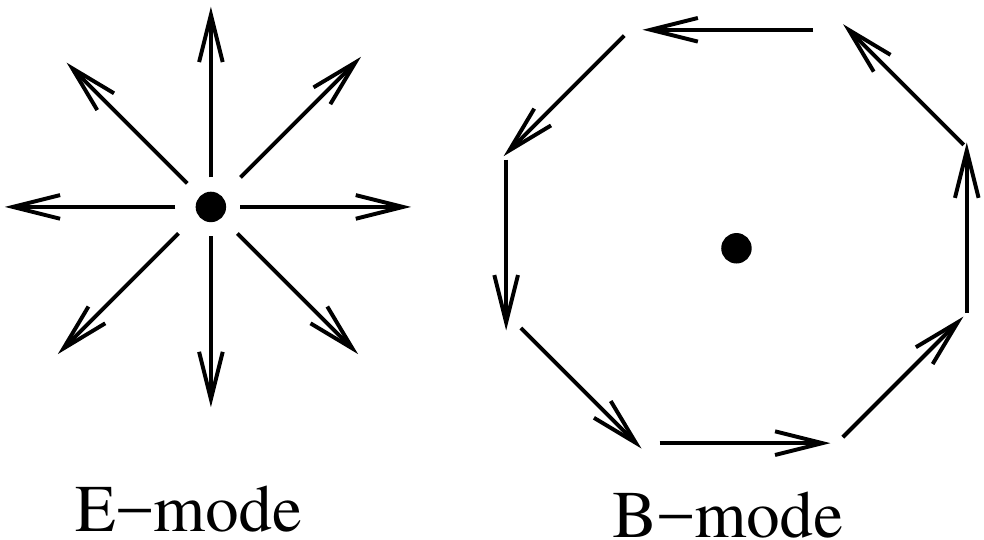}}}
\caption{Left: constraints on $r$ versus $n_s$ from ref.\ \cite{Ade:2015lrj}.
Right: $E$- and $B$-mode polarizations.}
\label{fig:planck-r-ns}
\end{figure}

In addition to the scalar perturbation, gravity waves get quantum
fluctuations during inflation with the same amplitude as the inflaton
field,
\be
	\delta (h_{\mu\nu})_k \sim {H\over 2\pi}
\ee
but with no further modulation by factors like $H/\dot\varphi$ since
there is no potential for the graviton.  The gravitational potential
wells produced by these fluctuations give a separate contribution to
the Sachs-Wolfe effect, beyond those coming from the density
perturbations.  However, they are only effective at large scales,
$\ell \ll 100$ (corresponding to $k^{-1} \gg 100\,$Mpc), because at
smaller scales, these tensor fluctuations re-enter the horizon before
$t_{\rm rec}$ and get Hubble damped, just like any other kind of
radiation.  The tensor modes that are not Hubble-damped can boost the CMB power in the
blue-shaded region of fig.\ \ref{fig:cobe}(right).  The power
spectrum associated with them is denoted by
\be
	P_t = {2\over m_p^2}\left(H\over 2\pi\right)^2 \equiv
	A_t\left(k\over k_*\right)^{n_t-1}
\label{Ateq}
\ee
The tensor-to-scalar ratio is
\be
	r = {A_t\over A_s} = 16\epsilon
\ee
which roughly quantifies the relative contributions from tensors
and scalars to the $C_\ell$'s at low $\ell$.  The lack of evidence
for any such contribution results in the upper bound 
\cite{Ade:2015lrj}
\be
	r < 0.1
\ee
where the reference scale is smaller, $k = (500\,{\rm Mpc})^{-1}$,
appropriate to the region of $\ell$-space where the tensor
contribution can
be appreciable.  The limit on $r$ can be translated into an upper
bound on the energy scale of inflation, since eq.\ (\ref{Ateq})
depends only on $H^2 = V/3m_p^2$ and not $V'$:
\be
	V_*^{1/4} \lesssim 2\times 10^{16}\,{\rm GeV}
\label{Vsbound}
\ee

The combined constraints on $n_s$ and $r$ are shown in fig.\ 
\ref{fig:planck-r-ns}(left) taken from ref.\ \cite{Ade:2015lrj}.  These
results illustrate the importance of the separate measurements of the
CMB {\it polarization} that Planck has carried out.  The polarization is
conventionally separated into two independent components, the $E$-mode
which is curl-free and the $B$-mode which is divergenceless, illustrated in
fig.\ \ref{fig:planck-r-ns}(right).  $E$-modes are produced by scalar
$\delta\rho/\rho$ fluctuations, while $B$-modes are only produced by tensor
perturbations (or foregrounds dominated by thermally emitting dust in the galaxy).  
Planck has observed  $E$-modes through their cross-correlations with temperature
fluctuations, $\langle \delta T\,E\rangle$,  and their autocorrelations,
$\langle
E\,E\rangle$, and inclusion of these data strengthen the constraints.
The $\langle B\,B\rangle$ correlations remain a holy grail of CMB
detection, since they would provide more direct evidence of primordial inflationary
tensor fluctuations than $r$.  There was excitement when BICEP2 claimed such a
detection \cite{Ade:2014xna}, but this turned out to be dust 
\cite{Ade:2015tva}.

\subsection{Example: chaotic inflation}
We now work through a specific example to show what is needed to test a
model against the data, namely chaotic inflation \cite{Linde:1983gd}.
(The name originates from a picture wherein the universe starts in a
disordered state, far from the minimum of the potential, with energy
density near the Planck scale.  Inflation can get started in
any region of size of several Planck lengths if it fluctuates into a somewhat
homogeneous state.  Once inflation starts, inhomogeneities are quickly
damped, and the inflating regions becomes much larger than those where 
inflation has not yet begun.)  For simplicity, the potential is taken to
be monomial,
\be
	V(\varphi) = \lambda m_p^4(\varphi/m_p)^p
\ee
with $p>0$.  The Hubble parameter is then
\be
	H = \sqrt{\lambda\over 3}\, m_p\left(\varphi\over m_p\right)^{p/2}
\ee
and the slow-roll equation is
\be
	3 H \dot\varphi = 3 H^2 {d\varphi\over dN} = -V' = -\lambda\, p\, m_p^3
	(\varphi/m_p)^{p-1}
\label{sreq2}
\ee

We can define $N$ to be the number of $e$-foldings until the end of
inflation, assuming $\varphi=\varphi_e$ at this time.  Then eq.\ (\ref{sreq2})
can be integrated,
\be
	N = -\int_{\varphi}^{\phi_e} d\varphi\, {3H^2\over V'} = 
	{1\over 2 p\, m_p^2}\left(\varphi^2-\varphi_e^2\right)
\label{Nsol}
\ee
Since $V'/V = p/\varphi$ and $V''/V = p(p-1)/\varphi^2$, the slow-roll parameters
are 
\be
	\epsilon = {p^2\over 2}\left(m_p\over\varphi\right)^2,\qquad
	\eta = p(p-1)\left(m_p\over\varphi\right)^2
\ee
We see that superPlanckian field values are necessary to justify slow roll.
The spectral index is then
\be
	n_s-1 = -p(p+2)\left(m_p\over\varphi\right)^2
\ee
which is always negative, called a red-tilted spectrum.

For most interesting values of $p$, inflation will end when $\epsilon=1$
(occurring before $\eta=1$), giving
\be
	\varphi_e \cong {p\over \sqrt{2}}\,m_p
\ee
We can then use (\ref{Nsol}) to reexpress $n_s$ as
\be
	n_s-1 = -{p+2\over (2N + p/2)}  \cong -{(1+p/2)\over N}
\label{nseq2}
\ee
The approximation is valid since we will be evaluating (\ref{nseq2})
at
horizon crossing of the reference scale $k_*$, when $N\gg 1$.

To compare to Planck data, we must evaluate $n_s-1$ using the value 
$N_*$ corresponding to horizon crossing of the relevant mode $k_*$,
eq.\ (\ref{Nstar}).  The second term on the right-hand-side is
$\ln k_*/a_0H_0 = 5.4$, but the third term depends upon $\lambda$, which
we have not yet determined, and the last one depends upon the reheat
temperature $T_{\rm rh}$, which is not known until we specify the theory
more completely to determine how reheating takes place.  To find $\lambda$
we need to use the normalization of the scalar power amplitude (sometimes
called the COBE normalization), eq.\ (\ref{cobenorm}), 
\be
	{V^{3/2}\over m_p^3\, V'} = {\sqrt{\lambda}\over p}
	\left(\varphi_*\over m_p\right)^{1+p/2} = 5\times 10^{-4}
\ee
which fixes
\be
	\lambda \cong {25\times 10^{-8}\, p^2\over (p(2 N_* + p/2))^{1+p/2}}
\ee
We see that the equation for $N_*$ has become transcendental, but with only
a weak log dependence of $N_*$ on the r.h.s.\ which can be solved by
iteration.  Knowing $\lambda$, we have information about the energy scale
of inflation,
\be
	V_* = (0.022\,m_p)^4 {p\over 2 N_*}
\ee
which we see is generically not far below the Planck scale, and in danger
of conflicting with the tensor bound (\ref{Vsbound}) unless $p$ is small.
This can also be seen in terms of $r$,
\be 	
	r = 16\epsilon = {4 p\over N_*} 
\ee
which is $0.07\,p$ at $N_*=55$, for example.

\begin{figure}
\centerline{\includegraphics[width=0.5\textwidth]{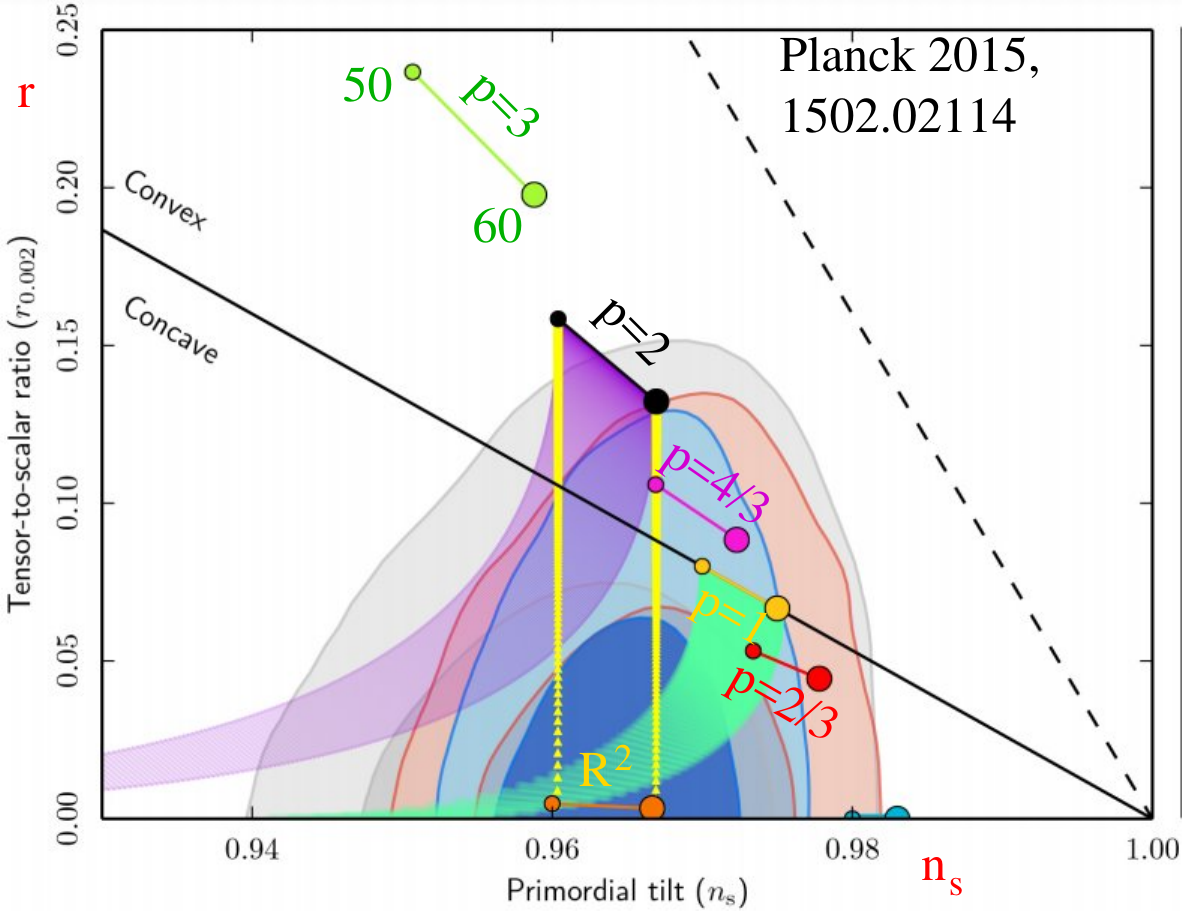}}
\caption{Predictions of chaotic inflation for $r$ versus $n_s$, overlaid on
the Planck allowed regions.}
\label{fig:planck-r-ns2}
\end{figure}

We still don't know what to take for $T_{rh}$, so the Planck collaboration
and many practitioners take it to be a free parameter, letting $N_*$ 
vary between 50 and 60 to allow for the uncertainty.  Given the nominal
dependence $N_*\sim (1/3)\ln T_{\rm rh}$ this might seem excessively 
generous and you are free to make more restrictive assumptions about 
$T_{\rm rh}$.    In any case, the resulting predictions for chaotic inflation
are shown as the diagonal line segments bounded by heavy dots in fig.\ 
\ref{fig:planck-r-ns2}.  Of the cases shown, only $p=2/3$ even lies in the
$2\sigma$ allowed region.  Such fractional powers may seem peculiar from
the point of view of renormalizable field theories, but can arise as an
effective description for large values of $\varphi$ in the context of
string-motivated axion monodromy models \cite{McAllister:2008hb}, for
example.

The best-fitting model indicated on fig.\ \ref{fig:planck-r-ns2} is
Starobinsky's $R^2$ inflation \cite{Starobinsky:1980te}, that can
be mapped onto a scalar field inflation model with potential of the form
\be
	V(\varphi) = \Lambda^4 \left(1 - \exp(-\sqrt{2/3}\,\varphi/m_p)\right)^2
\ee
(we will explain how, below; see eq.\ (\ref{staro})).
More generally, current Planck data prefer models with a convex potential.
These include Hilltop models \cite{Boubekeur:2005zm}
\be
	V = \Lambda^4\left(1 - (\varphi/\mu)^p + \dots\right)
\ee
and Higgs inflation, where the total action is
\be
	S = \int d^{\,4}x\, \sqrt{-g}\left({\cal L}_{\rm grav} + 
	{\cal L}_{\rm SM} + \xi R |H|^2\right)
\ee
As we will show at the end of this chapter, 
this can be transformed to scalar potential similar to that of
$R^2$ gravity by going to the Einstein frame.

\begin{figure}
\centerline{\includegraphics[width=0.5\textwidth]{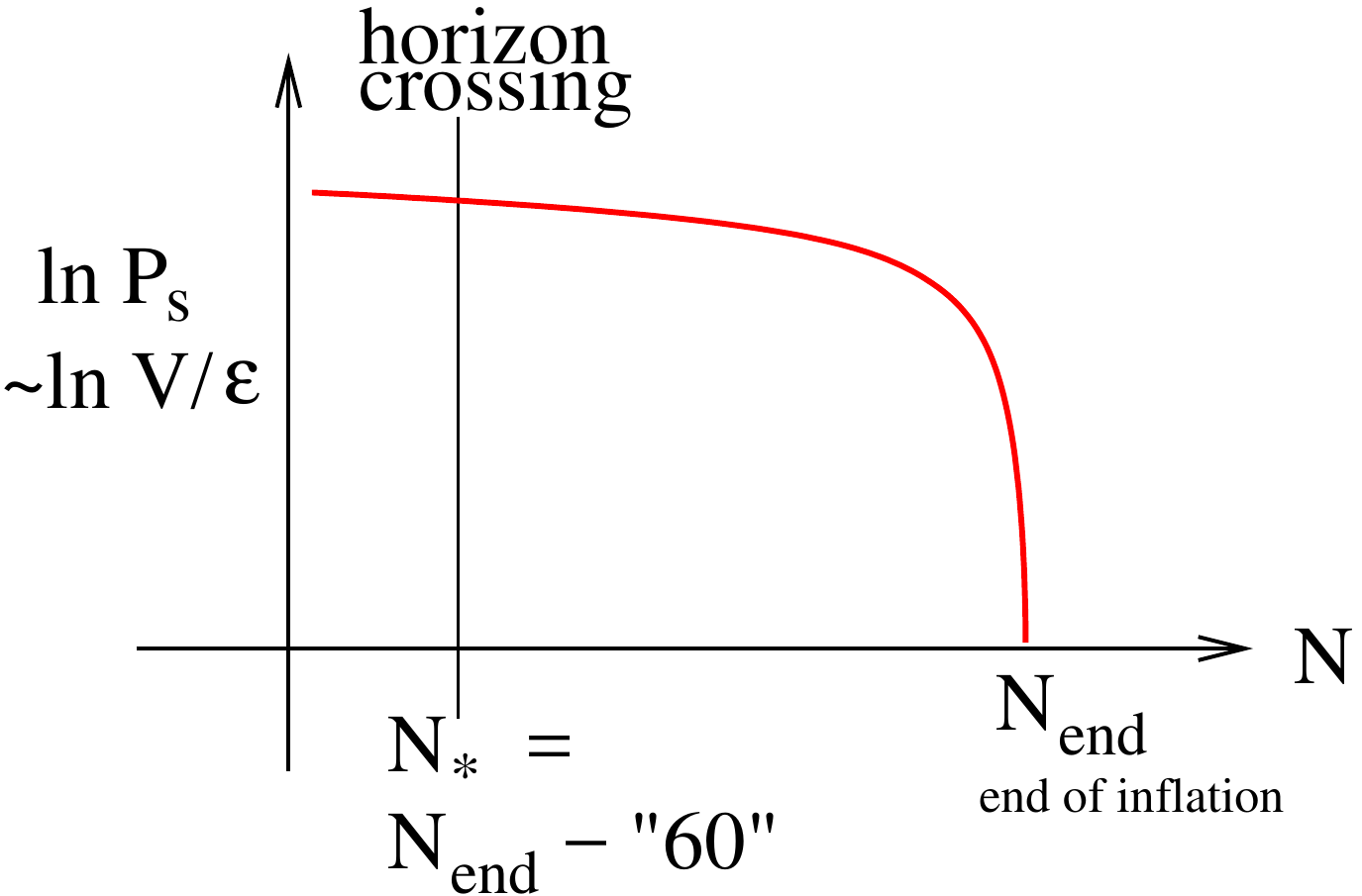}}
\caption{Typical dependence of the scalar power on the number of
$e$-foldings in a numerical solution.}
\label{fig:lnP}
\end{figure}

Sometimes one prefers to solve the coupled inflaton/Friedmann equations
numerically instead of using the slow roll approximation 
(see \cite{Cline:2006hu}) for details), particularly if the inflaton has
a noncanonical kinetic term $\sfrac12 f(\varphi)\dot\varphi^2$, or if there are several fields.  This can be done efficiently in
terms of the canonical field momentum
\be
	\pi = {\delta {\cal L}\over \delta\dot\varphi}= f(\varphi)\dot\varphi
\ee
Then
\be
	H = {1\over \sqrt{3}\, m_p}\left({\cal L}_{\rm kin} +
	V\right)^{1/2}
\ee
and the equations of motion can be written in first-order form
\bea	
	{d\pi\over dN} &\equiv& \pi' = -3\pi + {1\over H} {\partial\over
	\partial\varphi}\left({\cal L}_{\rm kin} -
	V\right)\nn\\
	{d\varphi\over dN} &\equiv& \varphi' = {\pi\over f H}
\eea
The end of inflation is easy to recognize since the fields start
oscillating around the minimum of the potential; this behavior can
typically be followed for numerous periods of oscillations before the 
Runge-Kutta step size
shrinks to zero and creates an exception in the code.  The exact expression for the spectral
index is
\be
	n_s-1 = {d\ln P_s\over d\ln k} = {d\ln P_s\over dN}
	\left(1 + {d\ln H\over dN}\right)^{-1}
\ee
where the correction term $(\dots)^{-1}$ is usually very close to unity.
The quantity $\ln P_s$ as a function of $N$ will resemble fig.\ 
\ref{fig:lnP}.  By evaluating it at $N_*$ (using eq.\ (\ref{Nstar}))
one can match to the observed spectrum.
	
\subsection{Filling in some details}

The previous section was meant to give an intuitive understanding of
inflation, and to allow you to start confronting your favorite model
against data.  Here we would like to sketch some of the details that 
would be needed for a deeper or more rigorous understanding.  Some useful
references for this material include \cite{Liddle:1993fq,Liddle:2000cg,
Lyth:2009zz,Linde:2007fr} and others we will cite below.

\subsubsection{Quantum fluctuations of the inflaton} 
First we want to explain the origin of the quantum fluctuation $\delta\varphi
= H/2\pi$.  This is straightforward to derive, by canonically quantizing 
a free scalar field in an FRW background, that we can approximate as
de Sitter space.  The expansion of $\varphi$ in Fourier modes looks the same
as in flat space,
\be
	\varphi(x) = \int {d^{\,3}k\over (2\pi)^{3/2}}\left(
	a_k \psi_k(x) + a^\dagger_k \psi^*_k(x)\right)
\ee
except that the mode functions now have a different time dependence,
\be
	\psi_k = e^{i\vec k\cdot\vec x} f_k(t)
\ee
where (using $a=e^{Ht}$ for a pure dS background),
\be
	\ddot f_k + 3H\dot f_k + \left(m^2 + k^2 e^{-2Ht}\right)f_k = 0
\label{fkeq}
\ee
If $\varphi$ is the inflaton then necessarily $m^2\ll H^2$ during inflation,
to have slow roll, and $k/a  = k e^{-Ht}> H$ before horizon crossing.  Hence we can
ignore the mass and solve (\ref{fkeq}) in the $m=0$ limit.  Defining
$z = ke^{-Ht}/H$, the solution can be written
\bea
	f_k &=& C z^{3/2}\left(J_{3/2}(z) + i Y_{3/2}(z)\right)\nn\\
	    &=& -C\sqrt{2\over\pi}\,z\left(i + z\right) e^{iz}
\label{fsol}
\eea
with a normalization constant $C$ to be determined.
The particular combination of the independent solutions $J_{3/2}$ and $Y_{3/2}$ was chosen
with hindsight, since we want this to agree with the usual Minkowski
space solution in the limit $H\to 0$, where up to an irrelevant phase
\be
	f_k \to C\sqrt{2\over\pi}{k\over H} e^{-i k t}
\ee
This shows that we took the right linear combination, and it fixes the 
normalization constant to be
\be
	C = {H\over k} \sqrt{\pi\over 4 k}
\ee

Now we can compute the r.m.s.\ fluctuations of $\varphi$ using the usual
property of the creation and annihilation operators, $\langle 0|a_k
a^\dagger_{k'}|0\rangle = \delta^{(3)}(\vec k-\vec k')$:
\be
	\left\langle 0|\varphi^2(x)|0\right\rangle = \int {d^{\,3}k\over (2\pi)^{3}}
	\left|f_k\right|^2  = 
	\int {d^{\,3}k\over (2\pi)^{3}}
	{H^2\over 2 k^3}\left( 1 + {k^2\over H^2}e^{-2Ht}\right)
\label{rmsfluc}
\ee
The second term is just the usual UV-divergent contribution already present
in Minkowski space, as can be seen from the fact that the factors of $H$
cancel out, and $e^{-2Ht}$ can be absorbed by rescaling $k\to k_{\rm phys}
e^{Ht}$.  We don't care about this term because it can be removed by
renormalization, and in any case its contributions are only important at 
distance scales that are much too small to be cosmologically relevant.
This first term however is new, and is associated with being in the dS
background.  The structure $d^{\,3}k/k^3$ shows that equal power is present
in the fluctuations from every logarithmic interval of $k$, and this
corresponds to a scale-invariant spectrum.

\begin{figure}
\centerline{\includegraphics[width=0.3\textwidth]{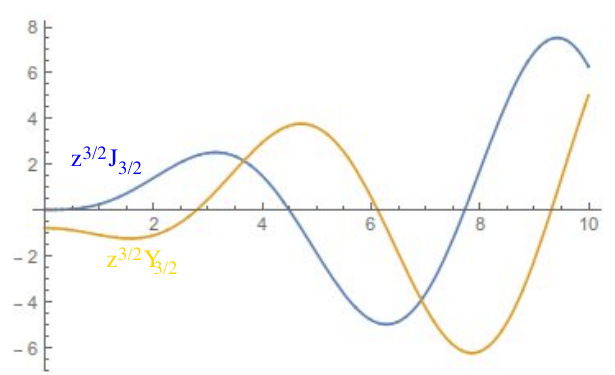}
	{\includegraphics[width=0.7\textwidth]{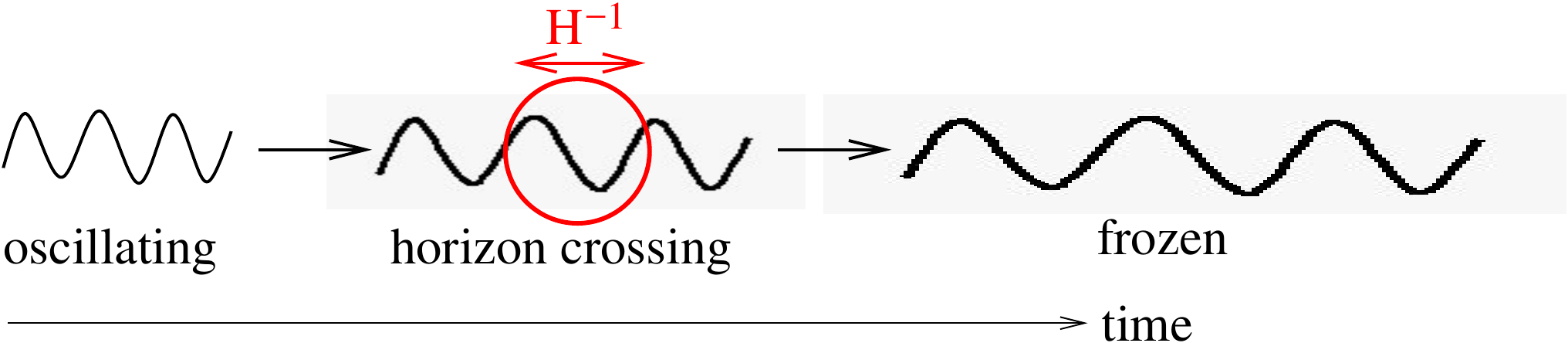}}}
\caption{Left: time-dependence of the inflation fluctuation modes,
as a function of $z = ke^{-Ht}/H$.  Right: illustration of the freezing 
of the mode functions following horizon crossing.}
\label{fig:flucs}
\end{figure}

The shape of the solutions (\ref{fsol}) can give some insight into the
significance of horizon crossing.   Fig.\ \ref{fig:flucs}(left) shows that
the mode functions stop oscillating at lates times (small $z$), starting around $z=1$,
which is when the corresponding wavelength goes outside the horizon.
Thereafter they remain ``frozen'' until later reentry into the horizon.
This behavior is illustrated more
qualitatively in fig.\ \ref{fig:flucs}(right).  It is argued \cite{Guth:1985ya}
that horizon crossing corresponds to the transition between quantum and
classical behavior of the fluctuations.  We can express the classical
field as
\be
	\varphi(x) = \int {d^{\,3}k\over (2\pi)^{3}}\, \varphi_k\, e^{ik\cdot x}
\ee
having a volume-averaged fluctuation
\be
	\langle\varphi^2\rangle = {1\over V} \int d^{\,3}x\, \varphi^2(x)
	= {1\over V} \int {d^{\,3}k\over (2\pi)^{3}} |\varphi_k|^2
\ee
Comparison with (\ref{rmsfluc}) shows that therefore $\varphi_k =
H\sqrt{V/2k^3}$, which leads to a scale-invariant power spectrum,
\be
	P_\varphi \propto k^3|\varphi_k|^2 \qquad\Longleftrightarrow\qquad
	\langle\varphi^2\rangle \sim \int {dk\over k}\, P_\varphi
\ee
The equal power in each logarithmic interval of scales implies that
every $e$-folding of inflation also contributes equal power.

\subsubsection{Cosmological perturbation theory}  Our description of fluctuations is
still over-simplified, since $\delta\varphi$ induces perturbations in the
metric that cannot be ignored.  To properly understand the interplay
requires cosmological perturbation theory, which is reviewed in the
references by Liddle and Lyth, as well as
\cite{Brandenberger:2003vk,Mukhanov:1990me,Kodama:1985bj}.
The general perturbation to the metric can be written as
\be
	\delta g_{\mu\nu} = \left(\begin{array}{c|c} 2\phi & -2 a^2 B_{,i}\\
	\hline
	-2 a^2 B_{,i} & 2(\psi \delta_{ij} + E_{,ij})a^2
	\end{array}\right)
\label{metricpert}
\ee
These perturbations can be decomposed in Lorentz scalars, vectors, and
tensors (gravity waves).  Vector perturbations always decay in expanding
space and we therefore neglect them.

By gauge transformations (diffeomorphisms), 
\be
	x^\mu \to x^\mu + \xi^\mu(x)
\ee
some of the functions in
(\ref{metricpert}) can be set to zero.  For example in the conformal
Newtonian, {\it a.k.a.} longitudinal gauge,
\be
	E_{,\ij} = B_{,i} = 0
\ee
and furthermore the perturbed Einstein equations imply $\psi=\phi$ as long
as there is no anisotropic stress, $T_{ij} = T_{ji}$.  Another popular
choice of gauge is the comoving one, in which the inflaton fluctuation is
defined to vanish (by demanding the surfaces of constant time have uniform
energy density), 
\be
	\delta\phi = 0, \quad B_{,i} = 0
\ee

An important quantity is the curvature perturbation  ${\cal R}$,
defined by generalizing the Friedmann equation to 
include the effects of inhomogeneity:
\be
	H^2(x,t) = {\rho(x,t)\over 3 m_p^2} + \sfrac23 {\nabla^2\over a^2}
	{\cal R}
\ee
It is related to the 3D curvature of the spatial surfaces on 
comoving foliations as
\be
	R^{(3)} = {4 k^2\over a^2}{\cal R}
\ee
in Fourier space, and it coincides with Bardeen's $\zeta$ variable
mentioned previously.  The utility of ${\cal R}$ is that it does not evolve
for superhorizon modes in single-field inflation (where entropy and
anisotropic stress perturbations vanish).  Therefore if one can compute
it at horizon exit, then the same value applies at the moment of horizon
reentry, at which time the equations of classical linear perturbation
evolution take over (to predict large scale structure or temperature
fluctuations).  

The expression for ${\cal R}$ in terms of metric perturbations generally
depends upon the choice of gauge, but it can be written in way that is
independent of the gauge \cite{Sasaki:1986hm}
\be
	{\cal R} = -\left(\Psi + {H\over \dot\phi}\Delta\varphi\right)
\ee
where 
\bea
	\Psi &\equiv& \phi + {1\over a}\left[(B-E')a\right]'\nn\\
	\Delta\varphi &=& \delta\varphi + (B-E')\varphi'
\eea
and prime denotes the derivative with respect to conformal time
$\eta$, with $dt = a\, d\eta$.  To rigorously compute the quantum
fluctuations of the inflaton, one must quantize ${\cal R}$ since the
canonically normalized field is a linear combination of metric and
inflaton perturbations (see also \cite{Mukhanov:1988jd}).  However the magnitude of this
fluctuation turns out to be the same as in the simplified approach, giving
a scalar power spectrum that is nearly scale invariant with amplitude
going as $H^4/\dot\phi^2$.

\subsection{Variations on the simplest inflation models}
Thus far we have assumed inflation is driven by a single scalar field with
nothing too exotic about its Lagrangian, and perturbative decays of the
inflaton after inflation.  To finish this lecture on inflation I would like
to mention some slightly more complicated scenarios that have been widely
studied.

\begin{figure}
\centerline{\includegraphics[width=0.4\textwidth]{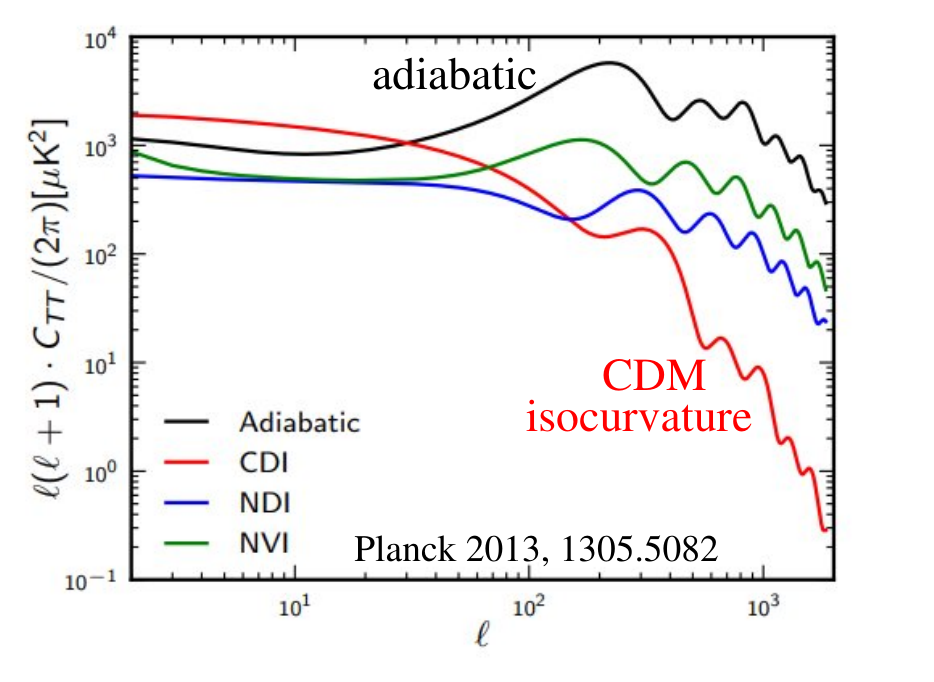}}
\caption{Comparison of CMB Doppler peaks predicted by pure isocurvature
versus adiabatic perturbations.  Adapted from ref.\ \cite{Planck:2013jfk}.}
\label{fig:isoc}
\end{figure}

\subsubsection{Isocurvature fluctuations}  The density perturbations we have
considered so far are called ``adiabatic,'' such the perturbations 
in separate components (baryons, cold dark matter, photons, neutrinos)
are related by
\be
	{\delta\rho_b\over \rho_b} = {\delta\rho_{\rm CDM}\over \rho_{\rm
CKM}} = \sfrac 34 {\delta\rho_\gamma\over \rho_\gamma}
	= \sfrac 34 {\delta\rho_\nu\over \rho_\nu}
\ee
The factor of $3/4$ can be understood from the fact that there is a single
temperature fluctuation controlling the densities:
\bea
	\rho_\gamma &\sim& T^4,\qquad \Longrightarrow 
	{\delta\rho_\gamma\over \rho_\gamma} = 4{\delta T\over T},\nn\\
	\rho_{\rm CDM} &\sim& m T^3,\qquad \Longrightarrow 
	{\delta \rho_{\rm CDM} \over \rho_{\rm CDM} }=  3{\delta T\over T}
\eea
and this is sourced by the curvature fluctuation ${\cal R}$.

In multifield inflation models, it is possible to have fluctuations of
different particle species that violate the adiabatic condition.
An orthogonal possibility is that particle number fluctuates between different
species in a way that keeps the total $\delta\rho$ and 
${\cal R}$ equal to zero.  In such a case, the entropy is perturbed, hence
these are known as entropy or isocurvature fluctuations.

In multifield inflation, at least two particles are approximately 
massless, $m_i^2\ll H^2$, but isocurvature perturbations can arise even if
only one field controls inflation, while a second light field 
eventually contributes significantly to the energy density of the universe.
An example is the axion.  Its isocurvature fluctuation can be quantified 
in terms of its number density, compared to the total entropy density 
$s$, through the parameter
\be
	S_a = {\delta(n_a/s)\over n_a/s} = {\delta n_a\over n_a} - 
	3{\delta T\over T} \cong {\delta n_a\over n_a}
\ee
which would vanish for an adiabatic perturbation.  The neglect of
$\delta T/T$ follows from the fact that $\delta \rho \sim m_a \delta n_a 
+\delta T n_\gamma =0$ for an entropy fluctuation \cite{Kolb:1990vq},
and $\rho_a\ll \rho$.  A very small temperature fluctuation can compensate
for $\delta n_a$ in the total energy density to keep $\delta\rho = 0$,
since only one degree of freedom is being compensated by many.

Running the CMB Boltzmann codes with isocurvature rather than adiabatic
fluctuations gives a pattern of acoustic peaks that look very different,
in strong disagreement with observations, as illustrated in fig.\ 
\ref{fig:isoc}.  Analysis shows that isocurvature modes can contribute no
more than about 7\% of the total perturbation.   Such considerations rule
out low values of the axion decay constant, depending upon the assumed
Hubble rate during inflation \cite{Ade:2015lrj}.

An interesting variation can occur if the isocurvature fluctuation is
able to decay later on and convert to adiabatic form, for example by
decay into radiation.  This opens the possibility that the inflaton
fluctuation could much smaller than normally assumed, and the perturbations
all arise from a second field, known as the curvaton \cite{Lyth:2001nq}.

\subsubsection{Nongaussianity}  The adiabatic fluctuations in standard inflation
behave as a Gaussian random variable, to a good approximation, with 
higher-point correlation functions being slow-roll suppressed:
\be
	\langle(\delta\phi)^3\rangle \sim V'''
\ee
The nonlinearity of gravity also induces nongaussian correlations, but
these are Planck-suppressed.  
The corresponding CMB temperature correlations $\langle(\delta T)^3\rangle$
are thus expected to be very small.  Any nongaussian correlations should 
ultimately arise from their counterparts in the curvature perturbation,
the bispectrum
\be
	\langle \zeta_{k_1} \zeta_{k_2}\zeta_{k_3}\rangle = f(\vec k_1,\vec
k_2)
\ee
which depends upon two independent wave vectors since translational
invariance implies $\sum \vec k_i=0$.  

In multifield models of inflation,
or single field models with complicated kinetic terms, significant
nongaussianity can arise.  A simple way of parametrizing it is to define
a nongaussian curvature perturbation contructed from the Gaussian one via
a nonlinearity parameter $f_{\rm NL}$ \cite{Komatsu:2001rj}
\be
	\zeta_{\rm NG} = \zeta + f_{\rm NL}\left(\zeta^2 -
\langle\zeta^2\rangle\right)
\ee
so that 
\be
	\langle\zeta\zeta\zeta\rangle \sim f_{\rm
        NL}\langle\zeta\zeta\rangle^2
\label{fNLts}
\ee
This gives a phenomenological way of characterizing the level of
nongaussianity that typical models like multi-field inflation might
predict, although the $k_i$-dependence (``shape'') of the bispectrum may 
not match the simple form predicted by (\ref{fNLts}) in a given model.
Planck currently constraints $|f_{\rm NL}| \lesssim 10$, depending upon
the shape.  Theorists continue to explore the possible implications of a 
future detection of nongaussianity; see for example \cite{Arkani-Hamed:2015bza}.

\subsubsection{Preheating}  Perturbative decay of the inflaton is a relatively
inefficient reheating mechanism, in the sense that typically the reheat
temperature is suppressed, $T_{\rm rh} \ll \Lambda_i$.  More efficient
means of particle production can occur in the background of the oscillating
inflation, namely parametric resonance, which can occur much faster than
perturbative decay \cite{Dolgov:1989us,Traschen:1990sw,Kofman:1994rk,Kofman:1997yn}.  This mechanism can
work even if the inflaton is a stable particle, allowing for the
possibility that the inflaton could be the dark matter 
\cite{Kahlhoefer:2015jma}.  For certain kinds of inflaton potentials
({\it e.g.,} $\varphi^n$ with $n>4$ during the reheating stage), 
conventional reheating may be particularly inefficient, and superseded by
gravitational particle production, a mechanism that only relies upon the
change in the time-dependence of the scale factor between inflation and 
conventional FRW expansion \cite{Ford:1986sy}.

\subsection{Current outlook}
The inflationary models that currently give the best fit to data seem to be
ones having a nonminimal coupling of the inflaton to gravity:
\be
	{\cal L}_J = \sqrt{-g_J}\left(\sfrac12 R_J(m_p^2 + \xi\varphi^2)
	 + \sfrac12(\partial\varphi)^2 - V(\varphi) 
	\right)
\ee
where the subscript $J$ denotes that we are in the Jordan frame, in which
the gravitational constant varies with time due to the evolution of
$\varphi$.  Cosmology is best understood by going to the Einstein frame
via a Weyl rescaling of the metric,
\be
	g_E^{\mu\nu} = {g_J^{\mu\nu}\over \Omega^2},\qquad
	\Omega^2 = 1 + \xi\varphi^2/m_p^2
\ee
This induces a complicated kinetic term for $\varphi$, and the canonically
normalized inflaton $\chi$ is related to $\varphi$ by
\be
	{d\chi\over d\varphi} = \left(\Omega^2 + 6\xi^2
\varphi^2/m_p^2\over
	\Omega^4\right)^{1/2}
\ee
In the Einstein frame, the Lagrangian reads
\be
	{\cal L}_E = \sqrt{-g_E}\left(\sfrac12 m_p^2 R_E 
	+ \sfrac12(\partial\chi)^2 - {V\over\Omega^4}\right)
\ee
The nice thing about this is that it makes any renormalizable potential
$V$ convex, favored by Planck data.  In particular if $V\sim \varphi^4$ at 
large field values, $V/\Omega^4$ becomes flat.  This feature enables allows
one to identify the standard model Higgs as the inflaton 
\cite{Bezrukov:2007ep}, or to rescue models of chaotic inflation with 
large values of the exponent \cite{Linde:2011nh}.

Even the best fitting model, Starobinsky's $R^2$ inflation, can be
understood in this way, by introducing a dimensionless auxiliary scalar field to
the Einstein-Hilbert action, with the Lagrangian \cite{DeFelice:2010aj}
\be
	{\cal L} = \sfrac12 m_p^2(1+ \varphi)R - \sfrac34 M^2 m_p^2(\varphi-1)
\label{staro}
\ee
Integrating out $\varphi$ trivially leads to the $R^2$ term in the
gravitational action,
\be
	{\cal L} \to \sfrac12 m_p^2 \left(R + {R^2\over 6 M^2}\right)
\ee
But instead of integrating it out, one can transform
to the Einstein frame, which generates a kinetic term for $\varphi$
and gives the canonically normalized inflaton a potential of the form 
\be
	V(\chi) \sim m_p^2 M^2 \left(1 - e^{-\sqrt{2/3}\chi/m_p}\right)^2
\ee
with the predicted spectral index and tensor ratio
\be
	n_s-1 \cong -{2\over N_*},\qquad r \cong {12\over N_*^2}\approx
10^{-4}
\ee

\subsection{Exercises}
1. {\bf Scalar field in FRW background.}\\
(a) Write the Lagrangian for a scalar field in a FRW gravitational
background and show that its equation of motion is
\[
	\ddot\varphi + 3 H \dot\varphi - {1\over a^2}\nabla^2\varphi
= -{\partial V\over \partial\varphi}
\]
where $H = \dot a/a$.\\
(b) Now consider a free massive scalar, with frequency $\omega \gg H$.
Using the ansatz
\[
	\varphi = \varphi_0\exp\left(-i\left[\int^t dt\, \omega - 
	\vec k\cdot\vec x\right]
	 -f(t)\right) + {\rm c.c.}
\]
show that $\omega = \pm\sqrt{(k/a)^2 + m^2}$ gives a solution, with
$f\sim \ln a$ if $m\ll H \ll k/a$ and $f\sim (3/2)\ln a$ if $m \gg
H$.\\
(c) The stress-energy tensor is
\[
	T_{\mu\nu} = {\delta S\over \delta g^{\mu\nu}} = 
	\partial_\mu\varphi\partial_\nu\varphi - \eta_{\mu\nu}
	{\cal L}
\]
Using the solution from (b), find $\langle T_{\mu\nu}\rangle$,
averaging over oscillations.  Average over directions of $\vec k$
to verify that $p=\rho/3$ if $m\ll k/a$ and $p\cong 0$ if $m\gg k/a$.
Also show that $\rho \sim 1/a^4$ or $1/a^3$ respectively, for these
two cases.\\
(d) A gas of relativistic scalar particles has $\rho = \pi^2 T^4/30$.
If the classical solution (b) gets thermalized into particles, how is
$T$ related to the parameters of that solution?  Repeat for a
nonrelativistic gas with $\rho\approx ({mT/2\pi})^{3/2}$.

\bigskip

2. {\bf Perturbative reheating.}  At the end of inflation, the Hubble
parameter is
\[
	H = {1\over \sqrt{3}\, m_p}\left(\rho_\varphi + \rho_r\right)^{1/2}
\]
where the oscillating and decaying inflaton has energy density
\[
	\rho_\varphi = {V_e\over a^3}e^{-\Gamma t}
\]
if $V_e$ was the scalar field energy density at the end of inflation
and $\Gamma$ is the decay rate.  (For convenience we take $a=1$ 
and $t\cong 0$ at
the end of inflation.)  $\rho_r$ is the energy density of the
radiation that is produced by the decays.  It satisfies the Boltzmann
equation
\[
	{d\over dt}\left( a^4 \rho_r\right) = a^4 \Gamma \rho_\varphi
\]
where the factor $a^4$ accounts for the redshifting and dilution of
the radiation in a comoving volume $a^3$.  At times before $1/\Gamma$,
it is a good approximation to ignore $\rho_r$ compared to $\rho_\varphi$
in the Friedmann equation, because of the redshifting, and also to 
approximate $e^{-\Gamma t/2}\cong 1$.  Use these approximations to 
simultaneously solve the two equations and show that $\rho_r \sim
(T/a)^4$ at late times $\gtrsim 1/\Gamma$, with the dependence on $V_e$ cancelling out.

\bigskip
3. {\bf Hilltop inflation.}  Analyze the hilltop inflation model,
$V = \Lambda^4(1-(\varphi/\mu)^2)$ in the slow-roll approximation,
assuming that horizon crossing of the relevant mode occurs at $N_* =
60$.  Hint: show that $\eta$ dominates the spectral index, but
$\epsilon$ controls the end of inflation.  Find $\mu$ and $\Lambda$
from the best fit to $n_s$ and $A_s$, as well as $\varphi_*$ (the value
of $\varphi$ at horizon crossing) and the 
predicted tensor-to-scalar ratio $r$.

\bigskip

4. {\bf Slow roll as attractor.}  The scalar field equation of motion
is second order and has two independent solutions, while the slow-roll
approximation is first order and has only one.  Find the two
independent solutions for the case where the initial condition is far from the 
slow-roll solution, and show that they quickly decay,
leaving only the slow-roll solution.  Hint: expand the EOM\ around
$\varphi = \varphi_{sr} + \delta\varphi$ to linear order in $\delta\varphi$ and
show that it satisfies
\be
 \ddot\delta\varphi + (3+\epsilon)H\dot\delta\varphi -
3(\epsilon-\eta)H^2\delta\varphi = \sfrac13(\epsilon-\eta)V'
\ee 
where $H$ and $V'$ are evaluated
at $\varphi_{sr}$.  Then solve it using problem 1, treating $H$ as 
approximately constant, and neglecting the small inhomogeneous term.

\bigskip

5. {\bf A step potential.}  Suppose $V$ has a sudden step when $\varphi$
crosses $\varphi_0$, 
\[
	V = \left\{\begin{array}{cc} a\varphi + V_2, & \varphi > \varphi_0\\
	b\varphi + V_1,& \varphi \le \varphi_0 \end{array} \right.
\]
(a) From the scalar equation of motion, derive an equation for the 
discontinuity in $\dot\varphi$ when $\varphi$ crosses $\varphi_0$.  \\
(b) Solve the slow-roll EOM\ on both sides of the step, and note
that $\dot\varphi$ does not have the right discontinuity.   However
a linear combination of the transient solutions you found in problem
4 can fix this problem; find it.  Note that it should be applied only
after crossing the step, since one can presume the field was slowly
rolling before it reached the step.\\
(c) The step leads to a modification of the power of the curvature or
density fluctuations at the scale corresponding to horizon crossing
when $\varphi$ passes the step.  What does it look like?

\section{Baryogenesis}

According to our timeline, fig.\ \ref{fig:timeline2}, the next likely 
important event of early cosmology may have been leptogenesis, or
baryogenesis, depending upon when the dark matter was produced.  In 
principle, the reheat temperature could have been too low for leptogenesis.
BBN tells us that $T_{\rm rh} \gtrsim 1\,$MeV at the lowest, but it is hard
to imagine creating the baryon asymmetry at such low temperatures.  In the
following highly abbreviated account, I will focus on leptogenesis and
electroweak baryogenesis as two popular theories for the origin of matter.

The present universe is observed to contain essentially only matter and
no antimatter, except for the rare antiparticles produced by cosmic rays.
There is an asymmetry between baryons and antibaryons (called the
baryon asymmetry of the universe, BAU) that can be expressed as
\be
	\eta = {n_B-n_{\overline B}\over n_\gamma} =
	\left\{\begin{array}{rl} [5.8-6.6]\times 10^{-10},&\hbox{from
BBN}\\
	(6.09\pm 0.06)\times 10^{-10},&\hbox{from CMB}\end{array} \right.
\ee
or
\be
	{n_B-n_{\overline B}\over s} = {\eta\over 7.04}
\ee
using the entropy density for reference.  In the context of inflation
it would be hard to imagine this asymmetry arising as an initial condition,
since entropy was generated during reheating, but any preexisting baryon
asymmetry was only diluted by the expansion of the universe, by a factor
of $e^{-3N}$.  It is therefore assumed that the BAU was created after inflation. 

\subsection{History} 

Sakharov  is acknowledged as the first to seriously consider baryogenesis,
in 1967 \cite{Sakharov:1967dj}, and three necessary conditions for creating the BAU
are attributed to him:  $B$ violation, departure from thermal equilibrium,
and violation of $C$ and $CP$.  In the paper, he did not spell out these
conditions as being necessary, but rather put forward a rather vague 
model of superheavy particle decay leading to a baryon asymmetry,
mentioning that the decays would violate thermal equilibrium, and noting
the need for $B$ and $CP$ violation.  There is no clear statement about the
necessity of all these ingredients.  It was only around 1978-79 that their
status started to be clarified.  Ref.\ \cite{Ignatiev:1978uf} presented a
model of $B$- and $CP$-violating out-of-equilibrium decays of a second Higgs
doublet, in which the three ingredients are clearly stated in the first
paragraph, and Sakharov is cited.  (In 1979 Sakharov himself wrote a paper
that was similar in spirit, in that it enunciated the conditions at the outset,
and cited the 1967 work, but in the context of GUTs.)

Outside of the Soviet Union, it took slightly longer to appreciate the need
for going out of thermal equilibrium.  Yoshimura \cite{Yoshimura:1978ex},
probably unaware of Sakharov's work, proposed that decay of GUT bosons
could produce the asymmetry, but he overlooked the out-of-equilibrium
criterion, and  only by neglecting a cancelling contribution obtained a
nonzero result. Its necessity was rigorously proven in ref.\
\cite{Toussaint:1978br} (also unaware of Sakharov) in a proposal for
baryogenesis via black hole evaporation.  That reference also clearly
shows the need for $C$, $CP$ and $B$ violation.

We will proceed by commenting on the three requirements.

\subsection{The three laws}

\subsubsection{$B$ violation}  Although baryon number is a symmetry of the SM 
Lagrangian, it is violated at the quantum level through the triangle
anomaly.  The baryon number current is
\be
	J_B^\mu = \sfrac13 \sum_f \bar q_f\gamma^\mu q_f
\ee
where the sum is over quark flavors, and its divergence is 
\cite{tHooft:1976rip}
\be
	\partial_\mu J_B^\mu  =  -n_f{g_2^2\over
	64\pi^2}\epsilon^{\mu\nu\rho\sigma} W^a_{\mu\nu} W^a_{\rho\sigma}
\ee
for $n_f$ families, 
in the presence of a background SU(2)$_L$ gauge field with strength
$W^a_{\mu\nu}$.

At zero temperature, the  $B$-violating effects are
through nonperturbative instanton configurations involving the Higgs field
and the $W$ fields, whose action is $S\sim 8\pi^2/g_2^2\sim 187$.  
Hence the tunneling rate per unit volume is of order
\be
	{\Gamma_{\rm inst}\over V} \sim v^4 e^{-2S}\sim 10^{-160} v^4
\ee
where $v\sim 100\,$GeV represents the weak scale. Hence the number of
baryon decays in the observable universe, over its lifetime, would be
\be
	H_0^{-4} {\Gamma_{\rm inst}\over V} \sim 10^{-25}
\ee
if only zero-temperature transitions were relevant.  But at high
temperatures $T \gtrsim v$, electroweak symmetry is restored and there is
no barrier between the $N$-vacua of the SU(2)$_L$ gauge theory. 
Transitions between these vacua result in the creation of quarks and
leptons such that $B$ and lepton number $L$ both change by 3 units,
one per generation \cite{Kuzmin:1985mm}.  For details, see for example \cite{Cline:2006ts}.

At high temperatures, sphalerons are the nonperturbative static field
configurations that violate $B$ and $L$.  One can think of the sphaleron
as the top of the energy barrier separating the $N$-vacua, at temperatures
below the electroweak phase transition (EWPT).  Above the transition, when
the barrier disappears, the sphaleron is no longer a well-defined field
configuration, but the same kind of $B$- and $L$-violating transitions still
occur, mediated by field configurations that have the same topological
properties as the low-temperature sphaleron.  These transitions are no
longer exponentially suppressed since there is no tunneling; they are only 
suppressed by powers of the weak gauge coupling.  One can visualize the
sphaleron transitions by a Feynman diagram with 9 left-handed quarks and 
3 left-handed leptons in the external states, even though there is no
analytic expression for the corresponding amplitude, at high $T$.  Instead,
lattice studies have determined the rate per volume of these transitions
\cite{Bodeker:1999gx},
\be
	{\Gamma_{\rm sph}\over V} = (1.05\pm 0.08)\times 10^{-6}\, T^4
\ee

In a thermal volume $V=1/T^3$ (which contains on average one particle of
each type in the early universe plasma), the corresponding rate goes as
$T$, while the Hubble rate goes as $H\sim T^2/m_p$.  This implies that
sphalerons come {\it into} equilibrium as $T$ falls below $10^{13}$\,GeV,
and only go back out of equilibrium at the EWPT when the exponential
suppression comes into effect.

Therefore one of the criteria for baryogenesis is already present in the
SM, provided that $T_{\rm rh} \gtrsim 100\,$GeV following inflation.
We will focus on two paradigms for baryogenesis that take advantage of
this, but let's mention a few other possible sources of $B$ violation from
physics beyond the SM.  It seems likely that such interactions should
exist, since $B$ is only an accidental symmetry of the SM: the particle
content and gauge symmetries of the SM forbid any $B$-violating operator
of dimension $\le 4$.

In the SU(5) GUT there are gauge bosons $X$ and $X'$ with interactions of
the form
\bea
	X_\mu \bar l_l \gamma^\mu d_R^c, &&\quad
	X_\mu \bar u_R^c \gamma^\mu q_L,\quad X_\mu \bar q_L \gamma^\mu
	e_R^c\nn\\
	X'_\mu \bar L_l\gamma^\mu u_R^c,&&\quad 
	X'_\mu \bar d_R^c\gamma^\mu q_L
\eea
where the superscript $c$ denotes charge conjugation.  Clearly there is no
consistent way to assign baryon number to $X_\mu$ or $X'_\mu$.  This source
of $B$ violation provided an early example of baryogenesis through out of
equilibrium decays of the heavy bosons \cite{Nanopoulos:1979gx}.

In supersymmetry (SUSY) it is possible to write $B$-violating operators if
$R$-parity is relaxed.  A superpotential term of the form
\be
	\lambda'' UDD
\ee
gives rise to potential terms in which two of the right-handed superfields
are replaced by the corresponding quark, while the third becomes a squark.
If $L$-violating operators are for some reason forbidden, or sufficiently
small, then proton decay will be suppressed despite the presence of the 
$UDD$ interactions.

\subsection{Loss of thermal equilibrium} 
Departure from thermal equilibrium is needed to get a BAU, since any process
in equilibrium has the same rate as its time-reversed process, by definition.
The out of equilibrium decay of a heavy particle like the $X$ or $X'$ of
SU(5) GUTs is a good example.  Suppose that $X$ can decay in such a way as
to produce more $B$ than $\bar B$.  The inverse decays will simply undo
this if the decays are in equilibrium.  To quantify it, we must consider
the Boltzmann equation for $n_B$, which we define to be the difference in 
densities of baryons and antibaryons.  For simplicity, focus on the $X'$
decays, and assume that there are equal numbers of $X'$ and its
antiparticle $\bar X'$ decaying in the plasma.  There must be a $CP$
asymmetry in the decays, such that the net baryon number per decay is given
by
\be
	\epsilon = {\sfrac13\Gamma(X'\to lu) - \sfrac23\Gamma(X'\to \bar
q\bar d) - \sfrac13\Gamma(\bar X'\to \bar l\bar u) + 
	\sfrac23\Gamma(\bar X'\to qd) \over \Gamma(X'\to{\rm\ any})
	+ \Gamma(\bar X'\to{\rm\ any})}
\ee
The Boltzmann equation is
\bea
	{1\over a^3} {d\over dt}(a^3 n_B) 
	&=& \dot n_B + 3H n_B \nn\\
	&=& (n_{X'} + n_{\bar X'})
	\Gamma_{X'} \epsilon - \sum_{y,z\atop X'{\rm\,or\,}\bar X'}
	B_{y+z} n_y n_z \sigma_{y+z\to\atop X'{\rm\,or\,}\bar X'} v_{\rm rel}
\label{Beq}
\eea
where $y,z$ denote the different possible initial states of the inverse
decays.  
The bottom line contains the collision terms of the Boltzmann equation, 
and they cancel each other, by definition, when the decays (and inverse
decays) are in thermal equilibrium, giving $\dot n_B=0$.

 The form
of (\ref{Beq}) makes it mysterious that two such different looking terms
could exactly cancel each other.  The original form of the collision term
in terms of amplitudes ${\cal M}$ makes it clear however.  Consider the contribution
from a single channel, say $X' \to lu$.  We label $X'$ as particle 1,
$l$ and $u$ as 2 and 3.  Defining the Lorentz-invariant phase space element
as $d\Pi_i = d^{\,3}p/[(2\pi)^3(2E)]$, the contribution to the collision term is
\be
\delta{\cal C} = \prod_i \int d\Pi_i\left[f_1(1-f_2)(1-f_3) - f_2
f_3(1+f_1)\right]|{\cal M}|^2 (2\pi)^4 \delta^{(4)}(p_1-p_2-p_3)
\ee
where $f_i$ are the Bose-Einstein or Fermi-Dirac distribution functions,
$(e^{\beta E}\pm 1)^{-1}$, and we have included the effect of Pauli
blocking or Bose enhancement in the final states.  The term in square
brackets can be written as
\be
	[\dots ] = f_1 f_2 f_3 \left[ e^{\beta(E_2+E_3)} - e^{\beta E_1}
	\right]
\label{cancel}
\ee
using $1-f = e^{\beta E}f$ for fermions and $1+f = e^{\beta E} f$
for bosons.	Clearly (\ref{cancel}) vanishes by energy conservation,
imposed by the delta function.  But of course we have assumed that the
distribution functions are those corresponding to thermal equilibrium
to get this result.

If $X'$ decays out of equilibrium, its true distribution function is not
Bose-Einstein, which would imply its number density is
\be
	n = {g\over (2\pi)^3} \int {d^{\,3}p\over e^{\beta E}-1}
	\cong g\left(mT\over 2\pi\right)^{3/2} e^{-m/T}
\ee
for $T\ll m$.  ($g=3$ is the number of degrees of freedom for a massive
vector.)  Instead it is much larger, 
\be
	n\sim {g\zeta(3)\over\pi^2} T^3 e^{-\Gamma t}
\ee
that is, the same as a radiation degree of freedom, except for the
particles that disappeared due to decays.  Therefore we can, in some
approximation, ignore the contribution of the inverse decays to
the collision term to estimate the baryon asymmetry.

Instead of using $a^3 n$ as the dependent variable (which is not a bad
choice since it remains constant under the Hubble expansion, in the absence
of collisions), it is convenient to use the proportional quantity
called the {\it abundance}
\be
	Y_i = {n_i\over s}
\ee
where $s = (\rho+p)/T$ is the entropy density,
\be
	s = {2\pi^2\over 45} g_{*,s} T^3
\ee
with
\be
	g_{*,s} = \sum_i g_i \left(T_i\over T\right)^3 \times
	\left\{\begin{array}{ll} 1,& \hbox{bosons}\\
	\frac78,& \hbox{fermions}\end{array}\right.
\label{gstars}
\ee
which allows for different species to have temperatures $T_i$ different
from that of the photon (due to kinetic decoupling of that species).

If we ignore the inverse decay term then the Boltzmann equation for the
baryon abundance simplifies to
\be
	\dot Y_B \cong 2 Y_{X'}\, e^{-\Gamma t}\, \Gamma\epsilon
\ee
which can readily be solved to yield the final abundance
\be	
	\left.Y_B\right|_{t\to\infty} = 
	2\epsilon \left.Y_{X'}\right|_{\rm initial}
\ee
This is exactly as we would expect from the fact that each $X'$ or
$\bar X'$ decay leads on average to $\epsilon$ baryons.

To make a better estimate, we can keep the inverse decay term in the
equation, and quantify it using the detailed balance argument,
\be
	\dot Y_B = 2\epsilon\Gamma\left( Y_{X'}(t) - Y_{X'}^{\rm
eq}(t)\right)
\label{Beq2}
\ee
which implies the collision term must vanish when $X'$ is in
equilibrium.  Its actual abundance can be estimated as $Y_{X',i}e^{-\Gamma
t}$ as before (where the initial value is 
$Y_{X',i}\cong 3/g_{*,s}\sim 10^{-2}$, since $X'$ is approximately three
degrees of freedom out of $g_{*,s}$ total, $\sim 320$ for supersymmetric
SU(5) GUT), 
while the equilibrium abundance is approximately 
\be
	Y_{X'}^{\rm eq}(t) \cong 
	Y_{X',i}\left\{\begin{array}{lr} 1,& T>m\\
	\left(m\over T\right)^{3/2} e^{-m/T},& T<m
	\end{array}\right.
\ee

It is convenient to define a new time variable, $x=m/T$; then
\be
	t = {0.301\, M_p\over \sqrt{g_*}\, T^2}	 = 
	{0.301\, M_p\over \sqrt{g_*}\, m^2} x^2 = {x^2\over 2 H(m)}
\ee
where $M_p = 1.22\times 10^{19}\,$GeV is the unreduced Planck mass, and
$m$ is the mass of $X'$. The last equality follows from the fact that $H=1/2t$ during radiation 
domination, where $H(m)$ denotes the Hubble parameter when $T=m$.
One can integrate (\ref{Beq2}) to find that 
\be
	Y_B(\infty) = {\epsilon Y_{X'}\Gamma\over H(m)}\left[
	2 {H(m)\over\Gamma} - O(1)\right]
\ee
We leave this as an exercise.  It shows that the washout correction is
small as long as $\Gamma\ll H(m)$.  This is exactly the condition for
the decays to be out of equilibrium: the reaction rate must fall below
the expansion rate.

\subsubsection{$C,CP$ violation}  In the previous example the role of the asymmetry
parameter $\epsilon$ was clear.  It is the quantity that vanishes if charge
conjugation $C$ or its combination with parity, $CP$, is conserved.  Let's
recall how fields transform under these discrete symmetries.  A Dirac
fermion $\psi$ goes as
\bea
	P: && \quad\psi = \left(\psi_L\atop\psi_R\right) \to 
	\left(\psi_R\atop\psi_L\right)(t,-\vec x)\nn\\
	C: &&  \quad\psi \to \left(\sigma_2\psi_L^*\atop 
	-\sigma_2 \psi_R^*\right)(t,\vec x)
	\nn\\
	CP: && \quad\psi \to \left(\sigma_2\psi_R^*\atop 
	-\sigma_2\psi_L^*\right)(t,-\vec x)
\eea
while for a complex vector $X^\mu$,
\bea
	P: && X^\mu\to (X^0,-X^i)(t,-\vec x)\nn\\
	C : && X^\mu \to -(X^\mu)^*(t,\vec x)\nn\\
	CP: && X^\mu \to (-X^0,X^i)^*(t,-\vec x)
\eea
One can show that under $CP$, a Lagrangian allowing for
$X\to \bar\chi\psi$ decays transforms as
\be
	\lambda\bar\psi X^\mu\gamma_\mu \chi + \lambda^* \bar\chi
	\bar X^\mu\gamma_\mu \psi \quad\to\quad
	\lambda^*\bar\psi X^\mu\gamma_\mu \chi + \lambda \bar\chi
	\bar X^\mu\gamma_\mu \psi
\label{Clag}
\ee
Therefore one needs a complex coupling, $\lambda\ne\lambda^*$, to have 
$CP$ violation.  This is exactly how $CP$ violation comes into the standard
model Lagrangian, via the CKM matrix in the $W$ boson couplings to quark
currents.

While complex couplings are a necessary condition for $CP$ violation, they
are not sufficient.  Notice that at tree level, the decay rates
$\Gamma(X\to \psi\bar\chi)$ and $\Gamma(\bar X\to \bar\psi\chi)$ both
go like $|\lambda|^2$, where the phase is irrelevant.  In fact the phase is
unphysical in the simple model (\ref{Clag}) since it can be removed by
a field redefinition,
\be
	X^\mu \to e^{-i\theta} X^\mu	\hbox{\quad  or\quad  }
	\chi\to e^{-i\theta} \chi \hbox{\quad  or\quad  } 
	\psi\to e^{i\theta} \psi
\ee
where $\lambda = |\lambda|e^{i\theta}$.

\begin{figure}
\centerline{\includegraphics[width=0.5\textwidth]{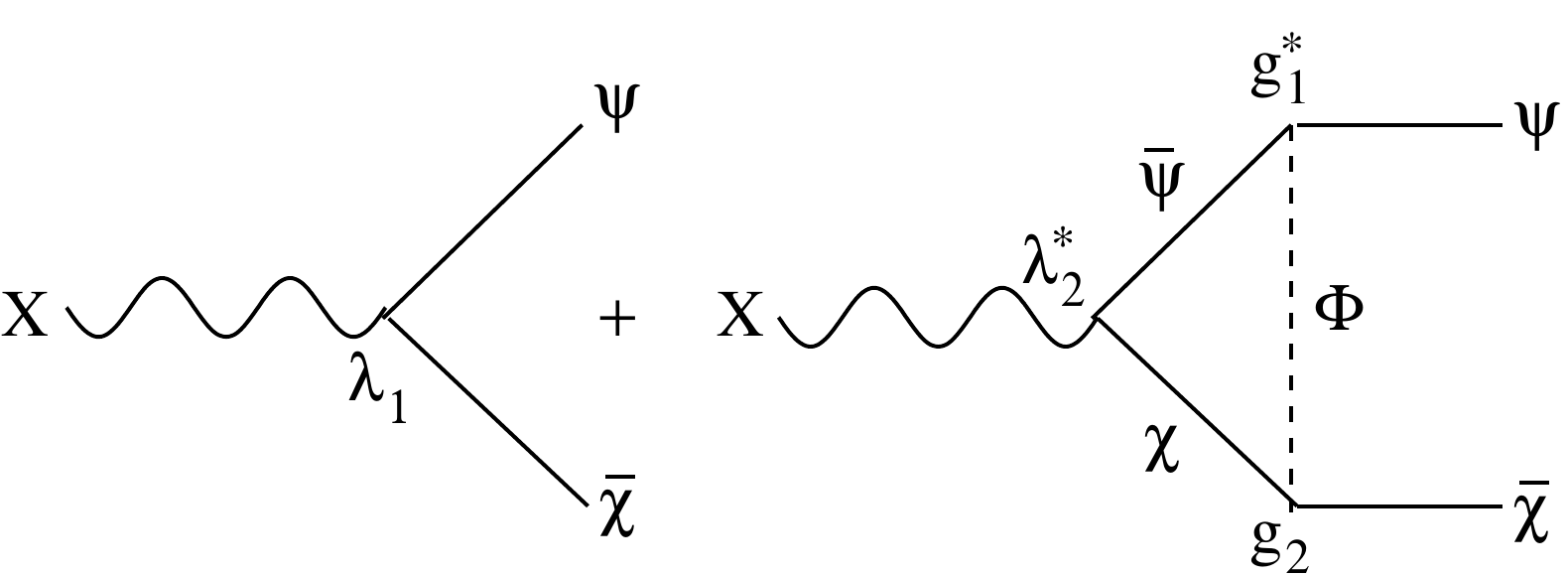}\hfil
\raisebox{5mm}{\includegraphics[width=0.25\textwidth]{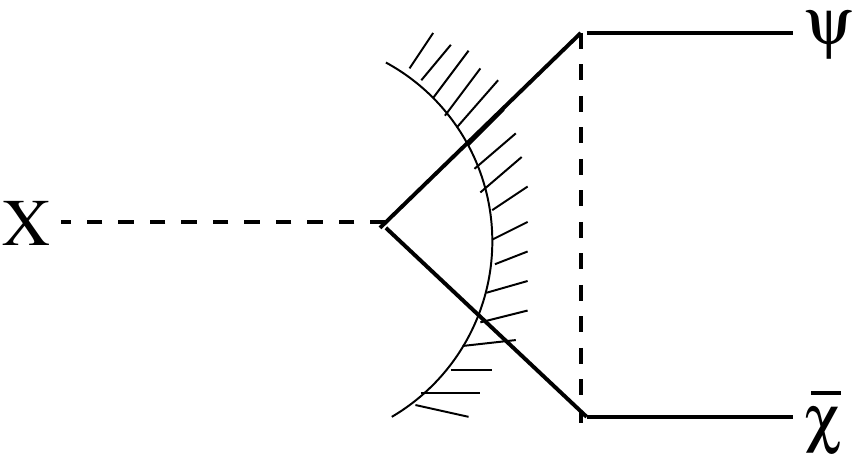}}}
\caption{Left: toy model for generating $CP$ asymmetry in $X\to\psi\bar\chi$ versus
$\bar X\to\bar\psi\chi$ decays.  Right: cut version of loop, for applying
the Cutkosky rule.}
\label{fig:toy-diags}
\end{figure}

To make a phase physical, we need more interactions---enough so that not
all phases can be removed by field redefinitions.  For example
\be
	\lambda_1\bar\psi\slashed{X}\chi + 
	\lambda_2\bar\psi\slashed{\bar X}\chi + g_1 \Phi\bar\psi\psi^c
	+ g_2\Phi\bar\chi\chi^c +{\rm h.c.}
\ee
where $\Phi$ could be a real or complex scalar.  Now consider the decay
$X\to\psi\bar\chi$ including one-loop corrections, as shown in fig.\
\ref{fig:toy-diags}(left).  The interference between tree and loop diagrams
produces the $CP$ asymmetry,
\be
	\epsilon = {|{\cal M}_{X\to\psi\bar\chi}|^2 - 
			|{\cal M}_{\bar X\to\bar\psi\chi}|^2
	\over |{\cal M}_{X\to\psi\bar\chi}|^2 + 
			|{\cal M}_{\bar X\to\bar\psi\chi}|^2}
\ee
because the loop diagram gets an imaginary part, predicted by the optical
theorem or unitarity of the $S$-matrix.  This tells us that the matrix
element has an imaginary part
\be
	2\,{\rm Im}\, {\cal M}_{X\to\psi\bar\chi} = 
	\sum_{\bar\psi,\chi\atop
	\hbox{spins}} \int d\Pi_{\bar\psi}\, d\Pi_{\chi} \,
	{\cal M}_{X\to\bar\psi\chi} \,
	{\cal M}_{\bar\psi\chi\to\psi\bar\chi}
\ee
The r.h.s.\ corresponds to the loop diagram shown in fig.\ 
\ref{fig:toy-diags}(right) with the internal propagators ``cut'' (put on
their mass shell) according to the Cutkosky rule \cite{Cutkosky:1960sp},
where the cut propagators are replaced by
\be
	{i\over p^2-m^2+i\epsilon} \quad\to\quad
	2\pi\,\theta(p_0)\, \delta(p^2-m^2)
\ee
This rule also applies to fermionic propagators once they have been
rationalized, $i(\slashed{p}-m + i\epsilon)^{-1} =
i(\slashed{p}+m)/(p^2-m^2+i\epsilon)$, {\it i.e.,} just multiply everything
by $(\slashed{p}+m)$.

As a consquence, the form of the tree plus loop contributions for the 
respective decays is
\bea	{\cal M}_{X\to\psi\bar\chi} \ \sim\ \lambda_1 + \lambda_2^*\, g_1^*
g_2\,(A+iB)\nn,\\
{\cal M}_{\bar X\to\bar\psi\chi} \ \sim\  \lambda_1^* + \lambda_2\, g_1
g_2^*\,(A+iB)\nn,\\
\eea
The important point is that there is no complex conjugation of $A+iB$,
which arises purely from the kinematics of the loop.  We then infer
\be
	\epsilon \cong {iB\over|\lambda_1|^2}
	(\lambda_1\,\lambda_2\, g_1\, g_2^* - {\rm c.c.})
	= -{2B\over |\lambda_1|^2} {\rm Im}(\lambda_1\,\lambda_2\, g_1\, g_2^*)
\label{toyCP}
\ee
In this example, the $CP$-violating phase that is invariant under field
redefinitions is the argument of $\lambda_1\lambda_2 g_1 g_2^*$, so indeed
we needed all four couplings nonvanishing to have $CP$ violation in this 
toy model.

\begin{figure}
\centerline{\includegraphics[width=0.25\textwidth]{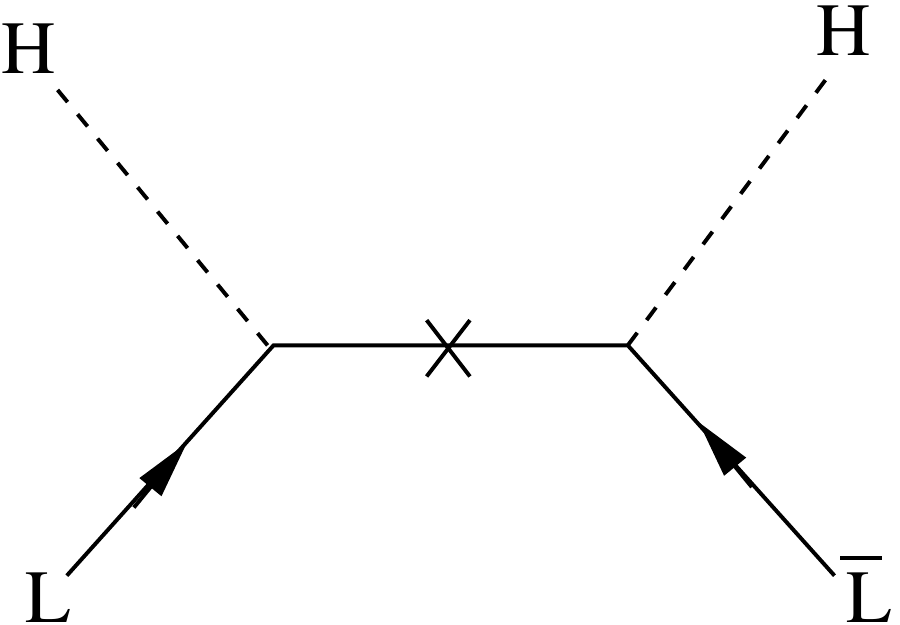}}
\caption{Neutrino mass operator induced by integrating out heavy sterile
neutrinos.}
\label{fig:seesaw}
\end{figure}

\subsection{Leptogenesis}
Leptogenesis is one of the most popular models for explaining the baryon
asymmetry, because it requires hardly any new physics ingredients beyond
those needed to explain neutrino masses by the seesaw mechanism.  This is
most simply accomplished by introducing several 
heavy right-handed neutrinos $N_i$, singlets under the SM gauge group, that
couple to lepton doublets and the Higgs via
\be
	y_{\nu,ij}\bar L_i \tilde H N_{R,j} + {\rm h.c.}
\ee
where $\tilde H = i\tau_2 H^* = (H^{0*}, -H^-)^T$.  The heavy neutrinos
have Majorana masses (without loss of generality we can work in the mass
eigenbasis for the $N_i$s):
\be
	\sfrac 12 M_j \bar N_j N_j^c
\label{MM}
\ee
Integrating out the heavy $N_j$ produces the dimension-5 Weinberg operator
\be
	(y_\nu M^{-1} y_\nu^T)_{ij}(\bar L_i\tilde H)(\tilde H^T L^c_j)
\ee
as shown in fig.\ \ref{fig:seesaw}.  Electroweak symmetry breaking then
gives the light neutrino mass matrix
\be
	m_{\nu,ij} = v^2 (y_\nu M^{-1} y_\nu^T)_{ij}
\ee
with $v=246\,$GeV.

The cross in fig.\ \ref{fig:seesaw} denotes an insertion of the heavy 
neutrino mass in the internal propagator, to explain the clash of arrows
showing the flow of lepton number.  The Majorana mass term (\ref{MM}) is
the only interaction in the theory that violates lepton number (by two
units), and so it must be involved.  Therefore only the $M/(p^2-M^2)$
part of the propagator $(\slashed{p}+M)/(p^2-M^2)$ contributes.  Of course
$p^2$ is negligible at low energy so we can neglect it in the denominator.
There are also analogous loop diagrams with no mass insertions or
clashes of arrows, that correct the real part of the amplitude, but
do not contribute to $CP$ violation.

\begin{figure}
\centerline{\includegraphics[width=0.6\textwidth]{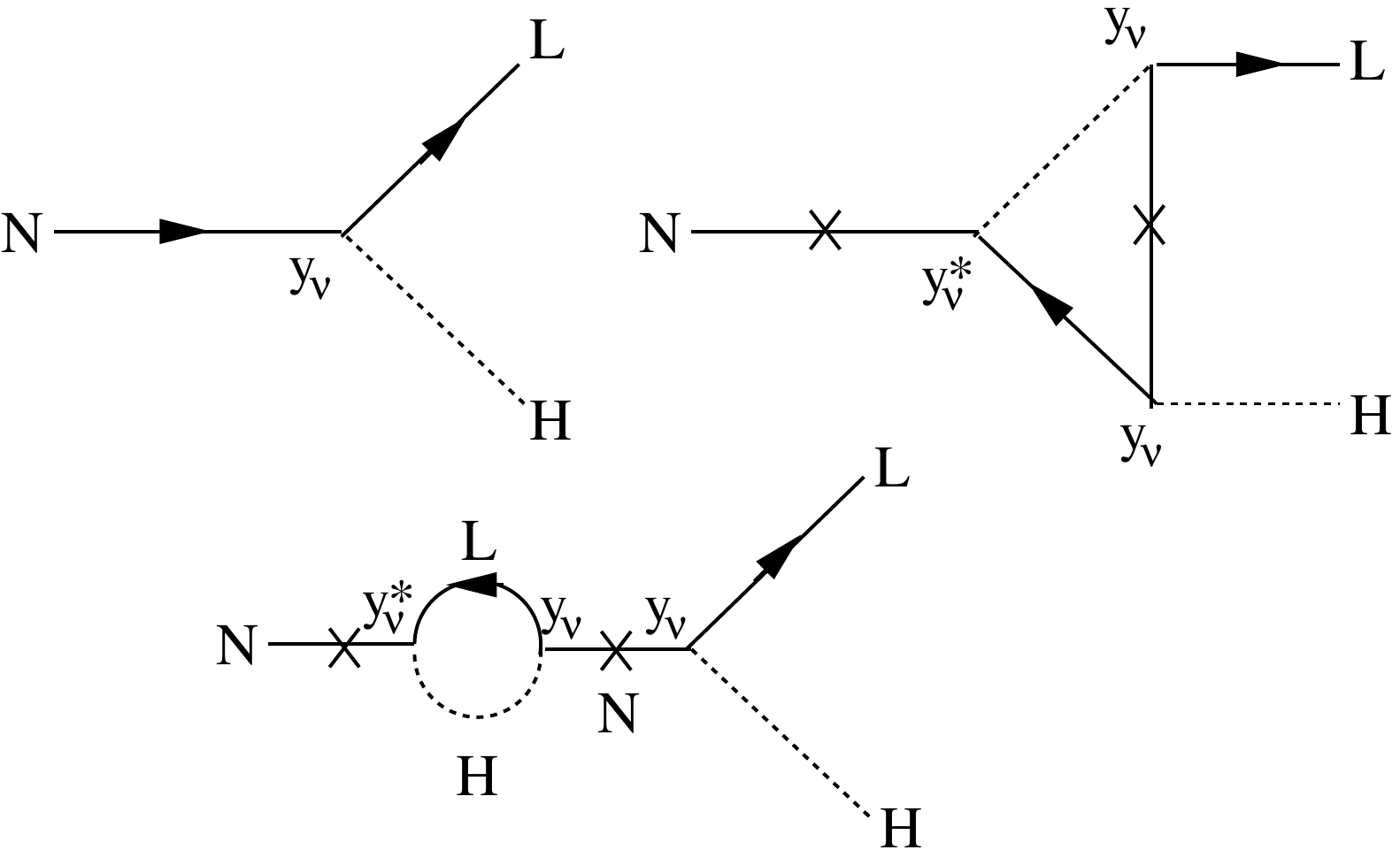}}
\caption{$CP$ asymmetric decay of heavy sterile neutrinos for leptogenesis.}
\label{fig:leptogen}
\end{figure}

The out-of-equilibrium decays of $N_j\to L_i H^c$ can create a lepton
asymmetry by the mechanism illustrated in our toy model.  The relevant
diagrams are given in fig.\ \ref{fig:leptogen}.  This produces a lepton
asymmetry, which sphaleron interactions will partially convert into
the baryon asymmetry, with $Y_B\sim -\sfrac13 Y_L$.  There are many 
useful reviews of leptogenesis, including 
\cite{Buchmuller:2004nz,Chen:2007fv}.  I will give a
brief synopsis of the main results here.

First, the $CP$ asymmetry for decays of $N_j$ can be computed from the
diagrams in fig.\ \ref{fig:leptogen} as in the toy model, by interfering
the tree-level and cut loop diagrams.  The result takes the form
\be
	\epsilon_j = -{3\over 8\pi} \sum_{i\neq j} {{\rm Im}[
	(y_\nu^\dagger y_\nu)_{ij}^2]\over (y_\nu^\dagger y)_{jj}}\,
	f\!\left(M_j\over M_i\right)
\ee
where $f$ is the loop function.  In the case of hierarchical heavy neutrino
masses, with $M_1\ll M_{2,3}$, $f\to M_1/M_i$.  It is obvious that we need
at least two families of $N_j$ since if $i=j=1$ there is no imaginary part.
By considering all possible forms for the neutrino Dirac mass matrix,
Davidson and Ibarra derived a famous bound in the hierarchical case
\cite{Davidson:2002qv},
\be
	|\epsilon_1| \le {3\over 16\pi} {M_1\over v^2}(m_3-m_2)
\label{DI}
\ee
in terms of the light neutrino mass eigenvalues $m_i$.  
Atmospheric neutrino oscillations fix $m_3^2-m_2^2 \sim (0.05\,{\rm
eV})^2$, so if the light $\nu$ mass spectrum is also hierarchical,
the bound (\ref{DI}) implies
\be
	|\epsilon_1| \lesssim {M_1\over 2\times 10^{16}\,{\rm GeV}}
\ee

If $M_1\ll M_{2,3}$, we expect that $N_1$ decays will give the dominant
contribution to the lepton asymmetry, because of $\Delta L=2$
{\it washout} processes mediated by $N_1$ exchange, shown in fig.\ 
\ref{fig:seesaw} where $N_1$ is the virtual state.  This interaction tends
to relax the asymmetry from $N_{2,3}$ decays toward zero until $T< M_1$, when it starts to
go out of equilibrium.  By this time the heavier neutrinos are long gone.
But this washout process also reduces the asymmetry from the $N_1$ decays
and must be included in the Boltzmann equations to get an accurate result.
One can consider two coupled equations, one for the lepton abundance
and one for that of $N_1$, which take the form \cite{Buchmuller:2004nz}
\bea
	{dY_{N_1}\over dx} &=& (-D + S)(Y_{N_1}- Y_{N_1}^{\rm eq})\nn\\
	{dY_{L}\over dx} &=& -\epsilon_1 D(Y_{N_1} - Y_{N_1}^{\rm eq})
	- W Y_L
\label{Beqlepto}
\eea
Here $D$ is the decay rate, $S$ is the rate of $\Delta L=1$ scatterings
(for example $NL\to t Q_3$ mediated by $H$ exchange in the $s$-channel),
and $W$ is the washout rate from both $\Delta L=1$ and 2 scatterings.

The out-of-equilibrium criterion for the $N_1$ decays can be quantified
in terms of an ``effective neutrino mass''  $\tilde m_1$
and an ``equilibrium neutrino mass'' $m_*$,
\be
	\tilde m_1 \equiv {v^2\over M_1}(y^\dagger_\nu y_\nu)_{11}
	\qquad<\qquad m_* \equiv {16\pi^{5/2}\over 3\sqrt{5}}\sqrt{g_*} {v^2\over M_p}
	\ \cong\  10^{-3}\,{\rm eV}
\ee
$\tilde m_1$ is 
only ``effective'' because it has the wrong kind of indices (belonging to
the heavy right-handed neutrinos) to be part of the light neutrino mass
matrix.
We emphasize that the inequality need not be literally satisfied; instead
one will pay a penalty in the produced asymmetry going as $m_*/\tilde m_1$
if it is not, on which we will elaborate shortly.

Recall our naive estimate in the toy model that
\be
	Y_L \sim {\epsilon\over g_{*,s}}
\ee
based on the idea that each $N_1$ decay yields on average $\epsilon$
leptons.  This ignores the effect of washout, but it can be corrected by
introducing an ``efficiency factor'' $\kappa$, 
\be
	Y_L \sim \kappa{\epsilon\over g_{*,s}}
\ee
which can be calculated in a given model by numerically solving the
Boltzmann equations.  One finds that $\kappa$ can be fit to an approximate
formula \cite{Buchmuller:2004nz}
\be
	\kappa \cong 0.02\left(0.01\,{\rm eV}\over \tilde m_1\right)
\label{keq}
\ee
if $\tilde m_1\gtrsim 10^{-3}$\,eV, which is known as the ``strong washout
regime.''  In this case, $N_1$ is decaying not far out of equilibrium, so
the washout processes are very effective, and the lepton asymmetry is
accordingly reduced.  Why would one want to work in this inefficient
regime?  It has the advantage of offering the nice approximation
(\ref{keq}), and moreover it is more predictive because the final lepton
asymmetry is not sensitive to initial conditions, including the reheat
temperature, as is the case in the weak washout regime.  And besides, 
great efficiency is not needed because the baryon asymmetry is quite small.

Putting these results together, we find the correct magnitude of the 
BAU if $M_1\gtrsim 3\times 10^{10}$\,GeV and $(y_\nu^\dagger
y_\nu)_{11}^{1/2}\gtrsim 0.002$.  These have the right order of magnitude
to agree with the observed mass difference $m_2^2-m_1^2=(0.0086\,{\rm
eV})^2$.  Hence leptogenesis seems to be a very natural and plausible
theory of baryogenesis.  Its main weakness is that it does not
make any unambiguous predictions for low-energy observables.  For
example, there are 6 unremovable phases in $y_{\nu,ij}$ if there are three
families of $N_j$, but only 3 phases in the PMNS matrix that describes
light neutrino mixing.

There exist variants of standard leptogenesis that are more testable.
One possibility is to use a more complicated version of the seesaw
mechanism, called inverse seesaw  \cite{Mohapatra:1986bd} that allows
a natural explanation of the light neutrino masses using a much lower
scale of heavy Majorana masses.  Thus the new physics needed for
leptogenesis can come down to testable scales \cite{Blanchet:2010kw}.
Another way is to have nearly degenerate $M_i$.  Then the loop function
$f(M_1/M_i)$ can be resonantly enhanced, to values of order $y_\nu^{-2}$,
allowing $\epsilon_j\sim 1$ and $M_i$ at the TeV scale.  This is known as
{\it resonant leptogenesis} \cite{Pilaftsis:2003gt}.  I leave as an
exercise to show that for quasi-degenerate $M_i$,
\be
	\epsilon_i \sim \sum_j{{\rm Im}(y_\nu^\dagger y_\nu)^2_{ij}
	\over (y^\dagger_\nu y_\nu)_{ii} (y^\dagger_\nu y_\nu)_{jj}}
	\, {\delta M^2_{ij} M_i\Gamma_j\over (\delta M^2_{ij})^2
	+ M_i^2\Gamma_j^2}
\ee 
where $\delta M^2_{ij} = M^2_i - M^2_j$ and $\Gamma_j$ is the decay 
rate of $N_j$.  Resonant leptogenesis makes predictions for low energy
lepton-violating processes like neutrinoless double beta decay, $\mu\to e
\gamma$, $\mu\to 3e$.  The TeV-scale heavy neutrinos might be produced at a
future $e^+e^-$ or $\mu^+\mu^-$ collider \cite{Pilaftsis:2005rv}.

\subsection{Electroweak baryogenesis}
 
 A more testable scenario relies on the electroweak phase transition (EWPT)
being first order, to achieve the out-of-equilibrium condition.  Bubbles of
true vacuum, with $\langle H\rangle = v/\sqrt{2}$ in their interior,
nucleate during such a transition, and the sphaleron rate $\Gamma_{\rm
sph}$ is exponentially  suppressed inside the bubbles.  If $v(T)/T\gtrsim
1$ at the time of nucleation, $\Gamma_{\rm sph}$ is small enough to avoid
washout of a baryon asymmetry produced by $CP$-violating interactions at
the bubble walls.   In the  standard model, the EWPT is a crossover transition,
not first order.  New physics at the weak scale is needed to make it first
order.

Moreover new sources of $CP$ violation beyond the SM are needed to produce a large enough 
BAU.  A simple argument involves the Jarlskog determinant
\cite{}, which is an
invariant measure of the $CP$-violating phase in the SM, analogous to the 
r.h.s.\ of (\ref{toyCP}) in our toy model,
\be
	J = {\rm det}\left[m_u^2,m_d^2\right]
	\sim \sin\delta \times f(|V_{CKM}|,m_{u_i}^2-m_{d_j}^2)
\ee
where $m^2_u$ and $m^2_d$ are the mass matrices of the up- and down-type
quarks, $\delta$ is the phase in the CKM matrix, and $f$ is a function
of the magnitudes of $V_{CKM}$ matrix elements and quark mass eigenvalues.  This quantity has
dimension 12, so the dimensionless measure of $CP$ violation should be
$J/v^{12}\sim 10^{-20}$ since $v=246\,$GeV is presumably the relevant scale.  This
is too small of course.  A valiant attempt to lower the relevant scale was
made in ref.\ \cite{Farrar:1993sp,Farrar:1993hn}, but this turned out to 
be spoiled by quantum decoherence by finite temperature scattering
\cite{Gavela:1993ts}.  

\subsubsection{The $CP$ asymmetry}

For electroweak baryogenesis (EWBG), the $CP$ violation needs to lead to a $CP$ 
asymmetry $\delta f_{CP} = f_L-f_R$ in some species of SM fermions: for
example an excess of left-handed (LH) versus right-handed top quarks, in the
symmetric phase outside of the bubble where they are massless and it makes
sense to talk about chirality \cite{Cohen:1990it,Cohen:1990py}. Sphalerons,
which interact only with the LH particles, would like to reduce the excess
in LH particles (minus their antiparticles).  This necessarily converts
some $O(1)$ fraction of $\delta f_{CP}$ into a baryon asymmetry.  
For example, suppose that $\delta f_{CP}$ is intially an excess in $t_L$:
$\delta f_{t_L} = -\delta f_{t_R} = \delta f_{CP}$.  Sphalerons can 
reduce $\delta f_{t_L}$ by roughly $1/2$ by partially converting it into
$\bar d_L$ and $\bar s_L$ asymmetries, for example, 
that couples the sphaleron to each generation.  
The net result would be
\be
	\delta f_{t_L} \to \delta f_{CP}/2,\quad \delta f_{d_L} = 
	\delta f_{s_L} = - \delta f_{CP}/2
\ee
This arrangement increases the entropy, and produces a baryon asymmetry since
the net baryon number is now proportional to
\be
	\delta f_{t_L} + \delta f_{d_L} + \delta f_{s_L} + \delta f_{t_R}
 = \left(\sfrac12 - \sfrac12 - \sfrac12 - 1\right) \delta f_{CP}
\ee
Charge is conserved by producing the neutrinos in this example.  This argument is
heuristic; we will give proper equations below.

\begin{figure}
\centerline{\includegraphics[width=0.5\textwidth]{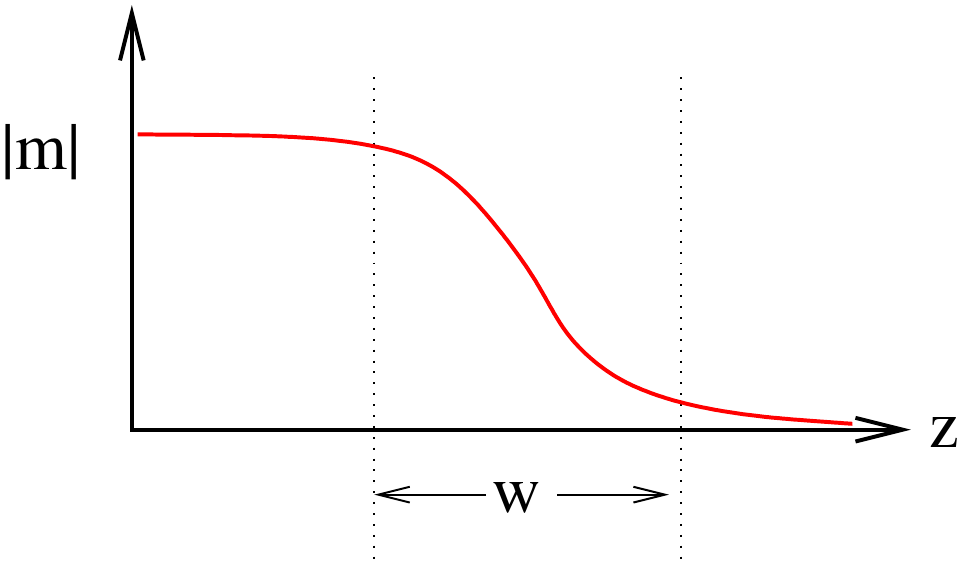}}
\caption{Spatially dependent fermion mass as a function of distance
transverse to a bubble wall during a first-order EWPT.}
\label{fig:mass}
\end{figure}

Computing the $CP$ asymmetry $\delta f_{CP}$ is not easy in EWBG, and it is
also controversial because the community is split between two formalisms.
The one I prefer, because it is known to be a controlled approximation,
is the semiclassical force approach \cite{Joyce:1994fu,Joyce:1994zt}.
It usually involves a fermion whose mass varies spatially across the bubble
wall (taken to be the $z$ direction), with Lagrangian
\be
	{\cal L} = \bar q \left(i\slashed{\partial} - [m_r(z) + i
	m_i(z)\gamma_5]\right) q
\label{Lmass}
\ee
We can rewrite the mass term in the form
\be
	m(z) = |m| e^{i\theta(z)\gamma_5}
\ee
imagining that the mass comes from the Higgs mechanism so that
\be
	|m(z)| = y h(z)
\ee
in the bubble wall, with some Yukawa coupling $y$.  The $z$-dependence
is illustrated in fig.\ \ref{fig:mass}.  $\theta(z)$ is a $CP$-violating
phase that arises from new physics.

One can solve the Dirac equation associated with (\ref{Lmass}), in an
expansion in derivatives of $\theta$, and from that obtain the dispersion
relation for the fermion in the background Higgs field.  An eigenstate of
energy has a spatially varying momentum: since the particle mass increases
as it goes into the broken phase, its momentum must decrease to conserve
energy.  This gives a $CP$-conserving force on the particle, $F = \dot p$.
But at first order in $\theta$ one finds a $CP$-violating component in the
force, that acts oppositely on left- versus right-handed states and
particles versus antiparticles.  The total force at this order is
\cite{Cline:2000nw}
\be
	F \cong -{(m^2)'\over 2E} \pm {(m^2\theta')'\over 2E^2} +
O(\theta^2)
\ee
where the first term is  $CP$-conserving and the second is $CP$-violating.

The semiclasscial force can induce the $f_{CP}(z)$ asymmetry in the
distribution function for the quark, which then biases sphaleron
interactions.  To calculate this effect, we need to put the effect of a
force back into the Boltzmann equation.  It comes via the Liouville
operator on the left-hand-side,
\be
	\left({d\over dt} + \vec v\cdot\vec\nabla + {\vec F\over m}\cdot
	\vec\nabla_p\right) f = {\cal C}[f]
\ee
This is a partial differential-integral equation which is intractable.  We
usually just take the first moment of it
 by integrating over momenta to get an equation for the particle densities.
But in the present case that is not an adequate approximation because the
particles are disturbed away from kinetic as well as chemical equilibrium
by the forces in the bubble wall \cite{Joyce:1994zt}.  

It turns out to be
sufficient to perturb the distribution function around the equilibrium one
by introducing a small $z$-dependent 
chemical potential $\mu(z)$, to keep track of the local number density, plus an
extra perturbation $\delta f_{\rm kin}$ whose exact form is unknown, but which is
assumed to be an odd function of the $z$ component of the particle's
velocity $v_z$ \cite{Cline:2000nw}.  Then we can take one additional moment, by integrating over
momentum weighted by $v_z$.  This introduces an dependent variable called the 
velocity perturbation,
\be
	u = \left\langle {p_z\over E}\, \delta f_{\rm kin}\right\rangle
\ee
Therefore each fermion interacting with the wall has associated with it
four dimensionless functions, $\xi_\pm = \mu_\pm/T$ and $u_\pm$, labeled by helicity, and four
first-order coupled diffusion equations, of the general form
\bea
	K_1 v_w\, \xi_\pm' + K_2 v_w\, u_\pm + u'_\pm
-\sum_{ijk}\Gamma_{ijk}(\xi_i,\xi_j,\xi_k) &=& 0  \nn\\
	K_3 \xi'_\pm + K_4 v_w\, u'_\pm + K_5 v_w\, u_\pm + \Gamma_\pm \xi_\pm 
	&=& \pm S
\label{diffeqs}
\eea
where we have linearized in the wall velocity $v_w$, and $K_i$ are
functions of $|m|/T$ \cite{Fromme:2006wx} (beware that I am not following
the same numbering scheme for the $K_i$ here).  The three-particle
interactions are represented by $\Gamma_{ijk}(\xi_i,\xi_j,\xi_k)$ which is
a rate times a linear combination of the appropriate chemical potentials.
The $\Gamma_\pm$ term represents the helicity-flipping rate inside the
bubble (and also outside, for species that have a bare mass term), 
which damps the $CP$ asymmetry.  The source term $S$ is due to the classical
force, and takes the form
\be
	S = v_w K_6(m^2\theta')'
\ee
The network (\ref{diffeqs}) should be solved numerically to get an accurate
estimate, since the solutions tend to change sign in the vicinity of the
wall, and large cancellations can occur in their local contributions to the
rate of baryon production.

Once the $CP$ asymmetries of the left-handed quarks and leptons 
are known, we can find the rate of baryon
violation by sphalerons by integrating
\be
	\dot n_B = -\frac32 {\Gamma_{\rm sph}\over T}\sum_{i=1}^3
	\sfrac12\left(3 \mu^i_{u_L} + 3 \mu^i_{d_L} + \mu^i_{e_L} + \mu^i_{\nu_L}
	\right)
\label{eq:gsph}
\ee
where the sum is over generations.  The extra factor of $1/2$ is
missing in many references in the literature (in some cases, both
factors of $1/2$).  It is more properly
understood as summing over chemical potentials of each doublet,
rather than each member of a doublet, since a single sphaleron
interaction can only produce one member at a time.  The normalization
in Eq.\ (\ref{eq:gsph}) is correct for the conventional definition of 
$\Gamma_{\rm sph}$ as the rate of Chern-Simons number diffusion
\cite{Moore:1996qs}, as
measured in lattice studies \cite{Bodeker:1999gx}.  

  The baryon density is then
\be
	n_b = \int dt\, \dot n_B = \int_{-\infty}^\infty {dz\over v_w}\dot
n_B \cong \int_0^\infty {dz\over v_w} \dot n_B
\label{nbeq}
\ee
by changing variables from time to space via wall trajectory $z = v_w t$.
The last approximation takes the sphaleron rate to vanish inside the bubble
(the bubble is so large by this time that it is well-approximated as being
planar, and we take $z<0$ to be the interior region) and the wall to be
infinitesimally thin, but it is better to weight the integral by the local
sphaleron rate, that smoothly interpolates to its value in the broken phase
\cite{Moore:1998ge}, and integrate over both regions.

\begin{figure}
\centerline{\includegraphics[width=0.4\textwidth]{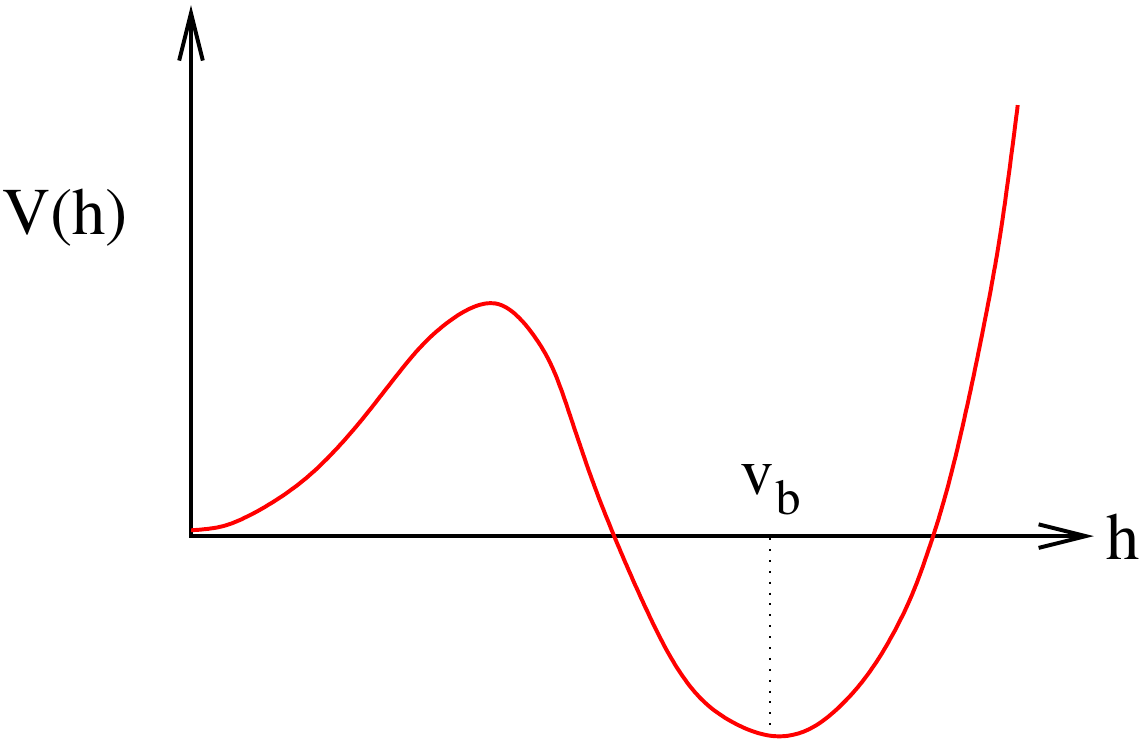}\hfil
\includegraphics[width=0.4\textwidth]{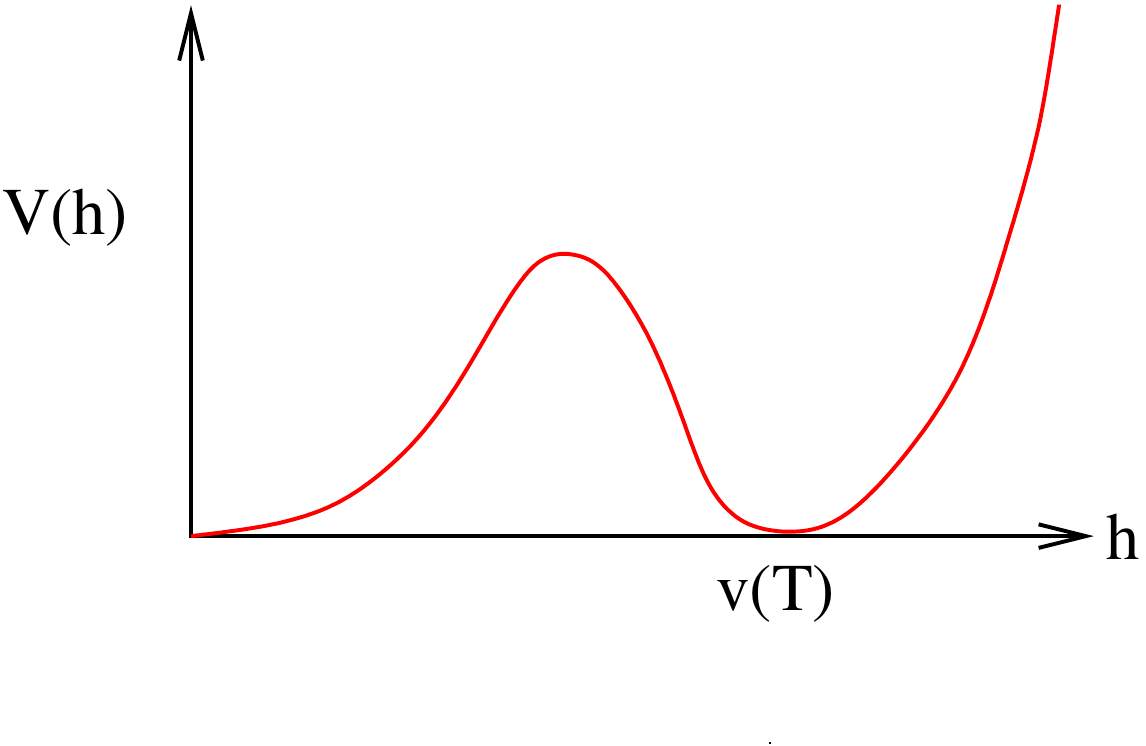}}
\caption{Left: form of the Higgs potential at the nucleation temperature
$T_n$, consistent with geting a
first-order phase transition.  Right: the corresponding picture at the
critical temperature $T_c > T_n$.}
\label{fig:PT}
\end{figure}

\subsubsection{Getting a first-order EWPT}

So far we have described the means of violating $CP$, but we still need to
fulfill the out-of-equilibrium condition for $\Gamma_{\rm sph}$ inside the
bubbles.  One possibility that I will not discuss here is known as
``cold electroweak baryogenesis,'' where the Higgs and gauge fields
are far from equilibrium during preheating after inflation, and the
reheat temperature is below the electroweak scale 
\cite{GarciaBellido:1999sv}.  Instead I will assume a high reheat
temperature.  Then the out-of-equilibrium criterion 
relies upon the EWPT being first order.  In the simplest
models, the new physics should make the Higgs potential look like
fig.\ \ref{fig:PT} at high temperatures, with a barrier separating the 
$h=0$ (unbroken) and $h=v_b$ (broken) phases.  The criterion for baryon
number to not relax back too much toward zero inside the bubbles is
\cite{Moore:1998swa}
\be
	{v_b\over T_b} \gtrsim 1.09
\label{sphbnd}
\ee
where $T_b$ is the temperature at which baryogenesis takes place.  Typically 
$T_b = T_n$, the nucleation temperature of the bubbles, which is lower than
the critical temperature $T_c$ where the two vacua are degenerate.  To
determine $T_n$ one must find the rate per unit volume of bubble nucleation and equate it
to $H^4$ \cite{Linde:1977mm,Linde:1981zj}.

The traditional method of generating the barrier in the potential was
through finite-temperature effects from the ``one-loop''
thermal correction to the
potential,
\be
	V_T(h) = T\sum_i \pm \int{d^{\,3}p\over (2\pi)^3} \ln\left(1 \mp
	e^{-\beta{\sqrt{p^2+m_i^2(h)}}}\right) \quad
	\left({\hbox{bosons}\atop\hbox{fermions}}\right)
\ee
summed over all particles whose masses depend upon the Higgs field $h$
(here $\beta = 1/T$).
This expression can be easily derived from statistical mechanics
by recalling that the free energy is $F = -T\ln Z$ and the partition function
is
\be
	Z = \prod_{\vec n} \sum_{j=0}^N (e^{-\beta E(\vec n)})^j,\quad
	N = \left\{{1,\ \hbox{fermion}\atop \infty,\ \hbox{boson}}\right.
\ee
and replacing the sum over modes that arises after taking the logarithm
by $\sum_{\vec n}\to \int d^{\,3}n = L^3 \int d^{\,3}p/(2\pi)^3$ in a box
of volume $L^3$.  

At high temperatures, one can expand $V_T$ in powers of $m_i^2/T^2$.  The
contribution from bosons goes as
\be
	V_T =\sum_i n_i\left({T^2 \over 24} m_i^2(h) - {T\over 12\pi}
	\left(m_i^2(h)\right)^{3/2} + O(m_i^4)\right)
\ee
where $n_i$ counts the number of degrees of freedom ({\it e.g.,} 3
polarizations for a massive gauge boson)
and the cubic term is the single nonanalytic contribution in the expansion.
It plays the crucial role of generating the barrier if the particle gets
all of its mass from the Higgs mechanism, $m_i^2 = \sfrac14 g^2 h^2$
for example, like the SU(2)$_L$ gauge bosons.  Then \cite{Anderson:1991zb}
\be
	V(h) \cong\left({n\over 24} g^2 T^2 - \frac12\lambda v^2\right) h^2
	- {n g^3\over 96\pi}T h^3 + {\lambda\over 4}h^4
\ee
At the critical temperature $T_c$, pictured in fig.\ \ref{fig:PT}(right), 
this takes the form
\be
	V(h) = {\lambda\over 4}h^2\left(h -v_c\right)^2
\ee
where
\be
	v_c = {n g^3\over 48\pi\,\lambda} T_c
\ee
Using the SM values $\lambda=0.13$, $g=0.65$ and $n=3\times 3$ for the 
3 massive vector bosons (approximating $g'=0$ for simplicity), we find
$v_c/T_c = 0.13$, well below the bound (\ref{sphbnd}), given that
the ratio does not change very much between $T_c$ and $T_n$.  We would need
$\lambda$ and consequently $m_h$ to be smaller, $m_h < 43\,$GeV, to satisfy
the sphaleron bound.
Lattice studies show that in fact for $m_h>80\,$GeV, there is no first
order transition at all, but rather a smooth crossover 
\cite{Rummukainen:1998as}.

Therefore new particles coupling to the Higgs are needed to enhance the
strength of the EWPT.  Notice that if $m_i^2(h) = m_0^2 + g^2 h^2$ with
a bare mass term $m_0$ not coming from electroweak symmetry breaking
(EWSB), then the ``cubic'' term is
not really cubic and this reduces its effectiveness.  It is difficult
to add new particles with $m_0=0$ since they cannot be much heavier than
the weak scale while remaining perturbatively coupled.  Moreover, resumming the
temperature corrections tends to generate $m_0 \sim gT$ making this an even
more generic problem. 

\begin{figure}
\centerline{\includegraphics[width=0.4\textwidth]{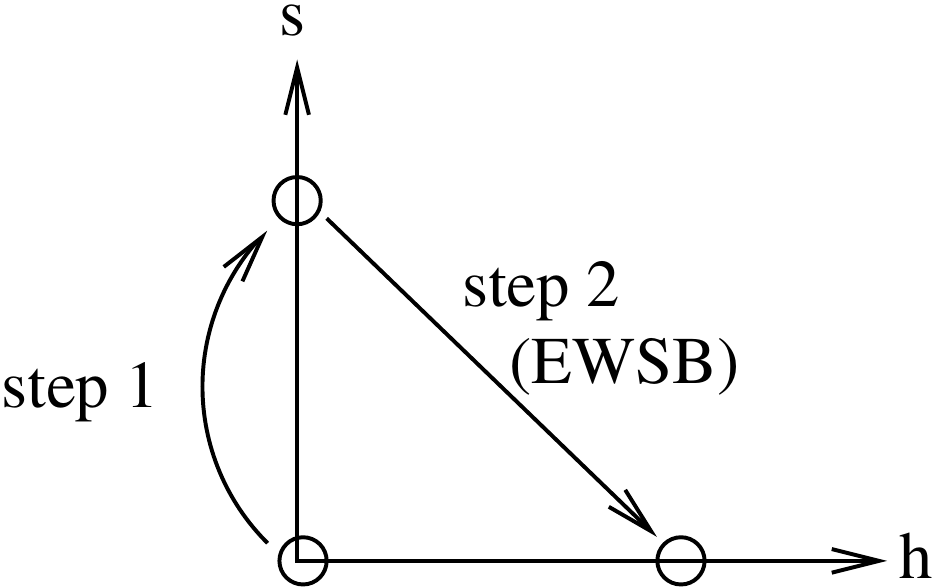}}
\caption{A two-step phase transition that can have a large tree-level
barrier between the false vacuum $h=0$ and the true one where electroweak
symmetry is broken.}
\label{fig:2step}
\end{figure}

A more robust way of strengthening the EWPT is to imagine a two-step phase
transition involving a second field $S$ that is a singlet under SU(2)$_L$.
The sequence of symmetry breaking is shown in fig.\ \ref{fig:2step}: at very
high temperature both $S=0$ and $h=0$; as $T$ decreases $S$ gets a VEV, 
and finally 
the electroweak transition occurs (during which $S$ may go to zero, but
this is not strictly necessary).  The tree-level potential is
\be
	V = V_{SM} + \lambda_s (s^2-w^2)^2 + \lambda_{hs} h^2 s^2
\ee
where the last term provides a barrier between the $h=0$ false minimum
and the EWSB true minimum.  This setup has the advantage that the barrier
is already present at tree level and does not rely upon feeble
finite-temperature effects that are suppressed by $g^2/12\pi$.  One needs a
moderate tuning to make the two minimum somewhat degenerate so that the
finite temperature effect is just to reverse the relative heights of the
two vacua.  In this way one can relatively easily engineer a strong phase
transition \cite{Espinosa:2011ax}.

\subsubsection{The wall velocity}

I would like to finish this lecture by commenting upon an outstanding issue
that is typically treated in a rough way because it is difficult to compute
from first principles: the bubble wall velocity $v_w$.  The predicted
baryon asymmetry can be somewhat sensitive to its value, especially if it
becomes too large.  If $v_w > c/\sqrt{3}$, the sound velocity in the
plasma, then baryogenesis essentially shuts off because there is no time
for a $CP$ asymmetry to diffuse in front of the wall.  The support of the
integral (\ref{nbeq}) goes to zero.  This behavior is not evident from the
diffusion equations as written in (\ref{diffeqs}) since they are linearized in 
$v_w$, and to my knowledge nobody has tried to quantify this for
fast-moving walls.  Such fast walls arise if the phase transition becomes
very strong, with large supercooling $(T_n\ll T_c)$, which is an
interesting limit because it leads to observable gravity waves 
\cite{Caprini:2015zlo}.  Typically bubbles that give observable gravity
waves are not compatible with baryogenesis, but exceptions can be found
\cite{Huang:2016cjm}.

In a vacuum, an expanding bubble wall would quickly accelerate to the speed
of light.   To compute the wall velocity in a plasma, one must balance the
outward force on the wall, due to the energy difference between the two 
vacua,
against the force of friction from particles interacting with the wall.  It
requires self-consistently determining the shape of the wall at the same
time, which is numerically challenging, since a stationary solution for the
wall only exists if one guesses the right value of $v_w$.  Most studies of EWBG therefore 
treat $v_w$ as a free parameter.  It has been computed for the SM 
\cite{Moore:1995si} (before $m_h$ was known to be too heavy for a first 
order EWPT) and the MSSM \cite{John:2000zq}, typically giving small values $v_w\cong
 (0.05-0.1)c$.

\subsection{Exercises}

1. {\bf Out of equilibrium decay.} Carry out the integration of the Boltzmann equation for the
toy model of baryogenesis from out-of-equilibrium decay of an
$X$ boson, with $CP$ asymmetry $\epsilon$ between decays
$X\to \bar\Psi\chi$ versus $X\to \Psi\bar\chi$, 
to evaluate the ``washout'' contribution from the
inverse decays.  In an exact treatment, the baryon number generated
should go to zero as $H(m_X)/\Gamma_X \to 0$.  Why does 
the present simplified approach go wrong in this respect?

\bigskip

2. {\bf Cut diagrams.} (a) Using the Cutkosky rule for the internal 
propagators, compute the imaginary part of the
self-energy $\Sigma_I$ at one loop for a heavy neutrino $N_i$ due to exchange
of the Higgs boson and a lepton doublet $L_j$, considering them to
be massless compared to $M_i$.  You should find that 
$\Sigma_I \sim i \slashed{p} P_\R$ where $p$ is the 4-momentum of the
decaying $N_i$ (feel free to work in its rest frame) and $P_\R =
(1+\gamma_5)/2$ is the chiral projection operator onto right-handed
states.\\
(b) Compare the
magnitude of $\Sigma_I$ to the tree-level decay rate of $N_i$. How
does it appear in the correction to the propagator? \\
(c) If you never did a similar exercise before, compute the imaginary
part directly from the $i\epsilon$ prescription in the full
propagators.  Do the loop integral first; the imaginary part can be
found from the Feynman parameter integral, using the fact that
$\ln(z)$ has a branch cut along the negative $z$ axis, and is $\pm
i\pi$ just above or below the cut.

\bigskip

3. {\bf $CP$ asymmetry from decays.}  Compute the part of the
CP asymmetry $\epsilon_i$ for leptogenesis coming from the
interference between the tree and one-loop self-energy diagrams.
The following steps will help.\\
(a) Using charge conjugation, one can show that
\[
	\bar L_i \tilde H P_\R N_{j} = \bar N_{j} \tilde H^{\sss
T}P_\R L^c_i, \quad \bar N_j \tilde H^\dagger P_\L L_i = 
	\bar L_i^c \tilde H^* P_\L N_j
\]
where $P_{\L} = (1-\gamma_5)/2$ and 
$L^c$ is the charge conjugated lepton doublet.  We have used that
$N = N^c$ since $N$ is a Majorana particle.  This provides a
convenient way of writing down the amplitudes with the lepton number
flow going the ``wrong'' way, {\it i.e.,} the tree versus loop
diagrams.  Write the amplitudes for the tree-level decay $N_j\to H^*
L_i$ and from the one-loop $N$ self-energy diagram.  (Why do we not
care about the one-loop $L$ self-energy diagram?)  As a check, your
loop contribution should pick out the $M_k/((\slashed{p} + \Sigma_I(p))^2-M_k^2)$ part of the
virtual $N_k$ propagator.  Note that we included the one-loop
self-energy to regulate the IR-divergence in case $p^2 \cong M_k^2$,
and used the fact that it goes like $\slashed{p}$ from problem 2.\\
(b) Use the result of problem 2 to evaluate the imaginary part of the
loop diagram.  The $P_\R$ in $\Sigma_I$ can be simplified since it is
acting on a right-handed external state.\\
(c) Square the amplitude to find the imaginary part, and divide by the
tree-level result to obtain $\epsilon$ in terms of the Yukawa
couplings, mass-squared differences, and decay rates $\Gamma_i$
as in our discussion of resonant leptogenesis.  The $\gamma_0$ 
(or $\slashed{p}$) acting
on the spinor for the decaying $N$ from the loop diagram can be 
disposed of using the Dirac equation for $N$ in its rest frame (or any frame).

\bigskip

4. {\bf Affleck-Dine baryogenesis \cite{Affleck:1984fy}.\footnote{Beware of
egregious typos if you read the paper.}}  Suppose $\phi$ is a complex field
that carries baryon number 1, with Lagrangian
\[
	|\partial\phi|^2 - m^2|\phi|^2 - i\lambda(\phi^4 - {\phi^*}^4)
\]
The small quartic interaction violates baryon number, but not CP.
(We can define the action of $CP$ on $\phi$ as $\phi \to
e^{i\pi/4}\phi^*$.)
Instead, $CP$ is ``spontaneously'' broken by the initial condition of 
the field, $\phi = i\phi_0$ (purely imaginary) and $\dot\phi = 0$.\\
(a) Write the equation of motion for $\phi$, assuming that the
solution is spatially homogeneous.\\
(b) Let $\phi = \phi_1 + i \phi_2$, and suppose that $\lambda
\phi_0^2\ll m^2$ so that the interaction term is a small perturbation,
and to zeroth order can be neglected, in particular for $\phi_2$
because at $t=t_0$ the imaginary part of the interaction term vanishes
in the equation of motion.
Then one can just 
solve the free field equation in an expanding background.  Use the
results you found in problem 1 of set 1 to construct the solution
for $\phi_2$ that satisfies the initial condition at $t=t_0$, assuming
that $m\gg H$ and the universe is radiation dominated.\\
(c) At $t=t_0$ the interaction term for $\phi_1$ does not vanish, so 
we expect the amplitude of $\phi_1$ to be of order $\lambda\phi_0^3$.
But at late times, this inhomogeneous source term dies away due to 
Hubble damping, and we can again solve the noninteracting equation.
Thus we can also estimate $\phi_1$ using problem 1 of set 1.  However
there is in general a phase difference $\delta$ between $\phi_1$ and $\phi_2$
due to the early-time evolution, that turns out to be of order 1
(after solving numerically).
Write down the appropriate estimate for $\phi_1$.\\
(d) The baryon number current is given by 
\[
	J_B^\mu = -i\phi^* \overset{\leftrightarrow}{\partial^\mu} \phi
\]
and the baryon number density is $n_B = J_B^0$.  Show that initially
$n_B=0$, but at late times it is nonzero.  How does the
out-of-equilibrium requirement come in?

\section{Dark Matter}
As mentioned in the introduction, many aspects of dark matter were treated
by other lecturers at this school.  My focus will be to make you aware of
different classes of dark matter models, that can be distinguished by their
respective production mechanisms.  Thermal freezeout was covered in the 
lectures of P.\ Fox, but I will add some complementary  observations,
before going on to briefly discuss asymmetric dark matter, freeze-in,
primordial black holes, and ultralight scalars/axions.  But first, a little
more history.

\subsection{Brief history of dark matter}
\label{bhdm}

Often the history of science as recounted by its practitioners is
oversimplified, since this results in a better story, or at least one that
is easier to tell.  I will do the same thing, claiming that  
in the case of dark matter it is really
fair to credit Fritz Zwicky with originating the idea in his 1933 paper
\cite{Zwicky:1933gu}, where he also coined the term, in German.  
See ref.\ \cite{Bertone:2016nfn} for the unsimplified story.  Oort is
sometimes given precedence \cite{1932BAN.....6..249O}, based on studies of stellar
dynamics around the Milky Way disk, but what he discovered is now known to
be nonluminous baryons rather than dark matter.

Neither the term nor the concept caught on very quickly.  Zwicky used it
again, this time in English, in his 1937 paper \cite{1937ApJ....86..217Z},
where he also proposed gravitational lensing as a way of mapping the dark
matter distribution!  However the term does not seem to appear again in the
literature until 1979 in the review of Faber and Gallagher 
\cite{Faber:1979pp}, which was also the most influential paper 
to start citing Zwicky's 
original 1933 work.  Before that, ``missing mass'' or ``missing matter'' were in use.  By this time the
concept was established thanks to the galactic rotation curves measured by
V.\ Rubin and collaborators \cite{1970ApJ...159..379R} around 1970,
although H.\ Babcock had already made a convincing such measurement for
the Andromeda galaxy in 1939 \cite{1939LicOB..19...41B}!

From the particle physics perspective, some highlights in the history of
dark matter were:

\begin{itemize}

\item 1980: de Rujula and Glashow proposed massive  neutrinos ($m_\nu\sim
24$\,eV)
as a dark
matter candidate \cite{DeRujula:1980mgi}.

\item 1983: White, Frenk and Davis showed that neutrino
dark matter (with mass of order 30 eV or less)
erases cosmological structure
on short distance scales, corresponding to the streaming length,
predicting large voids that are not observed 
\cite{White:1988fb}.  The idea that dark matter must be cold was 
thus born: DM should be nonrelativistic by the time the particle horizon
contains a mass comparable to a galaxy.

\item 1984: Steigman and Turner coined the term ``weakly interacting
massive particle'' (WIMP) \cite{Steigman:1984ac}, which became a popular
paradigm for dark matter; 
but they had in mind
decaying particles rather than dark matter.  (Already in 1977, Lee and
Weinberg had shown that heavy neutrinos annihilating with a weak scale
cross section would overclose the universe if their masses were less than 
about 2\,GeV \cite{Lee:1977ua}).

\item 1985: Goodman and Witten adapted an idea of Drukier and Stodolsky
(for detecting MeV-scale neutrinos) \cite{Drukier:1983gj} to the 
detection of 
weakly interacting DM particles  \cite{Goodman:1984dc}.  Among the DM
candidates they mentioned are axions, magnetic monopoles,
sneutrinos, 
photinos, and exotic QCD bound states (which seems to have also influenced
the list of candidates mentioned in \cite{Kolb:1990vq}), reminding us of 
a saying, {\it plus \c{c}a change, plus c'est la m\^eme chose}, except for
the magnetic monopoles, not currently a popular candidate. 

\end{itemize}

Although hot dark matter has long been ruled out, the intermediate
possibility of warm dark matter (WDM) is less clear.  The canonical
example is a sterile neutrino that was in thermal equilibrium, and becomes
nonrelativistic at $T\sim m\sim 1\,$keV.  This is the temperature at which
$\rho H^{-3} \sim 10^{12}\,M_\odot \sim$ mass of the Milky Way.  Lighter
DM would erase galaxy-sized structures and be classified as hot.
WDM with mass at the keV scale has a streaming length of 
$\sim 0.3\,$Mpc \cite{Viel:2005qj}.

Warm dark matter can help to address some problems of the cold dark matter 
scenario from $N$-body gravitational simulations, that give central density
profiles of DM halos going as \cite{Navarro:1996gj}
\be
	\rho(r)\sim {1\over r}
\label{NFW}
\ee
This is too cuspy compared to observations of rotation curves, and 
is known as the
cusp-core problem.  A further problem is that CDM simulations 
predict too many dwarf satellite galaxies orbiting Milky
Way-like halos (the ``missing satellites'' problem), and also too
many highly luminous secondary galaxies (the ``too big to fail'' problem);
for reviews see ref.\ \cite{Weinberg:2013aya,Bullock:2017xww}.  The origin
of globular clusters is also mysterious within CDM.  

But
WDM does not seem to solve all problems simultaneously, giving too much
suppression of satellites at a mass $\sim 2\,$keV that would solve the
cusp/core problem.  Moreover, damping of power in the matter fluctuations
at short scales would suppress Lyman-$\alpha$ absorption, leading to
the recent constraint $m> 3.5\,$keV \cite{Irsic:2017ixq}.  Thus WDM
is becoming increasingly cold, and less motivated by an ability to address
the small scale structure problems.

\subsection{Thermal freezeout}

The most popular mechanism for DM to attain its relic density is thermal
freezeout, $\chi\chi\to f\bar f$ annihilations into SM particles, going out
of equilibrium at a freezeout temperature quantified by $x_f = m_\chi/T_f
\cong16,\,28,\,30$ for $m_\chi = 0.1\,$GeV, 100\,GeV,  100\,TeV. It is a
fairly generic mechanism, assuming that the DM was in thermal equilibrium
at early times.  It works for self-conjugate DM as well as DM with a
conserved charge, in which case $\chi\bar\chi\to f\bar f$ is the relevant
interaction, and it is assumed there is no significant asymmetry between
$\chi$ and $\bar\chi$ abundances (this is the subject of  asymmetric dark
matter, below).

The basic result of thermal freezeout is that the relic density scales as
\be
	Y_\chi \sim {1\over \langle\sigma v\rangle m_\chi m_p}
\label{Ychi1}
\ee
where $\langle\sigma v\rangle$ is the thermally averaged annihilation cross
section times relative velocity.  The corresponding annihilation rate is
\be
	\Gamma = n_\chi\langle\sigma v\rangle
\label{Geq1}
\ee
and the annihilations go out of equilibrium at a temperature corresponding
to
\be
	\Gamma \sim H \sim {T^2\over M_p}
\label{Geq2}
\ee
Hence
\be
	Y_\chi = {n_\chi\over s} \sim {H\over \langle\sigma v\rangle s}
	\sim {1\over \langle\sigma v\rangle M_p T}
\ee
to be evaluated at the freezeout temperature $T_f$.
To determine $T_f$ we recall that the particle density goes as
$n_\chi \sim (m_\chi T)^{3/2} e^{-m_\chi/T}$ before it falls out of
equilibrium.  Then eqs.\ (\ref{Geq1},\ref{Geq2}) lead to the estimate
\be
	x_f \equiv {m_\chi\over T_f} \sim 
	\ln\left(M_p m_\chi \langle\sigma v\rangle\right)
	- \sfrac12\ln(x_f)
\label{xfeq}
\ee
which can be solved by iteration.  This explains the logarithmic 
dependence on $m_\chi$ of $x_f \sim 16-29$ alluded to above, and our
improved estimate
\be
	Y_\chi \sim {x_f\over \langle\sigma v\rangle M_p m_\chi}
\ee
This refines the rougher estimate (\ref{Ychi1}) by a number
$x_f\sim 20$.

Now let's compare this prediction with observation.  The Planck
collaboration determines
\be
	\Omega_{\rm CDM} = {\rho_{\rm CDM}\over \rho_{\rm crit}} = 0.258
\ee
while $\rho_{\rm CDM} = m_\chi Y_\chi s$ and $s=2891/$cm$^3$.  Solving for
the cross section gives
\be
	\langle\sigma v\rangle \sim 10^{-9}{\rm GeV^{-2}}\sim 10^{-26}{\rm
cm}^3/{\rm s}
\ee
independently of $m_\chi$ (except for the log dependence in $x_f$).  
This is considered to be a typical weak scale cross section with
$\sigma \sim \alpha^2/(100\,{\rm GeV})^2$ and $\alpha\sim 10^{-2}$.
Thus one seems to get the right relic density, independent of $m_\chi$
just by having new physics near the weak scale.  This was called a
``striking coincidence'' in the review article \cite{Jungman:1995df}.
Later it came to be known as the ``WIMP miracle.''\footnote{This term was
introduced by Jonathan Feng at the SLAC Summer Institute lectures in 2001.
He was motivated by a sense of frustration that particle
theorists did not take the striking coincidence very seriously.
His strategy seems to have worked.  The name first appears in a research
article in 2008 \cite{Baer:2008uu}, when it was already in common use.}\ \ 

To be more quantitative, we need to solve the Boltzmann equation, similarly
to the case of $X'$ decays in our toy model of baryogenesis.  We define
$Y_\chi = n_\chi/s$ and assume $n_\chi = n_{\bar\chi}$ in case $\chi$
is not self-conjugate.  As before $x = m_\chi/T$ takes the place of the
time variable.  Then
\be
	\dot n_\chi + 3 H n_\chi - \langle\sigma v\rangle
\left(n_\chi^2-n_{\chi,eq}^2\right)
\ee
becomes
\be
	{dY_\chi\over dx} = - {xs\langle\sigma v\rangle\over H(m_\chi)}
\left(Y_\chi^2 - Y_{\rm eq}^2\right)
\label{Beq3}
\ee
where
\be
	Y_{\rm eq} = {45\over 4\pi^4}{g_\chi\over
g_{*,s}} x^2 K_2(x) \cong
{45\over 2\pi^4}\left(\pi\over 8\right)^{1/2} {g_\chi\over
g_{*,s}} x^{3/2} e^{-x} 
\ee
for a nonrelativistic particle with $g_\chi$ degrees of freedom (1 for a
real scalar, 2 for a complex scalar or Weyl fermion, 4 for a Dirac
fermion), $g_{*,s}$ was defined in eq.\ (\ref{gstars}), the entropy density
is 
\be
	s = {\rho + p\over T} \cong 0.44\,g_{*,s} {m_\chi^3\over x^3}
\ee
and the Hubble parameter 
\be
	H = 1.66\sqrt{g_*} {T^2\over M_p}
\ee
is evaluated at $T = m_\chi$, with $g_*$ defined similarly to 
(\ref{gstars}), but with $(T_i/T)^4$ instead of $(T_i/T)^3$, such that the
energy density is $\rho = (\pi^2/30)g_* T^4$.

The cross section is thermally averaged, and the $v$ appearing there is 
usually considered to be the relative velocity between the annihilating
particles, but there is some subtlety
in this identification that becomes relevant when the annihilations are
relativistic \cite{Cannoni:2013bza}, which is not the case here.  Since the
particles are highly nonrelativistic for thermal freezeout, it is a good
approximation to use Maxwell-Boltzmann statistics so that
\be
	\langle\sigma v\rangle \sim \int d^{\,3}p_1\,d^{\,3}p_2\,
	e^{-{E_1/T} - E_2/T} \sigma |\vec v_1-\vec v_2|
\ee
A proper treatment of $v$ in \cite{Gondolo:1990dk} allows this to be
expressed as a single integral over the Mandelstam $s = (p_1 + p_2)^2$ 
($= 4E^2$ in the center-of-mass frame),
\be
	\langle \sigma v\rangle = {1\over 8 m^4 T K_2^2(m/T)}
	\int_{4 m^2}^\infty ds\, \sqrt{s} (s-4m^2) K_1(\sqrt{s}/T)
	\sigma(s)
\ee

\begin{figure}
\centerline{\includegraphics[width=0.3\textwidth]{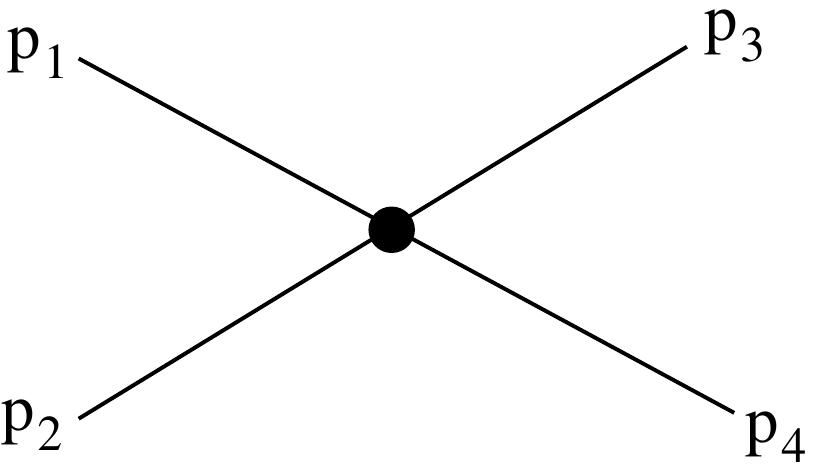}
\hfil\raisebox{0.7cm}{\includegraphics[width=0.3\textwidth]{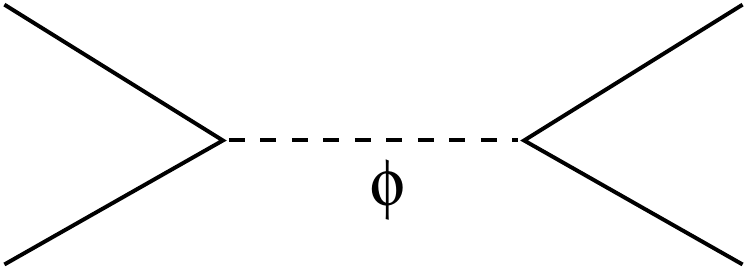}}}
\caption{Left: generic annihilation of dark matter particles (1 and 2) into
standard model particles (3 and 4).  Right: resonantly enhanced
annihilation.}
\label{fig:ann}
\end{figure}

For purposes of estimation it is often adequate (if one does not insist on great precision)
to ignore the thermal averaging and just
compute $\sigma v$ in the limit $v\to 0$, since for $s$-wave annihilation 
this is just a constant.  Then for the case of DM annihilating into 
equal mass particles, with particles labeled as $1+2\to 3+4$ as
in fig.\ \ref{fig:ann}(right),
\be
	\sigma = \int_{t_0}^{t_1} dt\,{|{\cal M}|^2\over 64\pi s\,
p_{1,\rm cm}^2} \cong
	{|{\cal M}|^2\,p_{3,\rm cm}\over 16\pi\, s\, p_{1,\rm
cm}}
\ee
in terms of the matrix element ${\cal M}$, the center of mass momenta of 
particles 1 and 3, and the integration limits
\be
	t_{0,1} = - (p_{1,\rm cm}\pm p_{3,\rm cm})^2\,;
\ee
See the kinematics review of 
\cite{Patrignani:2016xqp}.
Taking $v_{\rm rel} = 2 p_{1,\rm cm}/E = 2\sqrt{1-4 m^2/s}\cong 
 2 p_{1,\rm cm}/m$, $s\cong 4 m^2$ and $p_{3,\rm cm} = \sqrt{1-m_3^2/m^2}$,
we find
\be
	\sigma v_{\rm rel} \cong {|{\cal M}|^2\over 32\pi m^2}
	\sqrt{1-m_3^2/m^2}
\ee 
(times $1/2$ if particles 3 and 4 are identical). For simplicity we have
assumed that $m_4 = m_3$, which is usually the case.

The relative correction  from thermally averaging should
be of order $v^2 \sim T/m\sim 1/20$.  Of course this does not work if the
cross section is $p$-wave suppressed since then $\sigma v\to 0$ in the
limit of $v\to 0$.  Generally one can write
\be
	\langle \sigma v\rangle \cong \sigma_0 \left(T\over m\right)^n
\label{sigma0}
\ee
in the low-temperature limit, 
with $n=0$ for $s$-wave annihilation, $n=1$ for $p$-wave, {\it etc.} 
Another case where the simple estimate fails is when the cross section is
resonantly enhanced by an intermediate particle whose mass happens to be
close to twice that of the DM, as in fig.\ \ref{fig:ann}(right).  The
thermal averaging is then imporant since it allows $s$ to vary near the pole of
the propagator in
\be
	{\cal M} \sim {1\over s - m_\phi^2 + i m_\phi\Gamma_\phi}
\ee

\begin{figure}
\centerline{\includegraphics[width=0.5\textwidth]{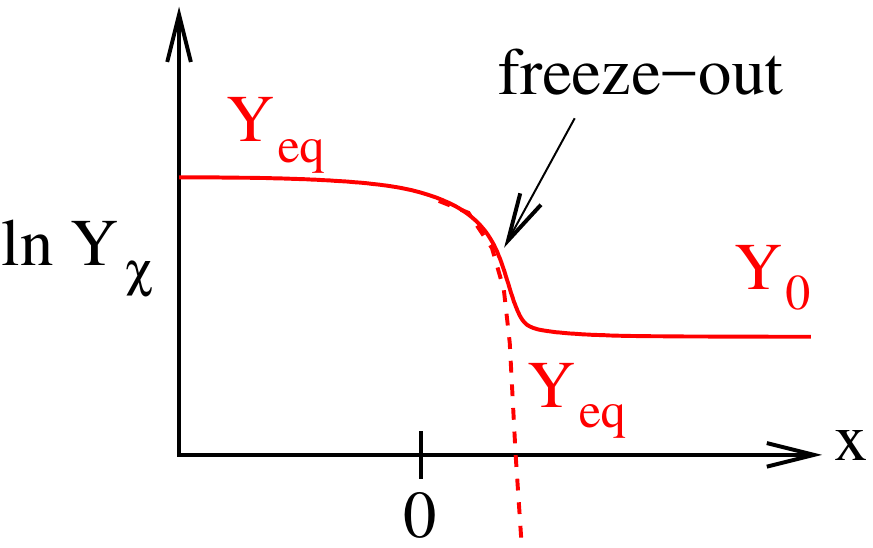}}
\caption{Solution of Boltzmann equation (solid curve), compared to the equilibrium 
abundance (dashed).}
\label{fig:freezeout}
\end{figure}

\begin{figure}
\centerline{\includegraphics[width=0.5\textwidth]{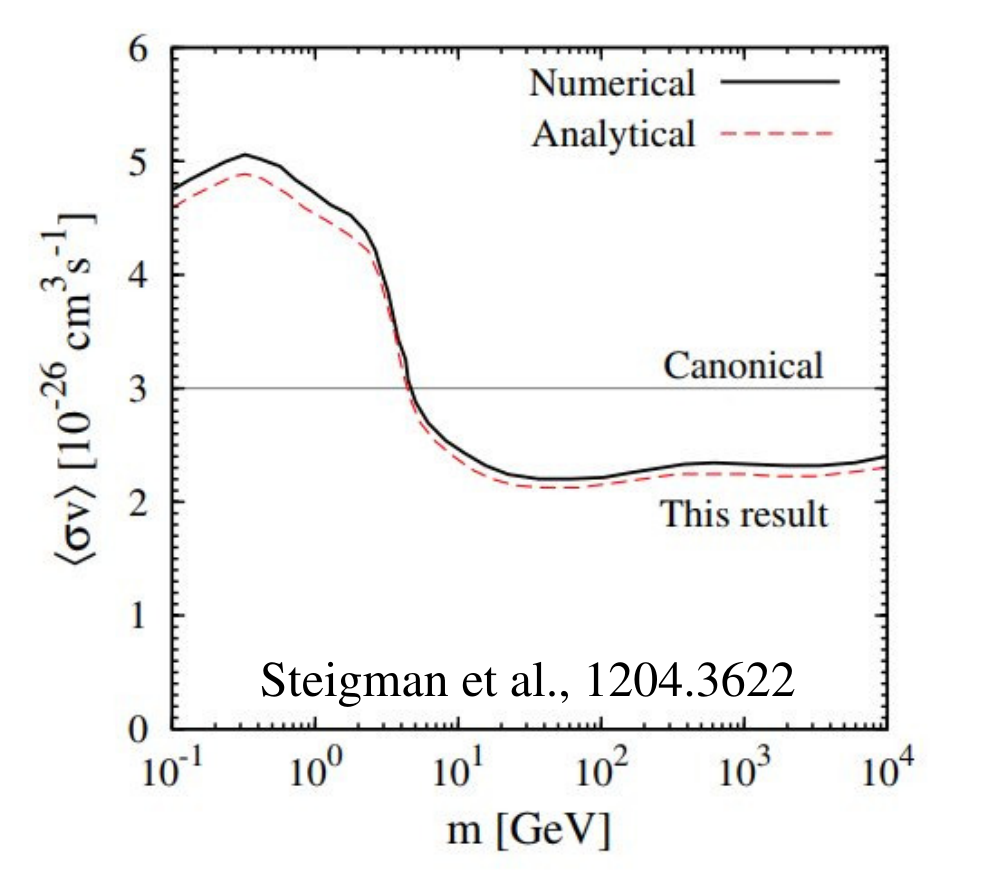}}
\caption{Value of $\sigma v$ needed to get the observed relic density,
for Majorana or self-conjugate scalar DM, reproduced from ref.\ 
\cite{Steigman:2012nb}.}
\label{fig:svrelic}
\end{figure}

To find the relic density, one must integrate the Boltzmann equation
(\ref{Beq3}), whose solution looks qualitatively like fig.\
\ref{fig:freezeout}.  But it is numerically challenging to follow the
whole evolution since the abundance changes very rapidly around the epoch
of freezeout, so one often tries to avoid writing code from scratch.
Publicly available codes like micrOmegas \cite{Belanger:2013oya}
and DarkSUSY \cite{Bringmann:2018lay} are commonly used.  

For a fast estimate, in the case of $s$-wave annihilation, one can simply
look up the value of the cross section for a given DM mass that yields
the observed relic density, \cite{Steigman:2012nb}, reproduced in 
fig.\ \ref{fig:svrelic}.  (An improved version
of this result can be found in fig.\ 3 of \cite{Bringmann:2018lay}.)
The ``canonical'' value of $3\times 10^{-26}\,$cm$^3$/s $\cong 25\times
10^{-10}\,$GeV$^{-2}$ assumes self-conjugate DM, like a Majorana fermion
or a real scalar.  For Dirac or complex scalar DM, one should multiply this
result by 2, as can be seen from comparing the rate of annihilation in
the two cases.  For self-conjugate DM, the rate of annihilation per
$\chi$ particle is
\be
	\Gamma = n_\chi \langle\sigma v\rangle
\ee
leading to the density $n_\chi \sim H/\langle\sigma v\rangle$.  But for
Dirac or complex scalar DM, it is
\be
	\Gamma = n_{\bar\chi} \langle\sigma v\rangle
\ee
where now the total DM density is given by $n_\chi + n_{\bar\chi}$.
The estimate $n_\chi \sim H/\langle\sigma v\rangle$ remains the same,
but $\langle\sigma v\rangle$ must be doubled to get the same value for the total density.
Notice that for Dirac DM, the total abundance is $Y_\chi + Y_{\bar\chi}$,
so the same Boltzmann equation (\ref{Beq3}) can be used for Dirac or
Majorana DM.

Finally, there is an approximate analytic solution to the Boltzmann
equation \cite{Scherrer:1985zt}, slightly improved upon in 
\cite{Cline:2013gha}.  The present abundance can be estimated as
\be
	Y_0 \cong \sqrt{45\, g_*\over \pi g_{*s}^2} {(n+1)x_f^{n+1}\over
	m_\chi M_p \sigma_0}
\ee
with $n$ and $\sigma_0$ defined in eq.\ (\ref{sigma0}), and
\bea
	x_f &\cong& \ln y_f - \sfrac12\ln\ln y_f\nn\\
	y_f &=& {g_\chi\over 2\pi^3}\sqrt{45\over 8 g_*} m_\chi M_p (n+1)
	\sigma_0\,;
\eea
compare with eq.\ (\ref{xfeq}).
This approximation scheme is valid in principle for any DM mass, so long
as freezeout occurs when $x_f\gg 1$ so the particle is nonrelativistic.
But there is an upper limit on $m_\chi$ from partial wave unitarity
\cite{Griest:1989wd},
\be
	\langle\sigma v\rangle \lesssim {4\pi\over m_\chi^2}\left\langle{1\over
	v_{\rm rel}}\right\rangle
\ee
assuming $s$-wave annihilation.  One cannot get a large enough cross
section to sufficiently suppress the relic density if $m_\chi$ becomes
too large.  Taking the modern value of $\Omega_{\rm CDM}$, this limit
is $m_\chi \lesssim 140\,$TeV, assuming self-conjugate DM.  Of course
this bound does not apply if many partial waves contribute, which would be
the case for composite DM made from weakly bound constituents, giving a
geometrical cross section (see for example ref.\ \cite{Harigaya:2016nlg}).

\subsection{Asymmetric dark matter (ADM)}
Although thermal freezeout seems like a generic mechanism, the fact
that baryonic matter gets its abundance from the matter-antimatter asymmetry 
makes it quite reasonable that dark matter could have a similar origin,
if it has a conserved number density analogous to baryon number.
This is a large subject that cannot be done justice in the little time I
have here; see ref.\ \cite{Zurek:2013wia} a comprehensive review.  I will 
not say anything about the specific mechanism for generating the dark
asymmetry, but simply assume its existence.

Whereas the density of baryonic antimatter in the universe is negligible, this need not be the case
in the dark sector, where there can be a significant symmetric component
to the density in addition to the dominant asymmetric component.  We can
define the two as
\bea
	n_{\rm sym} &=& n_\chi + n_{\bar\chi} - | n_\chi - n_{\bar\chi}|
	= {\rm min}(n_\chi, n_{\bar\chi})\nn\\
	n_{\rm asym} &=& | n_\chi - n_{\bar\chi}| \cong 
	{\rm max}(n_\chi, n_{\bar\chi})
\eea
For ADM, it is assumed that $n_{\rm sym}\ll n_{\rm asym}$.  Then indirect detection
signals from $\chi\bar\chi\to f\bar f$ in the galaxy will be suppressed
relative to thermal DM.  The question is, how much will the constraints 
be weakened?  One has to solve the Boltzmann equation again, but now taking
into account the conserved particle number in the asymmetric component.

\begin{figure}
\centerline{\includegraphics[width=0.65\textwidth]{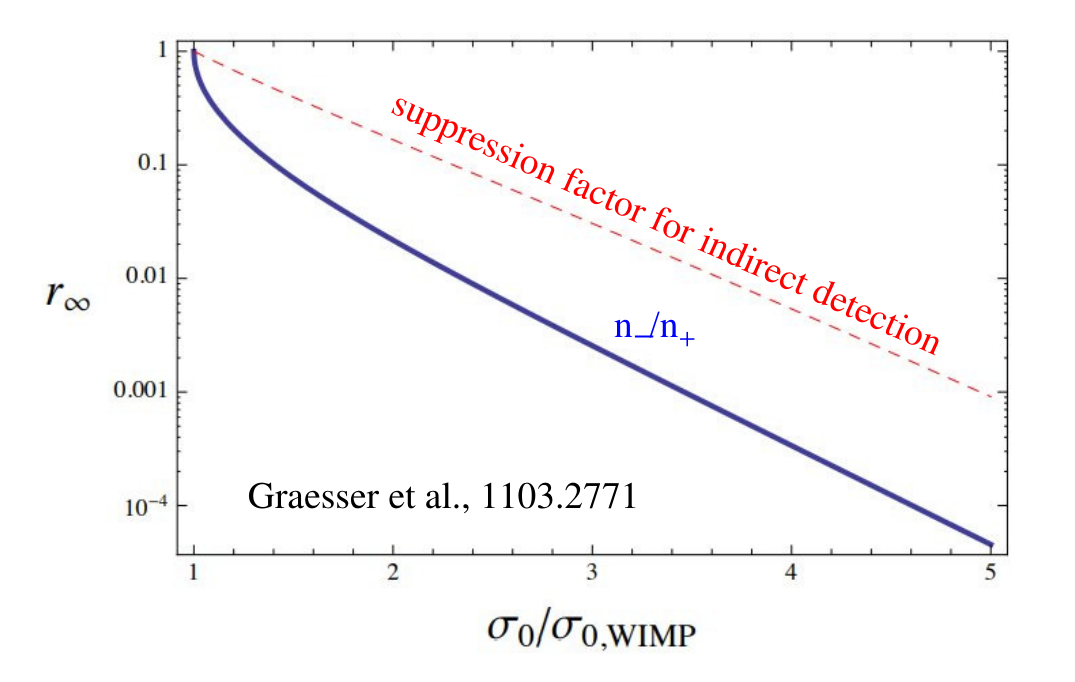}}
\caption{The suppression of the ratio of symmetric to asymmetric components
of asymmetric dark matter as a function of annihilation cross section
(solid curve) and the resulting suppression factor for indirect detection
signals (dashed), from ref.\ \cite{Graesser:2011wi}.}
\label{fig:adm}
\end{figure}

The fact that $n_{\rm asym}$ cannot be reduced by annihilations leads to
qualitatively different results for $n_{\rm sym}$ than in the case of
thermal DM, where the annihilations turn off once the
density falls below $n_{\rm eq}$.   For ADM, the density can no longer
fall like $n_{\rm eq}$ once it reaches $n_{\rm asym}$.  Therefore the 
annihilations continue longer than for thermal DM, and $n_{\rm sym}$
becomes much smaller than the corresponding thermal relic density.
This was worked out in ref.\ \cite{Graesser:2011wi}, with the result shown
in fig.\ \ref{fig:adm}.  The solid curve shows the ratio of abundances,
$n_{\rm sym}/n_{\rm asym}$ as a function of the cross section, in units of
the canonical value for thermal DM.  The dashed line shows the resulting
suppression factor in any indirect detection signal from annihilations
at late times.  Interestingly, for the case of $\chi\bar\chi\to e^+e^-$ or
other electromagnetically interacting final states,
 this leads to a {\it lower limit} on $\langle\sigma v\rangle$  
in order to sufficiently suppress annihilations \cite{Lin:2011gj},
\be
	\langle\sigma v\rangle\gtrsim \left\{ {1\times 10^{-25}\,{\rm cm}^3/{\rm s},
	\quad m_\chi = 10\,{\rm GeV}\atop 7\times 10^{-25}\,{\rm cm}^3/{\rm s},
	\quad m_\chi = 1\,{\rm MeV}}\right.
\ee

If the ADM is bosonic, there can be a different kind of indirect signal,
from its accumulation in neutron stars.  Bosons may achieve such a high
central density that a black hole can form that will consume the neutron
star.  This gives a limit on the cross section for scattering on nucleons
of  $\sigma_{\chi N}\lesssim 10^{-47}$\,cm$^2$  in a range of masses $m\chi
\in [5\,{\rm MeV} - 15\,{\rm GeV}]$, where direct detection constraints are
relatively weak.

\begin{figure}
\centerline{\includegraphics[width=0.25\textwidth]{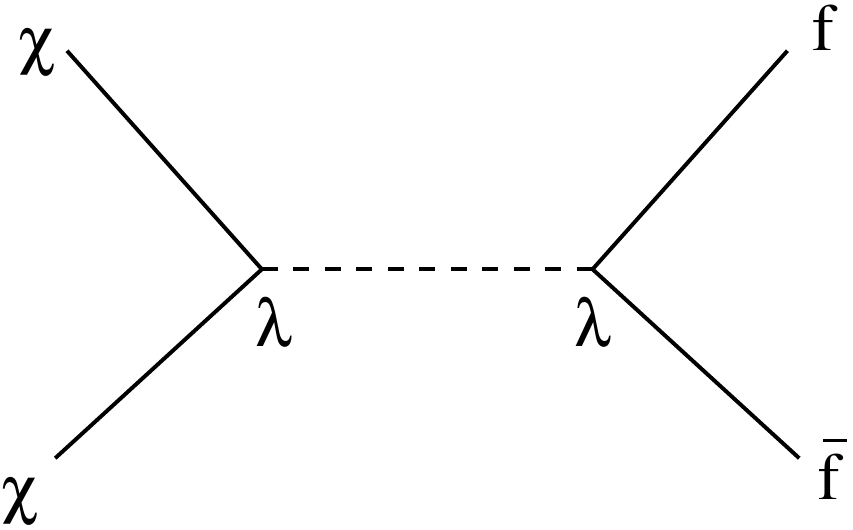}}
\caption{An annihilation process that is suppressed at high temperatures,
where freeze-in may be taking place.}
\label{fig:fi}
\end{figure}

\subsection{Freeze-in}  It is possible that DM interacts so weakly that it
never came into thermal equilibrium in the early universe.  Then initially
we would have $Y\cong 0$ and the Boltzmann equation would be approximately
\be
	{dY_\chi\over dx} \cong {x s\langle\sigma v\rangle\over H(m_\chi)}
	Y_{\rm eq}^2
\label{freezein}
\ee
$Y_\chi$ will then slowly approach $Y_{\rm eq}$ from below.  We can
estimate its present abundance by integrating  eq.\ (\ref{freezein}).
Unlike for thermal freezeout, it may not be a good approximation
to take $\langle\sigma v\rangle$ to be
constant  because the result, going as $\int dx\,x^2 K_2^2(x)$,
is dominated by $x\sim 0$, {\it i.e.,} high temperatures, and  there will
generally be some temperature dependence
in $\langle\sigma v\rangle$ at high $T$.  For example, consider the process
in fig.\ \ref{fig:fi} with $s$-channel boson exchange, which gives
$\langle\sigma v\rangle\sim \lambda^4 x^2/m_\chi^2$.  We find that
\be
	m_\chi Y_0 \sim 10^{-4}\lambda^4 M_p\cong 4.3\times 10^{-10}\,{
\rm GeV} 
\ee
in order to get the observed relic density, requiring $\lambda\sim
10^{-6}$.  This is a very weak coupling compared to that needed for
thermal freezeout.  To compare them, supposing $\langle\sigma v\rangle =
\lambda^4/m_\chi^2$ as $T\to 0$, and the cross section needed for thermal
freezeout is denoted by $\langle\sigma v\rangle_0$, then
\be
	{\langle\sigma v\rangle\over \langle\sigma v\rangle_0} \sim
	\left(0.1\,{\rm eV}\over m_\chi\right)^2
\ee
showing that for any reasonably heavy DM, the freeze-in cross section is
many orders of magnitude below that for freeze-out.

An interesting application of freeze-in is to DM candidates that
have only gravitational interactions, since these must be present
regardless of model-building choices \cite{Garny:2015sjg,Garny:2017kha}.
In this case $\langle\sigma v\rangle\sim T^2/M_p^4$ which leads to
$\int dx K_2(x)^2 \sim (T/m_\chi)^3$: the integral is dominated by the
high-temperature contribution, and is therefore sensitive to the reheat
temperature after inflation.  Dark matter of mass up to $\sim 10^{16}\,$GeV
can have the right relic density.  It is argued that even such heavy dark
matter could have observable signatures, since quantum gravitational
effects are  believed to break any global symmetries \cite{Banks:2010zn}, including those that
might stabilize dark matter.  Then, for example, fermionic DM could decay
through the operator $\bar\chi HL$, just like a heavy sterile neutrino,
suppressed by the action of a gravitational instanton via $e^{-S}$.

\begin{figure}
\centerline{\includegraphics[width=0.45\textwidth]{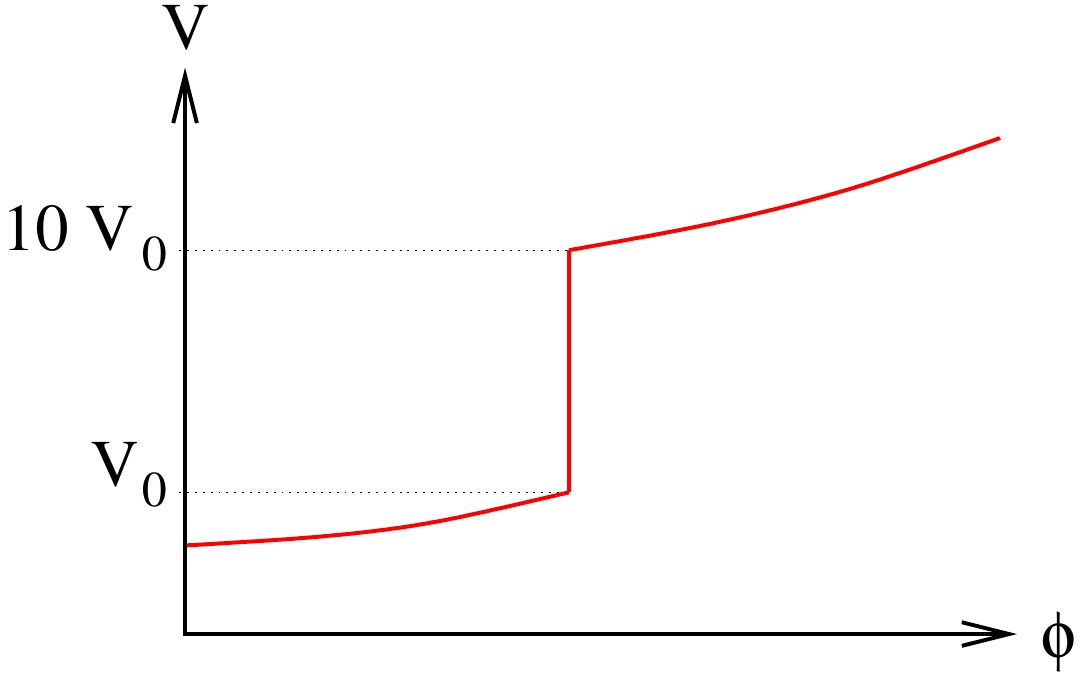}}
\caption{A step discontinuity in the inflaton potential, that
enhances power in fluctuations at a specific scale.  A factor of $\sim 10$
change in $V$ is required to get a factor of $\sim 10^7$ enhancement in the
power.}
\label{fig:step}
\end{figure}

\begin{figure}
\centerline{\includegraphics[width=0.85\textwidth]{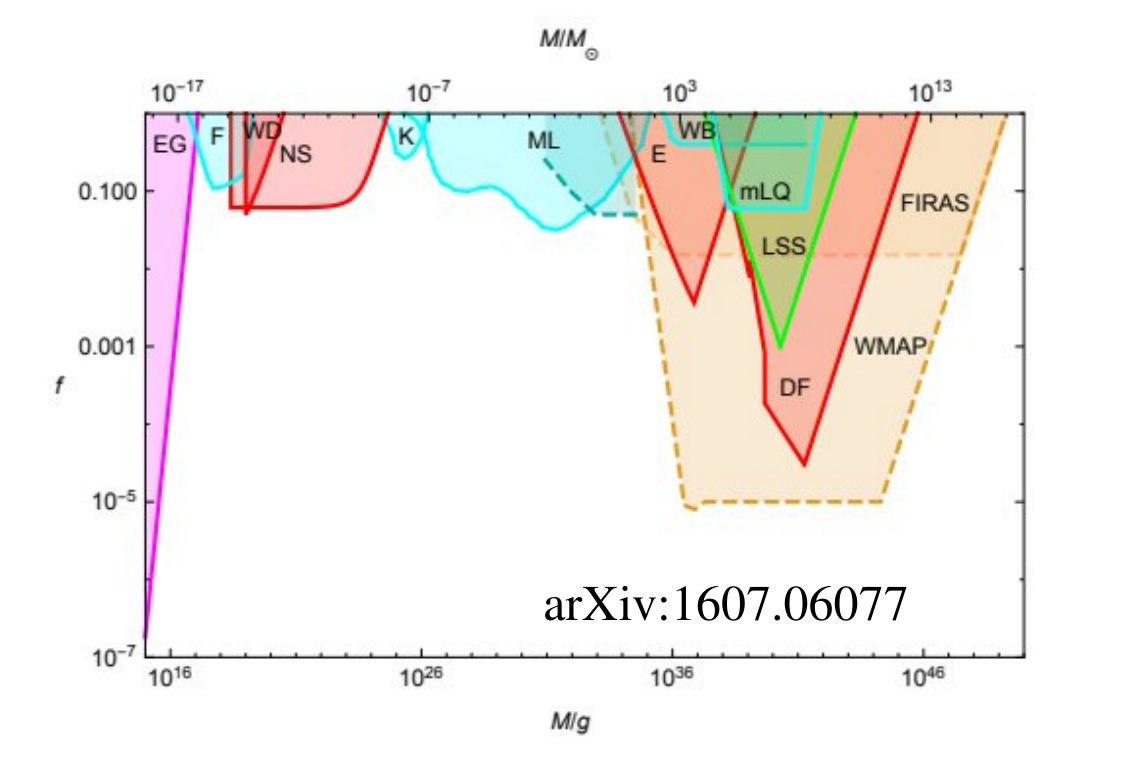}}
\caption{Upper limit on the fraction $f$ of DM in PBHs of a given mass,
assuming monochromatic distribution of masses, from ref.\ 
\cite{Carr:2016drx}. }
\label{fig:pbh-limit}
\end{figure}

\subsection{Primordial black holes}

The recent LIGO/VIRGO discovery of black hole mergers with masses of $\sim
30\,M_\odot$ \cite{Abbott:2016blz} has renewed interest in primordial
black holes (PBHs) as dark matter.   There are many constraints on PBH
dark matter, which ostensibly rule them out as being all of the dark
matter at any mass, assuming their masses are all the same
(monochromatic mass function).   Fig.\ \ref{fig:pbh-limit} from  ref.\
\cite{Carr:2016drx} shows the maximum fraction $f$ of the total dark
matter that is allowed in PBHs of a given mass.  Ref.\
\cite{Ricotti:2007au} found $f\lesssim 0.1$ for 30\,$M_\odot$ PBHs,
using 
FIRAS observations of the CMB spectral shape.  This would have been
distorted by X-rays from material accreting onto the PBHs.
Since the LIGO/VIRGO discovery, this constraint was reconsidered and shown
to be weaker in ref.\ \cite{Ali-Haimoud:2016mbv}, leaving room for
PBHs below 100\,$M_\odot$.  Complementary constraints from lack of
disruption of a star cluster in Eridanus disfavor the 30\,$M_\odot$
mass region \cite{Brandt:2016aco} but are subject to large
uncertainties.

It is sometimes said that PBHs are
a very conservative DM candidate because they require no new physics,
but this statement ignores the new physics that is probably needed to
produce PBHs with the observed relic density.  Hybrid inflation can 
produce PBHs, but in a mass range far below the LIGO/VIRGO region
\cite{GarciaBellido:1996qt}.  
A more exotic inflationary scenario
seems to be needed in order to produce
density fluctuations of sufficient power at wavelengths
$\lambda$ associated with the desired mass scale (quantified below).  
A nearly scale-invariant spectrum extending to these scales
(extrapolated from the COBE scale) has far too little power, since the
fluctuations are presumed to be Gaussian and one needs a large
amplitude $\delta\rho/\rho \sim 1$ to produce a black hole
\cite{Bringmann:2001yp}.  The simplest way to produce a spike in the
power at a given scale is to introduce a sharp step in the inflaton
potential \cite{Starobinsky:1992ts} (see fig.\ \ref{fig:step}), that
the inflaton crosses during the late stages of inflation, at the
moment when the scale $\lambda$ of interest first crosses the horizon.

\begin{figure}
\centerline{\includegraphics[width=0.85\textwidth]{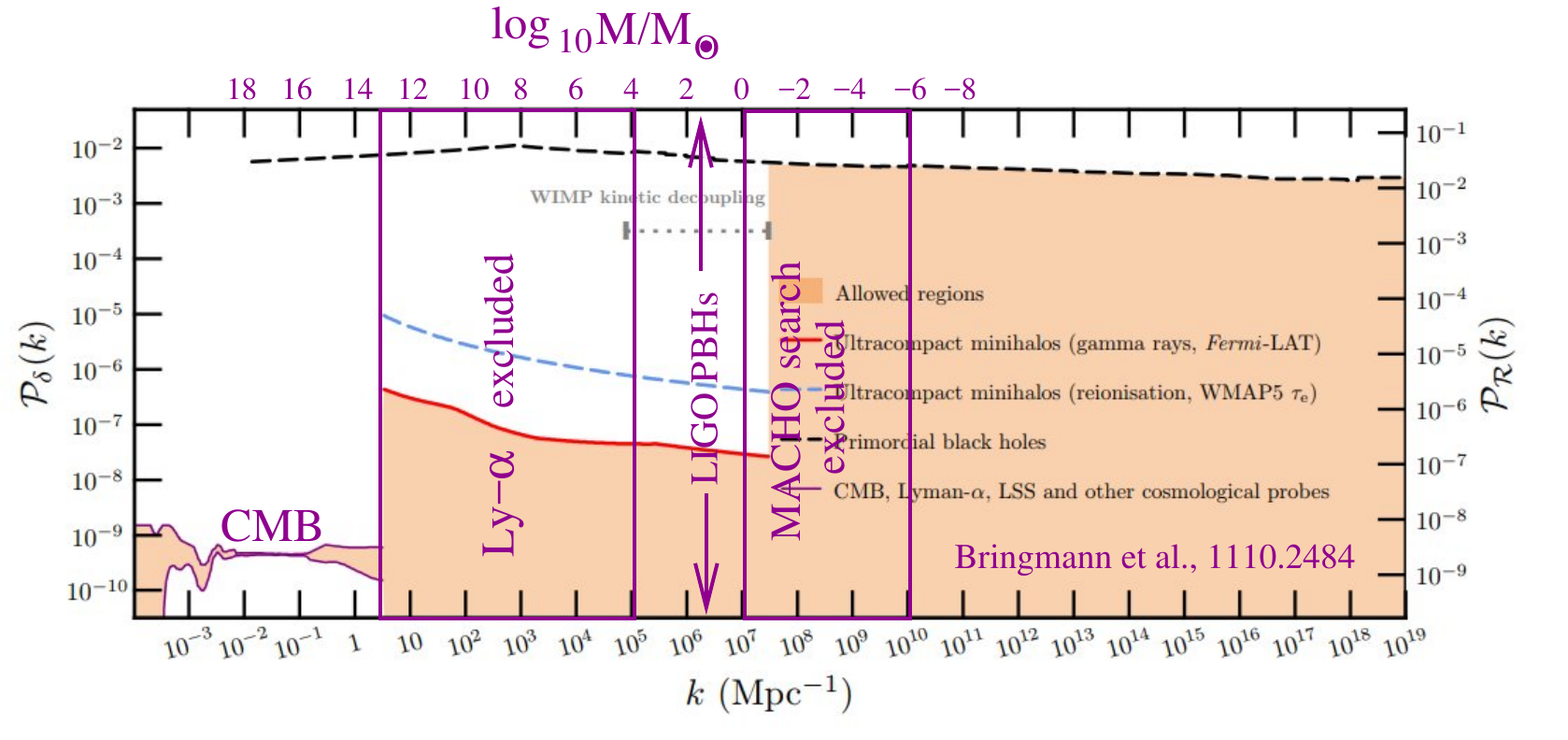}}
\caption{Constraints on power of $\delta\rho/\rho$ fluctuations versus
wavenumber, adapted from ref.\ \cite{Bringmann:2011ut}.  The corresponding black
hole mass on the top scale is estimated as $M_{\rm PBH}\sim 
\rho_{\rm crit}/k^3$.  See \cite{Carr:2016drx} for a summary of excluded
PBH mass ranges.}
\label{fig:pbh}
\end{figure}

In particular to get PBHs to be all of the DM, we need a boost of  order
$10^7$ times in the power spectrum ${\cal P}_\delta$ of  $\delta\rho/\rho$,
at the scale $k\sim 10^6\,$Mpc$^{-1}$, nearly a  million times smaller than
those probed by large scale structure,  $\sim 3$\,Mpc$^{-1}$.  The
situation is illustrated in fig.\  \ref{fig:pbh}, based on ref.\
\cite{Bringmann:2011ut}.  The dashed curve near the top shows the power
needed to produce PBHs with the right relic density, concentrated at 
a given mass scale.  (The shaded regions refer to constraints from
ultracompact minihalos, which can be ignored in the present
discussion.) I have overlaid on that figure a rough scale of
PBH masses and a few constraints neighboring the LIGO/VIRGO mass region,
 compiled by  ref.\ 
\cite{Carr:2016drx}).

The PBH mass corresponding to a fluctuation of physical wavelength
$\lambda$ is the mass contained in a horizon volume when that scale
crosses the horizon, $M_{\rm PBH} \sim \rho\lambda^3 \sim \rho/H^3$.  For most modes of
interest, horizon crossing occurs during radiation domination, and so
one must take into account the additional redshifting of density when relating
this mass to comoving scales.  The result is roughly \cite{Green:2004wb}
\be
	M_{\rm PBH} \sim 2\,M_{H,\rm eq} \left(k_{\rm eq}\over k\right)^2
	\cong 6\times 10^{13}{M_\odot \over \left(k\cdot {\rm 1\,
Mpc}\right)^2}
\label{MPBH}
\ee
where $M_{H,\rm eq} = 3.5\times 10^{17}\,M_\odot$ is the horizon mass
at equality and $k_{\rm eq} = 0.01\,$Mpc$^{-1}$ is the corresponding
wavenumber.  The power of $1/k$ is 2 instead of 3 because during
radiation domination, $\rho$ scales as $k^4$ rather than $k^3$.
Hence if the relevant scale crosses the horizon during radiation 
domination, there is an extra factor of $k$ to account for.

Fig.\ 
\ref{fig:pbh} demonstrates how fast the power in density fluctuations
has to rise as a function of wave number, to grow from its small value
well-constrained CMB region by seven orders of magnitude within six decades
of $k$.  A correspondingly sharp feature in the inflaton potential
would be needed.  
Not only should the feature be in a special location, but the
magnitude of the step (fig.\ \ref{fig:step}) must be tuned very precisely
to get the right relic density.  This is because the density fluctuations
are Gaussian, hence exponentially sensitive to the power.  This is
quantified in problem 4 below.  From the theoretical viewpoint, this makes
PBHs look like a peculiar dark matter candidate.

\subsection{Fuzzy or axion-like dark matter}

\begin{figure}
\centerline{\includegraphics[width=0.45\textwidth]{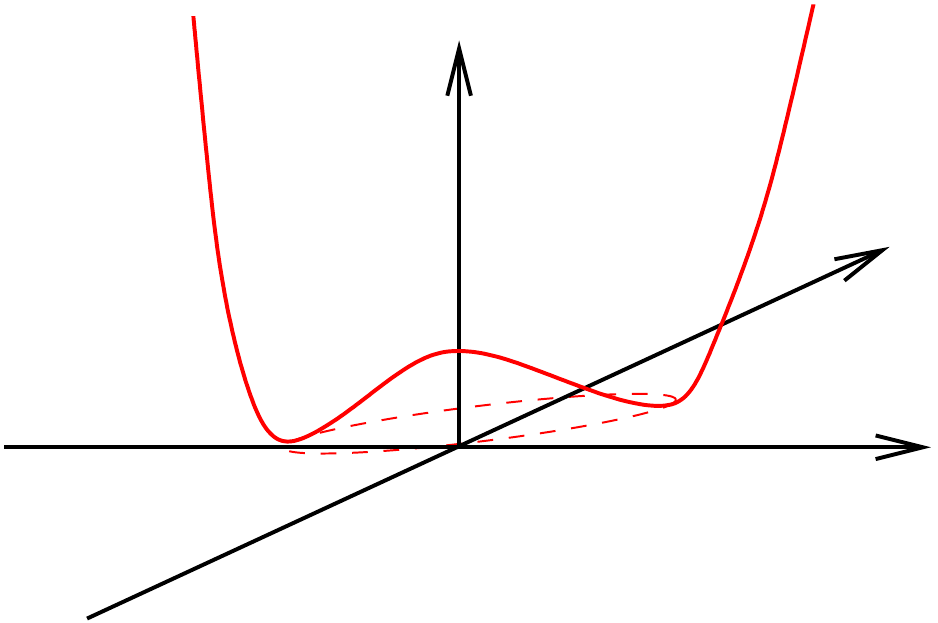}}
\caption{Potential for an axion-like DM candidate.}
\label{fig:axion}
\end{figure}

One of the earliest DM candidates, and still very popular, is a very light
scalar field, the axion \cite{Abbott:1982af,Dine:1982ah}, having a tilted wine-bottle
potential (fig.\ \ref{fig:axion}),
\be
	{\cal L} = \sfrac12 f^2 (\partial a)^2 - \Lambda^4\cos a
\ee
giving it a mass 
\be
	m_a = {\Lambda^2\over f}
\ee
Notice that we have taken $a$ to be dimensionless here so it is an angular
variable.
The basic idea (see lectures of A.\ Hook, this school)  is that the axion would have been a Goldstone boson of a
spontaneous broken symmetry, but the symmetry is explicitly broken by  
nonperturbative (instanton) effects, at a scale $\Lambda$ that might be 
suppressed by a small tunneling probability.  In the case of QCD there is
no such suppression because large instantons correspond to large running 
couplings, with small tunneling actions, and the favored axion mass range is $m_a\in [10^{-6},\,
10^{-2}]$\,eV \cite{Graham:2015ouw}.

On the other hand, string theory generically predicts many axion-like
particles whose masses are exponentially suppressed \cite{Hui:2016ltb}, and could
naturally be much lighter.  An interesting mass scale for cosmology is
$m\sim 10^{-22}\,$eV, whose corresponding de Broglie wavelength
$\lambda\sim$\,kpc coincides with the size of the central region of a Milky
Way-like galaxy.  Recall the core-cusp problem, discussed in section
\ref{bhdm}.  Such a large $\lambda$ would prevent central cusps on this
scale.  The cusp is ``fuzzed out,'' and the scenario is known as ``fuzzy 
dark matter'' \cite{Hu:2000ke}.  Despite the small mass, it is another form
of CDM since it is presumed to be too weakly coupled to have thermalized.

Axion-like particles (ALPs) can get their relic density from the {\it
misalignment mechanism}, if the potential term $\Lambda^4\cos a$ is
negligible at early times.  For example for the QCD axion, this term
is suppressed by powers of $T$ at high temperature \cite{Preskill:1982cy}.
Then during inflation, due to its quantum fluctuations, $a$ will take
a random initial value 
$a_i \in [0,\,2\pi]$, which becomes homogenous in a given
causal patch during the inflationary expansion.  As the universe cools,
eventually the $\Lambda^4\cos a$ becomes important: as soon as
$H$ falls below $m_a$, the axion starts oscillating around the
 minimum of the potential, with decaying amplitude
\be
	a(t) \sim a_i \left(R_i\over R\right)^{3/2} = 
	a_i \left(T\over T_i\right)^{3/2}
\ee
Here $R$ is the scale factor, and $R_i$ its value when $H(T_i)=m_a$.
At the corresponding temperature $T=T_i$, the energy density is of order $\rho_a\sim \Lambda^4$.  
Therefore the present density of axions is
\bea
	\Omega_a \sim {\Lambda^4\over \rho_{\rm crit}} \left(T_0\over
T_i\right)^3 &=& 0.28 \left(m_a\over 10^{-22}\,{\rm eV}\right)^{1/2}
	\left(f\over 4\times 10^{16}\,{\rm GeV}\right)^2\nn\\
&=& 0.28 \left(m_a\over 10^{-3}\,{\rm eV}\right)^{1/2}
	\left(f\over 7\times 10^{11}\,{\rm GeV}\right)^2 \,,
\eea
highlighting fiducial values corresponding to QCD-like axions and fuzzy DM,
respectively.

An important caveat is that we assumed the approximate global symmetry was
already broken at a scale above that of inflation.  If the phase
transition happens after inflation, then there can be important additional
contributions to the relic density from the formation and decay of
axion strings and domain walls.  It is difficult to reliably quantify these extra
contributions; simulations of the string network are needed.  See for
example ref.\ \cite{Klaer:2017ond}.

\subsection{Self-interacting dark matter (SIDM)}
It doesn't exactly fit into my general scheme of classifying DM models by
their production mechanisms, but it is important to mention the possibility
that dark matter may have strong self-interactions \cite{Spergel:1999mh},
since this has become a popular approach to solving the small-scale
structure problems of CDM that were discussed around eq.\ (\ref{NFW}); see \cite{Tulin:2017ara} for a review.

\begin{figure}
\centerline{\includegraphics[width=0.25\textwidth]{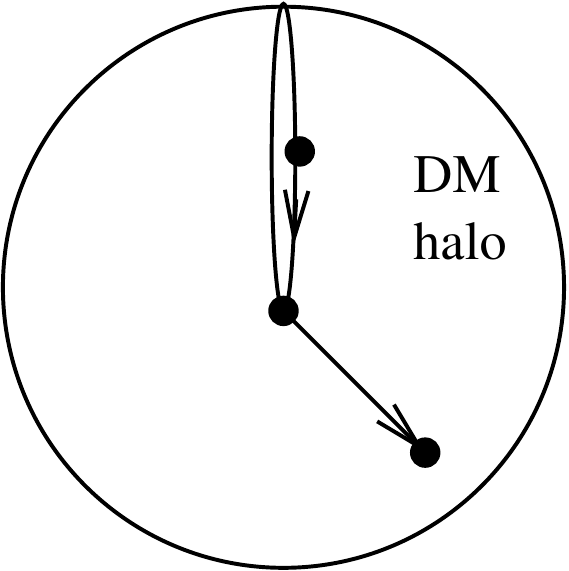}}
\caption{Coring of a cuspy halo by DM self-interactions.}
\label{fig:sidm}
\end{figure}

SIDM can produce cored DM profiles as illustrated in fig.\ \ref{fig:sidm}.
An energetic DM particle in an elliptical orbit may scatter with a
DM particle in the central region, giving it energy and allowing it to move
out of the central region.  Velocities are higher in the outer regions
\cite{1985ApJS...58...39B} so this provides a way of transferring energy
to the lower-velocity particles that would otherwise be trapped near the
center.  This is confirmed by $N$-body simulations incorporating
SIDM \cite{Vogelsberger:2012ku,Rocha:2012jg}.  Cross sections of order
\be
	{\sigma\over m}\sim 0.1\, {{\rm cm}^2\over {\rm g}} = 
	0.17\,{{\rm barn}\over {\rm GeV}}
\ee
are found to solve the cusp-core and other small-scale structure problems,
while being compatible with the Bullet Cluster constraint
\cite{Randall:2007ph}.  (For reference, nucleons in the real world have 
$\sigma/m\sim 20$\,b/GeV.)  
The combination $\sigma/m$ is relevant because the
scattering rate is
\be
	\Gamma = n\sigma v = \rho {\sigma\over m} v
\ee
and $\rho$ is fixed by $\Omega_{\rm CDM}$.

It is natural to expect self-interactions if the DM is part of a larger
hidden sector \cite{ArkaniHamed:2008qn}.  For example, DM could be charged
under a hidden U(1) gauge symmetry which could lead to strong,
velocity-dependent scattering.  DM could be in the form of dark atoms
in such a scenario \cite{Kaplan:2009de}, for which the 
geometric self-interaction
cross section can easily be very large \cite{Cline:2013pca}.

In one class of models, the self-interactions can be important for
determining 
the relic density: Strongly Interacting Massive Particle (SIMP) DM is 
a scenario where $\chi\chi\to f\bar f$ is absent or subdominant, and
instead one has strong $\chi\chi\chi\to \chi\chi$ annihilations, taking
the DM to be a scalar \cite{Hochberg:2014dra}.  The prototypical example is pions in a dark sector,
annihilating through a Wess-Zumino-Witten interaction 
\cite{Wess:1971yu,Witten:1983tw}
\be
	{\cal L}\sim {1\over f^5}\epsilon_{\mu\nu\alpha\beta}
	{\rm tr}(\pi\partial^\mu\pi
\partial^\nu\pi\partial^\alpha\pi\partial^\beta\pi)
\ee
in the notation of chiral perturbation theory \cite{Georgi:1985kw}.

We can repeat the order-of-magnitude estimate for the relic density
for $3\to 2$ annihilations.  The rate of annihilation per DM particle
 must now depend on the density squared,
\be
	\Gamma n^2\langle\sigma_{3\to 2} v\rangle \ =\  H(m_\chi) \ \sim\  {m_\chi^2\over M_p}
\ee
leading to 
\be
	n \sim {m_\chi\over \sqrt{M_p \langle\sigma_{3\to 2} v\rangle}}
\ee
Notice that $\sigma_{3\to 2}$ has different dimensions than a $2\to 2$
scattering cross section.  To define its thermal average, one should
examine the collision term in the Boltzmann equation; see for example
\cite{Kuflik:2017iqs}.  If for simplicity we imagine there is only one
dimensionful scale in the dark sector, so that 
$\sigma_{3\to 2} \sim 1/m_\chi^5$,  we find that $m_\chi \sim 1\,$GeV
is needed to achieve the observed relic density; see problem 7 below.

\subsection{Exercises}

1.  {\bf Annihilation kinematics.}  In the center of mass frame for
two annihilating DM particles, the velocities are $\pm p/E$ along
some axis.  Show that the relative velocity is $2\sqrt{1 - 4 m^2/s}$.
Generalize this to the case where the annihilating particles have
unequal masses, which would be relevant for coannihilating DM 
scenarios.   (For more about coannihilating DM, 
see ref.\ \cite{Griest:1990kh}.)

\bigskip

2. {\bf Thermal averaging.}  Suppose the DM annihilation cross section
has a resonant enhancement, so that (ignoring factors of order 1)
\[
	\sigma   = {\lambda^4 m_\chi^2\over
	\sqrt{1-4\, m_\chi^2/s}\left[(s-m_\phi^2)^2 + m_\phi^2\Gamma_\phi^2
	\right]}	
\]
where $\phi$ is the particle exchanged in the $s$ channel.  In the
narrow-width approximation for the resonance, you can treat the
Breit-Wigner factor as a representation of the delta function.
Use this approximation to evaluate the Gondolo-Gelmini thermally
averaged $\sigma v$.  Then take $4m_\chi^2 = m_\phi^2 - \epsilon^2$,
where $\epsilon \ll m_\chi$, and suppose that $T = m_\chi/30$ 
(which could represent the freezeout temperature) for
evaluating the $T$-dependent functions.   How does the
result compare to the naive procedure of evaluating $\sigma v$ at
$s = 4 m_\chi^2 = m_\phi^2$?

\bigskip

3. {\bf Freeze-in from above.}  Suppose that the DM particle $\chi$
was never in thermal equilibrium, but its initial abundance $Y_i$ (at
some initial $x_i$) was $\gg Y_{\rm eq}$.  If $Y(x)$ remains $\gg
Y_{\rm eq}$ throughout the evolution, we can solve the  Boltzmann
equation ignoring the $Y_{\rm eq}$ term.\\
(a)  Carry this out, and show
that in the regime where $Y$ changes by a large factor, its final
abundance is independent of $Y_i$.\\
(b)  Show that the time scale for $Y$
to reach its present value is typically much shorter than the
time scale governing $Y_{\rm eq}$.  Use this observation to derive 
a bound on the relevant parameters,  from the requirement that
$Y\gg Y_{\rm eq}$ at all times.\\
(c) A similar combination of parameters determines the relic density.
If this matches observations, what constraint must $m_\chi$ satisfy
in order for this version of freeze-in to work?

\bigskip

4. {\bf Primordial black holes.}  We will try to roughly estimate the
probability of producing a PBH of a given mass from inflationary 
density perturbations.  See astro-ph/0109404 for details.\\
(a)  The mean-squared relative mass fluctuation $\sigma^2 = \delta
M^2/M^2$ in a region of size $R$ can be obtained
from the power spectrum $P_\delta(k)$ for the density perturbation
$\delta\rho /\rho$ using the formula
\[
	\sigma^2_R = \left\langle\left(\delta M^2\over M^2\right)_R
\right\rangle = \left\langle\left(\int d^{\,3}r\, W_R(r) 
	\delta\rho(\vec x+\vec r)\over
	\rho \int d^{\,3}r\, W_R(r)\right)^2\right\rangle
\]
where $V$ is some large fiducial volume averaged over, 
$\rho$ is the mean density, and $W_R$
is a window function that selects a region of size $r\lesssim R$, for example
the top-hat $\Theta(R-r)$ or the Gaussian $\exp(-r^2/2R^2)$.  We will
take the Gaussian for simplicity.  By Fourier transforming everything
and using the definition of the power spectrum
\[
	\left\langle {\delta\rho_k\over\rho} {\delta\rho_{k'}\over
\rho}\right\rangle 
	=  \delta(\vec k + \vec k') {P_{\delta}(k)\over k^3}
\]
(corresponding to $\delta \rho_k$ being a Gaussian random variable),
show that 
\[
\sigma^2_R = {1\over 2\pi^2}\int {dk\over k}
	e^{-k^2 R^2} P_{\delta}(k)
\]
Here $P_\delta$ is normalized so that a scale-invariant spectrum 
would have $P_\delta$ constant. However remember that $P_\delta$ has an extra
power of $k^4$ since $\delta\rho_k/\rho\sim k^2{\cal R}_k$ (${\cal R}$
is the 3D curvature invariant, whose power is nearly scale-invariant).  For the present crude estimate, take
\[
	P_\delta(k) \sim A_s \left(k\over aH\right)^4 
\]
where $A_s \sim 10^{-9}$ is the amplitude of the scalar power
spectrum (in the region of $k$ measured by the CMB), and $k/aH$ is the physical wave number at the moment when
the comoving scale $k$ re-entered the horizon.\\
(b)  The relative mass fluctuation $\delta_M = (\delta M^2/M^2)_R$
in a region of size $R$ is also a Gaussian random variable, whose
variance we just estimated in part (a).  To form a black hole, we
need a rare fluctuation from the tail of the distribution such that
the fluctuation is large, $\delta_M\sim 1$.  Show that the probability
to have a fluctuation in the interval $\delta \in [\delta_1,\delta_2]$
is dominated by the lower limit,
\[
	P \sim {\sigma_R\over \sqrt{2\pi}\,\delta_1}
	e^{-\delta_1^2/(2\sigma_R^2)}
\]
If $\delta_1\sim 1$, a black
hole will form, whose mass is of order the total mass contained in the
region of size $R$, namely $M\sim \rho R^3$.  If $R$ is taken to
be a comoving
scale, then we can evaluate this today, taking $k\to 1/R$ in eq.\
(\ref{MPBH}).

 Assuming the scalar power is really
scale-invariant, with the COBE normalization, estimate the probability
to form a black hole of mass $\sim 30\,M_\odot$.\\
(c)  To increase the probability, one needs to assume that the
scalar power spectrum is much larger at the wave numbers $k\sim 1/R$
of interest.  How large must it be to get a probability of order 1
to form a $\sim 30\,M_\odot$ black hole?\\
(d) In problem 5 of set 1, you estimated the boost in power coming from
a step in the inflaton potential.  Translate your result from part (c)
to estimate how large the step in the potential must be.

\bigskip

{5. \bf Misalignment mechanism.}  Derive the result for the relic density
of an axion-like particle given in the lectures,
\[
	\Omega_a \sim 0.3 \left(m\over 10^{-22}\,{\rm eV}\right)^{1/2}
	\left(f\over 4\times 10^{16}\,{\rm GeV}\right)^2
\]

\bigskip

{6. \bf Self-interacting dark matter.}  Suppose dark matter is
a scalar particle with self-interaction $\lambda\phi^4$ and
$\lambda \sim 1$.  Find the mass that corresponds to a cross section
such that $\sigma/m \sim 0.1$\,cm$^2$/g.

 \bigskip

{7. \bf SIMP dark matter.}  Consider scalar DM with mass $m$ and 
a $3\to 2$ cross section of order $\sigma v \sim 1/m^5$.  Estimate
the mass that gives the right relic density.

\bigskip
{\bf Acknowledgments.}  I thank the students of TASI for perceptive
comments and questions that allowed me to improve these lectures.
Thanks to the following people for helpful
discussions or correspondence: H.\ Baer, R.\ Brandenberger, T.\ Bringmann,  
T.\ Degrand, J.\ Feng, W.\ Fischler, J.\ Garcia-Bellido, K.\ Kainulainen, R.\ Kolb, A.\ Liddle, D.\ Lyth, G.\ Moore,  
P.\ Scott, X.\ Tata. I especially thank M.\ Puel for pointing out
numerous
mistakes and typos.   This work was supported by the Natural Sciences and
Engineering Research Council (NSERC) of Canada.

\begin{appendix}
\section{TASI---recommended road bike rides}
A useful bike map for the Boulder area can be found  
\href{https://assets.bouldercounty.org/wp-content/uploads/2017/03/bike-map-2017.pdf}{here}.

The Boulder Creek bike path heading west is a good warm-up, though not 
going very far before becoming hard-packed dirt and then ending.  At this
writing, Fourmile Canyon Road  is closed to cyclists due to
reconstruction.  Boulder Canyon Drive is not recommended, having heavy
traffic, no shoulder, and a tunnel.  A longer ride can be found by taking
the sign for Settler's Park and heading toward 4th St.,  a designated
bikeway.  It ends at Linden Dr., which provides a 7.3\% climb for
masochists, with little reward on the descent since much braking is
required.

For a longer ride, Hygiene is a popular destination.  Avoid  Highway 119
even though the shoulder is wide enough; too much traffic. Boulder Creek
path to Pearl, then 61st is much better.  A fine loop that includes Hygiene
is along US 36, turning at Hygiene Road and returning to Boulder on
65th/63rd. The route north offers interesting possibilities for variations
that require much more climbing, mainly Olde Stage Road, with a quite
gratifying descent along Lefthand Canyon Drive, back to US 36 near
Altona.  

A more ambitious variation is Lee Hill Drive, connecting to Lefthand Canyon Drive.
The latter is recommended for a longer foray into the mountains as it is quite smooth
and has little traffic.  (At this writing, James Canyon Dr.\ is still damaged by the
flooding and becomes dirt at some point.)  
From the dorms, take Folsom/26th to Tamarack, Spotswood, 22nd,
Upland, 19th, and catch the bike trail after crossing Violet; it brings you to Lee Hill
Dr.

The climb up to NCAR is quite scenic, though not very long.  I followed it
up  by continuing south out of town on the Broadway trail, which turns into
Marshall Road, eventually turning into C170, a smooth highway with a good
shoulder and not too much traffic.   When C170 reaches US 36, you can cross
underneath it and return via the Turnpike bike path, giving a great view of
the mountains and very speedy descent.  At the bottom of the descent,
instead of crossing back underneath US 36 I turned right and found my way
to the South Boulder Creek bike path (heading north), which was worth the
trouble.  It joins up with the usual Boulder Creek path heading back toward
CU.  Get off at Folsom for a direct path to the dorms.

\end{appendix}

\bibliographystyle{JHEP}
\bibliography{tasi-bib}
\end{document}